\documentclass[11pt]{article}
\pdfoutput=1

\usepackage{latexsym}
\usepackage{chapterbib}
\usepackage{graphics,graphicx}
\usepackage{amssymb,amsmath,amsfonts,amstext}
\usepackage{hyperref}
\usepackage[bbgreekl]{mathbbol}
\usepackage{color,xcolor}
\usepackage{geometry}  
\usepackage{braket}
\usepackage{cancel}
\usepackage[utf8]{inputenc}

\usepackage{datetime}
\usepackage{slashed}
\usepackage{tikz}

\allowdisplaybreaks 

\def\){\right)}
\def\({\left( }
\def\]{\right] }
\def\[{\left[ }

\def\NO{\nonumber}

\newcommand{\be}{\begin{equation}}
\newcommand{\ee}{\end{equation}}

\def\bea{\begin{eqnarray}}
\def\eea{\end{eqnarray}}

\def\bal#1\eal{\begin{align}#1\end{align}}

\def\bald{\begin{aligned}}
\def\eald{\end{aligned}}

\def\bsub{\begin{subequations}}
\def\esub{\end{subequations}}

\def\beqx{\begin{displaymath}}
\def\eeqx{\end{displaymath}}

\newcommand{\bmat}{\left(\begin{array}}
\newcommand{\emat}{\end{array}\right)}

\def\a{\alpha}
\def\b{\beta}
\def\c{\chi}
\def\d{\delta}
\def\e{\epsilon}

\def\g{\gamma}
\def\h{\eta}
\def\j{\psi}
\def\k{\kappa}
\def\l{\lambda}
\def\m{\mu}
\def\n{\nu}
\def\o{\omega}
    
\def\p{\pi}

    \def\th{\theta}
\def\r{\rho}
\def\s{\sigma}
\def\t{\tau}
\def\x{\xi}
\def\z{\zeta}

\def\F{\Phi}
\def\G{\Gamma}
\def\J{\Psi}
\def\L{\Lambda}
\def\O{\Omega}
    
\def\P{\Pi}

\def\S{\Sigma}

\def\X{\Xi}

\def\ve{\varepsilon}

\def\ca{{\cal A}}
\def\cb{{\cal B}}
\def\cc{{\cal C}}
\def\cd{{\cal D}}
\def\ce{{\cal E}}
\def\cf{{\cal F}}

\def\ch{{\cal H}}

\def\cj{{\cal J}}
\def\ck{{\cal K}}
\def\cl{{\cal L}}
\def\cm{{\cal M}}
\def\cn{{\cal N}}
\def\co{{\cal O}}
\def\cp{{\cal P}}
\def\cq{{\cal Q}}

\def\cs{{\cal S}}
\def\ct{{\cal T}}

\def\cv{{\cal V}}
\def\cw{{\cal W}}
\def\cx{{\cal X}}

\def\bb#1{\ensuremath{\mathbb{#1}}} 

\def\bo{{\raise-.3ex\hbox{\large$\Box$}}}               
\def\pa{\partial}                                       
\def\face{{\raise.2ex\hbox{$\displaystyle \bigodot$}\mskip-2.2mu \llap {$\ddot
        \smile$}}}                                   
\def\>{\rangle}                                      
\def\<{\langle}                                      

\def\tx#1{\text{#1}}
\def\sbtx#1{{}_{\rm #1}}                           
\newcommand{\sub}[1]{\phantom{}_{(#1)}\phantom{}}    
\def\wt#1{\widetilde{#1}}                            
\def\Hat#1{\widehat{#1}}                             
\def\lbar#1{\ensuremath{\overline{#1}}}              
\def\leftrightarrowfill{$\mathsurround=0pt \mathord\leftarrow \mkern-6mu
        \cleaders\hbox{$\mkern-2mu \mathord- \mkern-2mu$}\hfill
        \mkern-6mu \mathord\rightarrow$}        
\def\dvec#1{\vbox{\ialign{##\crcr
        \leftrightarrowfill\crcr\noalign{\kern-1pt\nointerlineskip}
        $\hfil\displaystyle{#1}\hfil$\crcr}}}           
\def\diag{{\rm diag \,}}                                

\def\-{\hphantom{-}}

\def\bala#1\eala{\begin{align*}#1\end{align*}}

\begin{document}

\begin{titlepage}

\pagestyle{empty}

\begin{flushright}
 {\small SISSA 12/2017/FISI}
\end{flushright}
\vskip1.5in

\begin{center}
\textbf{\LARGE Supercurrent anomalies in 4d SCFTs}
\end{center}
\vskip0.2in

\begin{center}    
{\large Ioannis Papadimitriou\footnote{\href{mailto: ipapadim@sissa.it}{\tt ioannis.papadimitriou@sissa.it}}}
\end{center}
\vskip0.2in

\begin{center}
{\small SISSA and INFN - Sezione di Trieste, Via Bonomea 265,\\
34136 Trieste, Italy}
\end{center}
\vskip0.2in

\begin{abstract}
	
We use holographic renormalization of minimal $\cn=2$ gauged supergravity in order to derive the general form of the quantum Ward identities for 3d $\cn=2$ and 4d $\cn=1$ superconformal theories on general curved backgrounds, including an arbitrary fermionic source for the supercurrent. The Ward identities for 4d $\cn=1$ theories contain both bosonic and fermionic global anomalies, which we determine explicitly up to quadratic order in the supercurrent source. The Ward identities we derive apply to any superconformal theory, independently of whether it admits a holographic dual, except for the specific values of the $a$ and $c$ anomaly coefficients, which are equal due to our starting point of a two-derivative bulk supergravity theory. We show that the fermionic anomalies lead to an anomalous transformation of the supercurrent under rigid supersymmetry on backgrounds admitting Killing spinors, even if all superconformal anomalies are numerically zero on such backgrounds. The anomalous transformation of the supercurrent under rigid supersymmetry leads to an obstruction to the $Q$-exactness of the stress tensor in supersymmetric vacua, and may have implications for the applicability of localization techniques. We use this obstruction to the $Q$-exactness of the stress tensor, together with the Ward identities, in order to determine the general form of the stress tensor and $R$-current one-point functions in supersymmetric vacua, which allows us to obtain general expressions for the supersymmetric Casimir charges and partition function.

\end{abstract}

\end{titlepage}

\tableofcontents
\addtocontents{toc}{\protect\setcounter{tocdepth}{3}}
\renewcommand{\theequation}{\arabic{section}.\arabic{equation}}

\section{Introduction}
\setcounter{equation}{0}

Supersymmetric quantum field theories on curved backgrounds admitting a notion of rigid supersymmetry have attracted considerable interest in recent years, mainly due to localization techniques. Such techniques utilize the $Q$-exactness of certain operators on compact manifolds admitting Killing spinors, and can be used to compute various observables exactly, for any value of the couplings \cite{Pestun:2007rz}. 

This has motivated an extensive study of field theory backgrounds that support rigid supersymmetry, in various spacetime dimensions \cite{Festuccia:2011ws,Samtleben:2012gy,Klare:2012gn,Dumitrescu:2012ha,Liu:2012bi,Dumitrescu:2012at,Kehagias:2012fh,Closset:2012ru,Samtleben:2012ua,Cassani:2012ri,deMedeiros:2012sb,Hristov:2013spa} (see also \cite{Blau:2000xg,Kuzenko:2012vd} for earlier work). Following \cite{Festuccia:2011ws}, the approach to constructing a theory possessing rigid supersymmetry on curved backgrounds involves starting from a version of supergravity and sending Newton's constant to infinity, so that gravity becomes non-dynamical. Accordingly, somewhat different versions of rigid supersymmetry on curved backgrounds exist, depending on whether one starts from old minimal \cite{Festuccia:2011ws,Samtleben:2012gy,Liu:2012bi,Dumitrescu:2012at}, new minimal \cite{Klare:2012gn,Dumitrescu:2012ha,Cassani:2012ri}, or conformal supergravity \cite{Klare:2012gn,Cassani:2012ri}. 

All these approaches lead to consistent theories with rigid supersymmetry, and determine the {\em classical} Lagrangian, as well as the supersymmetry transformations of the classical fields. Moreover, one can apply the Noether procedure to this Lagrangian in order to derive the {\em classical} Ward identities that local gauge-invariant operators satisfy, reflecting the global symmetries of the theory. However, to obtain the {\em quantum} Ward identities one needs to compute the appropriate path integral on a general curved background in order to determine the global {\em anomalies}. Global anomalies for $\cn=1$ supersymmetric theories in $d=4$ have been classified using superspace cohomology arguments in \cite{Bonora:1984pn}, and have been computed explicitly in a number of examples, both in superspace \cite{McArthur:1983fk,Buchbinder:1986im}, and in component language \cite{Anselmi:1997am}.

However, both the rigid supersymmetry transformations on a curved background and the general {\em form} of the Ward identities, including the quantum anomalies, can also be determined through holographic techniques, as was done for gauge transformations and the axial anomaly in \cite{Witten:1998qj}, or for Weyl transformations and the trace anomaly in \cite{Henningson:1998gx}. In particular, starting with a suitable gauged supergravity in asymptotically AdS$_{d+1}$ space, one can derive the action of rigid supersymmetry on the sources of local gauge-invariant operators on the $d$-dimensional conformal boundary, as well as the {\em quantum} Ward identities these operators satisfy. The crucial point here is that the {\em form} of the Ward identities and of the global anomalies is determined by symmetries and is therefore universal. In particular, the same form of the Ward identities applies to a wider class of supersymmetric theories, which need not admit a holographic dual. 

Of course, the holographic calculation does not determine the elementary Lagrangian on any curved background, except possibly for terms that are protected by non-renormalization theorems. Moreover, even when specifying the theory in terms of gauge-invariant operators and their sources, there are two possible caveats in deriving the rigid supersymmetry transformations and the quantum Ward identities using holographic techniques. The first is that the version of rigid supersymmetry induced holographically on the conformal boundary by bulk (minimal) gauged supergravity corresponds specifically to that obtained from {\em conformal} supergravity \cite{Sohnius:1981tp} on the boundary through the Festuccia-Seiberg argument \cite{Klare:2012gn,Cassani:2012ri}\footnote{Although the transformations of the background fields induced holographically on the boundary do indeed correspond to those of conformal supergravity, the claim that the current multiplet induced holographically on the boundary is the conformal multiplet may not be quite correct. In fact, a simple counting of off-shell degrees of freedom for the holographic sources on the boundary shows that fermionic and bosonic degrees of freedom do not much. This suggests that the current multiplet induced holographically on the boundary should be compared with the standard current multiplets with the auxiliary fields integrated out, in which case they all coincide.}. However, this reflects the fact that the holographic Ward identities are expressed in terms of the operators at the ultraviolet fixed point, and apply to both conformal and massive theories. The second is that the holographic calculation leads to specific values for the anomaly coefficients, which generically do not apply to theories without a holographic dual, or even to holographic theories at weak coupling. For instance, starting from two-derivative supergravity in AdS$_5$ necessarily leads to $a=c$ in the resulting Ward identities on the conformal boundary \cite{Henningson:1998gx}. However, starting from a higher derivative supergravity theory in the bulk may lift this degeneracy.      
        
In this paper we focus on $\cn=2$ superconformal theories in $d=3$ and $\cn=1$ superconformal theories in $d=4$. Starting with minimal $\cn=2$ gauged supergravity in four and five dimensions respectively, we use holography to derive the general form of the quantum superconformal Ward identities on an arbitrary curved background, including an arbitrary fermionic source for the supercurrent. A complementary analysis for  $\cn=2$ gauged supergravity coupled to hypermultiplets is carried out in \cite{Song}. The minimal $\cn=2$ gauged supergravity describes holographically the current multiplet, consisting of the stress tensor, $\ct^{i}_a$, the $R$-symmetry current, $\cj^i$, and the supercurrent $\cs^i$. Their respective local sources are the vielbein $e^a_i\sub{0}$, a $U(1)$ gauge field $A_{(0)i}$, and a chiral gravitino $\J_{(0)+i}$, whose values specify the field theory background. Notice that in the presence of fermionic sources it is necessary to introduce the vielbein as a fundamental source, rather than the metric $g_{(0)ij}$. In the absence of fermion sources, however, one may work exclusively with the metric $g_{(0)ij}$ and its conjugate symmetric stress tensor $\ct^{ij}$.  
The superconformal Ward identities we obtain holographically are given in eq.~\eqref{WIDs}. For $\cn=1$ superconformal theories in $d=4$ they are\footnote{An earlier attempt at holographically deriving the supertrace Ward identity starting from minimal $\cn=2$ gauged supergravity in five dimensions was made in \cite{Chaichian:2003kr}. However, both the term involving the $R$-symmetry current and the contribution of the Ricci curvature to the supertrace anomaly were missed in that analysis.}   
\bal
&\hskip-0.1cm D_{j}(e^{a}_{(0)i}\ct_a^j-\lbar\cs^j\J_{(0)+i}-\lbar\J_{(0)+i}\cs^j)+\lbar\cs^j\cd_{i}\J_{(0)+j}+\lbar\J_{(0)+j}\overleftarrow{\cd }_{\hskip-0.08cmi}\cs^j+F_{(0)ij}\cj^j=\ca_{Mi}, \hskip-0.0cm\NO\\
&\hskip-0.1cm D_{i}\cj^i +i\sqrt{3}(\lbar\cs^i\J_{(0)+i}-\lbar\J_{(0)+i}\cs^i)=\ca_{R},\NO\\
&\hskip-0.1cm \cd_{i}\cs^i+\frac12\ct^i_a\G^a\J_{(0)+i}-\frac{i}{8\sqrt{3}}\cj^i(\G_{ij}-2g_{(0)ij})\G^{jpq}\cd_{p}\J_{(0)+q}=\ca_{S},\NO\\
&\hskip-0.1cm e_{(0)i}^a\ct_a^i-\frac12\lbar\J_{(0)+i}\cs^i-\frac12\lbar\cs^i\J_{(0)+i}=\ca_{W},\NO\\
&\hskip-0.1cm \G_{i}\cs^i-\frac{i\sqrt{3}}{4}\cj^i\J_{(0)+i}=\ca_{sW},\NO\\
&\hskip-0.1cm e^{i[a}_{(0)}\ct^{b]}_i+\frac14(\lbar\cs^i\G^{ab}\J_{(0)+i}-\lbar\J_{(0)+i}\G^{ab}\cs^i)=0,
\eal
where the local functions of the sources $\ca_{Mi}$, $\ca_R$, $\ca_S$, $\ca_W$ and $\ca_{sW}$ are given in eq.~\eqref{anomalies} and are related to the global anomalies of the theory. We have slightly simplified the notation here for the sake of brevity, but we refer to \eqref{WIDs} and appendix \ref{conventions} for the precise form of the Ward identities and a guide to our notation, respectively. For $\cn=2$ superconformal theories in $d=3$ the Ward identities are of the same form, except that the local terms $\ca_{Mi}$, $\ca_R$, $\ca_S$, $\ca_W$ and $\ca_{sW}$ are absent.

For a generic $\cn=1$ superconformal theory in four dimensions, the Weyl and $R$-current anomalies take the form
\be
\ca_W\sim c\cw^2-a\ce,\qquad \ca_R \sim (c-a)P[g\sub{0}]+(5a-3c)\e^{ijkl}F_{(0)ij}F_{(0)kl},
\ee
where 
\bal
\ce=&\;R^{ijkl}[g\sub{0}]R_{ijkl}[g\sub{0}]-4R^{ij}[g\sub{0}]R_{ij}[g\sub{0}]+R^2[g\sub{0}],\NO\\
\cw^2=&\;R^{ijkl}[g\sub{0}]R_{ijkl}[g\sub{0}]-2R^{ij}[g\sub{0}]R_{ij}[g\sub{0}]+\frac13R^2[g\sub{0}]-8F^{ij}_{(0)}F_{(0)ij}+\co(\J^2_{(0)}),
\eal
are respectively the Euler density and the {\em supersymmetrized} Weyl squared conformal invariant, while $P[g\sub{0}]$ is the topological Pontryagin density. In order to specify the numerical factors, besides the $a$ and $c$ anomaly coefficients, one would need to appropriately normalize the gauge field $A_{(0)i}$. The precise factors for the correctly normalized gauge field can be found e.g. in \cite{Cassani:2013dba}, which also corrected a sign in the original expressions for the anomalies in \cite{Anselmi:1997am}. As anticipated, our computation reproduces these anomalies but with $a=c$. In particular, we obtain the full superconformal invariant $\cw^2$, up to quadratic order in the supercurrent source.

The terms $\ca_{Mi}$ and $\ca_S$ that appear respectively in the diffeomorphism and supersymmetry Ward identities are related to the $R$-symmetry anomaly $\ca_R$ and take the form 
\be
\ca_{Mi} \sim \e^{jklp}F_{(0)ij}F_{(0)kl}A_{(0)p},\qquad 
\ca_{S} \sim \e^{iskl}F_{(0)sk}A_{(0)l}(\G_{ij}-2g_{(0)ij})\G^{jpq}\cd_p\J_{+q}.
\ee
The appearance of these terms in the Ward identities simply reflects the fact that the $R$-current operator $\cj^i$ is the {\em consistent} current (see e.g. \cite{Jensen:2012kj} for a recent overview). Writing the Ward identities in terms of the {\em covariant} (and gauge invariant) current $\cj\sbtx{cov}^i$, these terms get eliminated. In section \ref{dict} we confirm that the generating functional is invariant under local diffeomorphisms\footnote{I am grateful to Davide Cassani for pointing this out to me.} and so indeed $\ca_{Mi}$ does not represent an anomaly in the diffeomorphism invariance of the theory. In contrast, the term $\ca_S$ does lead to a non-invariance of the generating functional under local supersymmetry transformations, which will play a crucial role throughout our analysis. Finally, the terms $\ca_W$ and $\ca_{sW}$ represent anomalies under local Weyl and local superWeyl transformations, respectively. 

Our main result concerns the consequences of the two fermionic anomalies, namely the supersymmetry anomaly $\ca_S$ and the superWeyl anomaly $\ca_{sW}$, on the transformation of the supercurrent under rigid supersymmetry. In particular, we show that on backgrounds admitting a (conformal) Killing spinor $\z_+$, which satisfies the Killing spinor equation \eqref{Killing}, the anticommutator of the corresponding supercharge with the supercurrent is generically anomalous, namely 
\bal\label{supercurrent-anticommutator}
\{\lbar\cq[\z],\cs^i\} =-\frac12\ct^{ij}\G_{j}\z_{+}+\frac{i}{8\sqrt{3}}\G^{ijk}\(\G_{kl}-2g_{(0)kl}\)\z_+ D_{j}\cj^l+\frac{i}{2\sqrt{3}}\(\G^i_{l}-3\d^i_l\)\z_-\cj^l+\ca_\z^i.
\eal
The supercurrent anomaly $\ca^i_\z$, given explicitly in \eqref{supercurrent-var}, is related to the supersymmetry and superWeyl anomalies as
\be
\ca_\z^i\sim\frac{\d}{\d\lbar\J_{(0)+i}}\int d^4x\(\bar\z_+\ca_{S}+\lbar\z_-\ca_{sW}+\text{h.c.}\).
\ee
Moreover, we show through explicit examples that there are backgrounds admitting Killing spinors where all anomalies in the superconformal Ward identities are numerically zero, yet the supercurrent anomaly $\ca_\z^i$ is non-zero. In those cases, the supercurrent anomaly poses an obstruction to the $Q$-exactness of the operator     
\be\label{Q-exact-intro}
-\frac12\ct^{ij}\G_{j}\z_{+}+\frac{i}{8\sqrt{3}}\G^{ijk}\(\G_{kl}-2g_{(0)kl}\)\z_+ D_{j}\cj^l+\frac{i}{2\sqrt{3}}\(\G^i_{l}-3\d^i_l\)\z_-\cj^l.
\ee

It may be useful to point out that the anomalous transformation \eqref{supercurrent-anticommutator} of the supercurrent under rigid supersymmetry is reminiscent of the anticommutator of two fermionic generators of an $\cn=1$ superVirasoro algebra in two dimensions, namely
\bal
[L_m,L_n]=&\;(m-n)L_{m+n}+\frac{c}{12}m(m^2-1)\d_{m+n,0},\NO\\
[L_m,G_r]=&\; \frac12(m-2r)G_{m+r},\NO\\
\{G_r,G_s\}=&\; 2L_{r+s}+\frac{c}{12}(4r^2-1)\d_{r+s,0}.
\eal
Although in two dimensions the anomalous term in the transformation of the supercurrent survives in flat space, leading to a central extension of the superVirasoro algebra, in four dimensions the anomalous term is manifest only in curved space. However, even in four dimensions, the consequences of this anomalous transformation of the supercurrent should be visible in flat space correlation functions.    

We show that the anomalous transformation \eqref{supercurrent-anticommutator} of the supercurrent under rigid supersymmetry provides a resolution to two paradoxes discussed recently in \cite{Genolini:2016ecx}. Firstly, considering the class of supersymmetric backgrounds in eq.~\eqref{susy-background}, which were originally obtained in \cite{Klare:2012gn,Dumitrescu:2012ha}, we revisit the argument of  \cite{Closset:2013vra,Closset:2014uda,Assel:2014paa}, according to which the supersymmetric partition function on this class of backgrounds is independent of the complex functions $u(z,\bar z)$ and $w(z, \bar z)$. We show that the $u$ and $w$ variations of the supersymmetric partition function are indeed a linear combination of certain components of the the operator \eqref{Q-exact-intro}. However, the fact that the supercurrent anomaly is generically non-zero on these backgrounds means that this operator is not $Q$-exact, and hence the supersymmetric partition function is {\em not} invariant under deformations of the functions $u(z,\bar z)$ and $w(z, \bar z)$. Although our derivation of the superconformal Ward identities is holographic, the argument we use to show the non-invariance of the supersymmetric partition function is a {\em field theory} argument, using only the supercurrent anomaly. However, the result matches precisely the answer obtained in \cite{Genolini:2016ecx}, by explicitly evaluating the bulk on-shell action. We therefore conclude, that holographic renormalization is perfectly compatible with supersymmetry, and it is in fact the field theory assumption that the supersymmetric partition function is invariant that is not quite correct!       

The second puzzle concerns the BPS relation among the conserved charges in supersymmetric states and the supersymmetric Casimir energy, which has been discussed extensively in the recent literature \cite{Assel:2015nca,Martelli:2015kuk,Genolini:2016sxe,DiPietro:2016ond,Brunner:2016nyk,Genolini:2016ecx}. The BPS relation follows again from the $Q$-exactness of the operator \eqref{Q-exact-intro}, but the presence of a non-zero supercurrent anomaly implies that the BPS relation is also anomalous! We show that the supercurrent anomaly together with the Ward identities provide enough constraints to determine the one-point functions of the stress tensor and the $R$-current in any supersymmetric state in terms two arbitrary scalar functions. This general solution for the supersymmetric one-point functions allows us to obtain general formulas for the conserved charges for any BPS state, including the supersymmetric Casimir charges of the global vacuum, as well as the supersymmetric partition function. In fact, we show that there is a one-parameter family of consistently defined charges,  all of which satisfy the (anomalous) BPS relation. Moreover, the supersymmetric partition function is independent of the definition of the charges. 

The rest of the paper is organized as follows. In section \ref{model} we briefly review the minimal $\cn=2$ gauged supergravity in four and five dimensions. Section \ref{hamiltonian} sets the stage for the subsequent holographic analysis and discusses the radial Hamiltonian formulation of the bulk supergravity dynamics. Section \ref{algorithm} contains our main technical results, where we carry out the procedure of holographic renormalization in order to derive all boundary counterterms, both bosonic and fermionic. Moreover, we show that there is a two-parameter family of supersymmetric renormalization schemes that also preserve parity, corresponding to adding the two possible superconformal invariants \eqref{invariants}. In section \ref{dict} we use these boundary counterterms in order to define holographically the local current operators and derive the quantum superconformal Ward identities, including all quantum anomalies. In section \ref{backgrounds} we discuss the anomalous transformation of the supercurrent under rigid supersymmetry, and we revisit the dependence of the supersymmetric partition function on the functions $u(z,\bar z)$ and $w(z, \bar z)$, parameterizing the supersymmetric backgrounds found in \cite{Klare:2012gn,Dumitrescu:2012ha}. Finally, we use the anomalous transformation of the supercurrent, together with the Ward identities, in order to obtain general expressions for the one-point functions of the stress tensor and the $R$-current in any supersymmetric state, and to derive general expressions for the Casmir charges and the supersymmetric partition function. Our notation and a number of technical details are presented in three appendices.

\section{Minimal $\cn=2$ gauged supergravity}
\label{model}
\setcounter{equation}{0}

Our focus in this paper is on the minimal gauged supergravities in four and five dimensions \cite{Freedman:1976aw,Gunaydin:1983bi}, which we briefly revisit in this section, keeping terms up to quadratic order in the gravitino. The field content of minimal gauged supergravity in both four and five dimensions comprises of the gravity multiplet only, consisting of the metric $g_{\m\n}$, a complex Dirac gravitino $\J_\m$ (equivalently a pair of symplectic-Majorana gravitini) and the graviphoton $A_\m$.\footnote{It can be checked that the number of on-shell bosonic and fermionic degrees of freedom match in both four and five dimensions, respectively as $2+2=2+2$ and $5+3=4+4$.} 

The minimal gauged supergravity action is obtained by gauging the $U_R(1)$  subgroup of the $USp_R(2)\cong SU_R(2)$ $R$-symmetry group of $\cn=2$ Poincar\'e supersymmetry
and takes the form \cite{Freedman:1976aw,Gunaydin:1983bi}
\bal\label{action}
S=\;&\frac{1}{2\k^2}\int d^{d+1}x\sqrt{-g}\(\rule{0cm}{0.5cm}R-F_{\m\n}F^{\m\n}-2\L+c_1\;\e^{\m\n\r\s\l}F_{\m\n}F_{\r\s}A_\l\right.\NO\\
&\left.\hskip0.5in-\lbar\J_\m\G^{\m\n\r}(\overleftrightarrow{\nabla}_\n+2i\frak gA_\n)\J_\r-\frac{d-1}{\ell}\lbar\J_\m\(\G^{\m\n}+ic_2(\G^{\m\n\r\s} F_{\r\s}+2F^{\m\n})\)\J_\n\),
\eal
where $\k^2=8\p G_{d+1}$ is the gravitational constant and
\be
\L=-\frac{d(d-1)}{2\ell^2},\qquad \frak g=\frac 1\ell\sqrt{(d-1)(d-2)/2},
\ee
are respectively the cosmological constant and $U_R(1)$ gauge coupling in $D=d+1$ dimensions. As we verify in appendix \ref{bulk-susy}, the constants 
\be
c_1=-\frac{2\ell}{3\sqrt{3}}\d_{d,4}, \quad c_2=\frac{\ell}{\sqrt{2(d-1)(d-2)}},
\ee
are fixed by supersymmetry. Notice that $c_1$ is zero unless $d=4$, i.e. $D=5$. It should be emphasized that although we have written the action \eqref{action} for generic dimension $D=d+1$ in order to treat the cases $D=4$ and $D=5$ simultaneously, this action is not supersymmetric in dimensions other than four or five. The covariant derivative acts on the gravitino as
\be
\nabla_\m\J_\n=\pa_\m\J_\n+\frac14\o_{\m\a\b}\G^{\a\b}\J_\n-\G_{\m\n}^\r\J_\r,
\ee
where $\o_{\m\a\b}$ is the spin connection, $\G^\r_{\m\n}$ is the Christoffel symbol of the metric $g_{\m\n}$, and $\G^{\a\b}$ denotes the antisymmetrized product of two gamma matrices, as defined in appendix \ref{conventions}, where we explain our index notation and gamma matrix conventions.\footnote{We hope that the distinction between the Christoffel symbols $\G^\r_{\m\n}$ and the antisymmetrized products of gamma matrices will be clear from the context in the following.}  

The bulk action \eqref{action} must be complemented by the Gibbons-Hawking boundary terms \cite{Argurio:2014uca} 
\be\label{GH}
S\sbtx{GH}=\frac{1}{2\k^2}\int_{\pa\cm_\ve} d^dx\sqrt{-\g}\;(2K+\lbar\J_i\Hat\G^{ij}\J_j),
\ee
which are required for the Dirichlet problem of the action \eqref{action} to be well posed on a {\em finite} cutoff surface $\pa\cm_\ve$. However, in order for the variational problem to be well posed on a non-compact asymptotically locally AdS space, additional boundary terms are required \cite{Papadimitriou:2005ii}.  We will derive these systematically in section \ref{algorithm}. Moreover, the Gibbons-Hawking terms \eqref{GH} are required for the action to preserve supersymmetry on the boundary.  

The supersymmetry transformations that leave the action \eqref{action} off-shell invariant up to boundary terms are \cite{Freedman:1976aw,Gunaydin:1983bi}
\bal\label{SUSY-trans}
\d E^\a_\m=\;&\frac12\(\lbar\e\G^\a\J_\m-\lbar\J_\m\G^\a\e\),\qquad \d A_\m=\; ic_3\(\lbar\J_\m\e-\lbar\e\J_\m\),\NO\\
\d\J_\m=\;&\nabla_\m\e+ic_4\(\G_\m{}^{\n\r}-2(d-2)\d_\m^\n\G^\r\)F_{\n\r}\e-\frac{1}{2\ell}\(\G_\m-2i\ell\frak g A_\m\)\e,\NO\\
\d\lbar\J_\m=\;&\lbar\e\overleftarrow\nabla_\m+ic_4\lbar\e\(\G^{\r\n}{}_\m-2(d-2)\d_\m^\n\G^\r\)F_{\n\r}+\frac{1}{2\ell}\lbar\e\(\G_\m-2i\ell\frak g A_\m\),
\eal
where
\be
c_3=\sqrt{\frac{d-1}{8(d-2)}},\quad c_4=\frac{1}{\sqrt{8(d-1)(d-2)}}, 
\ee
and $E_\m^\a$ is the vielbein of the metric $g_{\m\n}$, i.e. $g_{\m\n}=\h_{\a\b}E^\a_\m E^\b_\n$. These transformations are correct only to linear order in the gravitino, corresponding to the fact that the action \eqref{action} is specified to quadratic order in the gravitino. In appendix \ref{bulk-susy} we check explicitly that the supersymmetry transformations \eqref{SUSY-trans} leave the action \eqref{action} invariant up to boundary terms, which we compute.

\section{Radial Hamiltonian formalism}
\label{hamiltonian}
\setcounter{equation}{0}

The first step in order to construct the holographic dictionary for the supergravity theory \eqref{action} is to formulate the dynamics in Hamiltonian language, with the radial coordinate emanating from the conformal boundary playing the role of Hamiltonian time. This formulation allows one to systematically construct the covariant local boundary counterterms required in order to render the variational problem well posed on the conformal boundary, and to derive the renormalized observables in the dual field theory, as well as the Ward identities they satisfy, including all related global anomalies.       

The radial Hamiltonian formulation of the dynamics singles out a radial coordinate $r$ emanating from the conformal boundary $\pa\cm$ in $\cm$ and describes the evolution of the induced fields on the constant radius slices $\S_r\cong \pa\cm$ under radial translations. This foliation of $\cm$ need only hold in an open neighborhood of $\pa\cm$, since the variational problem and the holographic dictionary require knowledge of the space of asymptotic solutions only. Equivalently, the holographic dictionary and the boundary counterterms require only ultraviolet data.    

The induced fields on the radial slices $\S_r$ are obtained through a standard ADM decomposition of the bulk fields, given in eq.~\eqref{ADM-decomposition} in appendix \ref{conventions}. This decomposition is somewhat more involved in the presence of fermion fields, since a Hamiltonian description of fermions requires an additional decomposition of the spinors according to {\em radiality} \cite{Freedman:2016yue}, defined in eq.~\eqref{radiality}. Further details on the ADM decomposition of spinor fields are provided in appendix \ref{conventions}.

\paragraph{Radial Lagrangian} Inserting the ADM decomposition \eqref{ADM-decomposition} in the supergravity action \eqref{action} leads to a number of bulk terms that involve only up to first order radial derivatives on the induced fields, as well as a number of boundary terms that are exactly canceled by the Gibbons-Hawking terms \eqref{GH}.\footnote{This is in fact a constructive argument for {\em deriving} the Gibbon-Hawking terms.} We can therefore write   
\be
S+S\sbtx{GH}=\int_{\S_r} dr\;L,
\ee
where the radial Lagrangian is given by 
\bal\label{rLagrangian}
L=\;&\frac{1}{2\k^2}\int d^dx\;N\sqrt{-\g}\Bigg\{K^2-K_{ij}K^{ij}-\frac{2}{N^2}\g^{ij}(\dot A_i-\pa_i a-N^k F_{ki})(\dot A_j-\pa_j a-N^l F_{lj})\NO\\
&-2\L+R[\g]-F_{ij}F^{ij}+\frac{1}{N}c_1\(4(\dot A_i-\pa_i a)\Hat\e^{ijkl}F_{jk}A_l+a\; \Hat\e^{ijkl}F_{ij}F_{kl}\)\NO\\
&\rule{0.cm}{.8cm}+\frac 2N(\dot{\lbar\J}_{+i}\Hat\G^{ij}  \J_{-j}+\lbar\J_{-i} \Hat \G^{ij} \dot\J_{+j} )+ (K+N^{-1} D_k N^k )\lbar \J_i\Hat\G^{ij}\J_j+\frac{1}{4N}e_{ak}\dot e_b^{\;k}\lbar \J_i \G\{ \Hat \G^{ij},\G^{ab}\}\J_j\NO\\
&+\frac2N \dot e_a^{(i} e_b^{j)}\lbar\J_i\G^{ab}\J_j+\frac{1}{2N}K_{kl}\(\(\lbar\J_r-N^i\lbar\J_i\)[\Hat\G^{kj},\Hat\G^l]\J_j-\lbar\J_j[\Hat\G^{kj},\Hat\G^l]\(\J_r-N^i\J_i\)\)\NO\\
&-\frac{2i(d-1)c_2}{N^2\ell}(\dot A_i-\pa_i a-N^j F_{ji})\(N\lbar\J_k\G\Hat\G^{ikl}\J_l-\lbar\J^i(\J_r-\J_l N^l)+(\lbar\J_r-N^l\lbar\J_l)\J^i\)\NO\\
&\rule{0.cm}{.8cm}+\frac{1}{4N}\lbar\J_i \(2(\pa_k N)\;[\Hat \G^{ij},\Hat \G^k]-(D_k N_l)\G\{\Hat \G^{ij},\Hat\G^{kl} \} \)\J_j-\frac{N^i}{N}(\lbar\J_j\G\Hat\G^{jk}\bb D_i\J_k-\lbar\J_j\overleftarrow{\bb D}_i\G\Hat\G^{jk}\J_k)\NO\\
&-\lbar\J_i\Hat\G^{ijk}\bb D_j\J_k+\lbar\J_i\overleftarrow{\bb D}_j\Hat\G^{ijk}\J_k-\frac 1N\lbar\J_k\overleftarrow{\bb D}_j\G\Hat\G^{jk}\(\J_r-N^i\J_i\)-\frac 1N\(\lbar\J_r-N^i\lbar\J_i\)\G\Hat\G^{jk}\bb D_j\J_k \NO\\
&+\frac 1N\lbar\J_k\G\Hat\G^{jk}(\bb D_j\J_r-N^i\bb D_j\J_i)+\frac1N(\lbar\J_r\overleftarrow{\bb D}_j-N^i\lbar\J_i\overleftarrow{\bb D}_j)\G\Hat\G^{jk}\J_k +\frac{2i\frak g}{N}(a- N^k A_k)\lbar\J_i\G\Hat\G^{ij}\J_j\NO\\
&-\frac{d-1}{N\ell}\(N\lbar\J_i\Hat\G^{ij}\J_j+(\lbar\J_r-N^i\lbar\J_i)\G\Hat\G^j\J_j+\lbar\J_j\Hat\G^j\G(\J_r-N^i\J_i)\)+2\lbar\J_i\J_j F^{ij}\NO\\
&-\frac{2i\frak g}{N}A_i\Big(N\lbar\J_k\Hat\G^{kil}\J_l+ (\lbar\J_r-N^k\lbar\J_k)\G\Hat\G^{ij}\J_j - \lbar\J_j\G\Hat\G^{ij}(\J_r-N^k\J_k)\Big)\NO\\
&-\frac{i(d-1)c_2}{N\ell}F_{kl}\Big(\(\lbar\J_r-N^i\lbar\J_i\)\G\Hat\G^{jkl}\J_j -\lbar\J_j\G\Hat\G^{jkl}\(\J_r-N^i\J_i\)+N\lbar\J_i\Hat\G^{ijkl}\J_j \Big)\Bigg\}.
\eal
In deriving this Lagrangian we have decomposed the bulk vielbein as in eq.~\eqref{bulk-vielbein} and have partially fixed the $SO(1,d)$ frame rotations by choosing the frame specified in eq.~\eqref{default-frame}. As we shall see shortly, this does not lead to any loss of generality since one can obtain the first class constraint associated with frame rotations by invoking the fact that there is no torsion.  

\paragraph{Canonical momenta} As mentioned above, in order to define the symplectic variables associated with the gravitino we need to use the radiality projectors 
\eqref{radiality} to decompose its transverse components as \cite{Kalkkinen:2000uk}
\be
\J_i=\J_{+i}+\J_{-i}.
\ee 
The variables $\J_{+i}$ and $\J_{-i}$ are symplectic conjugates and so one radiality can be treated as a generalized coordinate, while the other as the corresponding canonical momentum. We will adopt the convention that $\J_{+i}$ is the generalized coordinate and $\J_{-i}$ is (proportional to) the corresponding canonical momentum.

The canonical momenta conjugate to the vielbein $e_i^a$ on $\S_r$, the gauge field $A_i$, and the induced fermion fields $\J_{+i}$ and $\lbar\J_{+i}$ can be read off the radial Lagrangian \eqref{rLagrangian} and are given by
\bal\label{momenta}
\p_a^{\;i}=\;& \frac{\d L}{\d\dot e^a_{\;i}}=\frac{\sqrt{-\g}}{2\k^2}\( 2e_{a(j}\d^i_{k)}\Big(2\g^{j[k}\g^{l]m}K_{lm}-Y^{jk} \Big)-(e^{bi}Z_{ab}+e_{aj}P^{ji})\),\NO\\
\p^i=\;&\frac{\d L}{\d \dot A_i}=\frac{2\sqrt{-\g}}{\k^2}\(-\frac1N(\g^{ij}(\dot A_j-\pa_j a)-N_j F^{ji}+U^i)+c_1\;\Hat\e^{ijkl}F_{jk}A_l\) ,\NO\\
\p^{i}_\J=\;&-\frac{\d L}{\d\dot{\J}_{+i}}=L\frac{\overleftarrow{\d} }{\d\dot{\J}_{+i}}=\frac{\sqrt{-\g}}{\k^2} \lbar\J_{-j}\Hat\G^{ji},\qquad \p^{i}_{\lbar\J}=\frac{\d L}{\d\dot{\lbar\J}_{+i}}=\frac{\sqrt{-\g}}{\k^2} \Hat\G^{ij}\J_{-j},
\eal
where we have defined for convenience the fermion bilinears
\bal 
U^i=\;& \frac{i(d-1)c_2}{2\ell}\(N\lbar\J_j\G\Hat\G^{jki}\J_k+(\lbar\J_r-N^k\lbar\J_k)\J^i-\lbar\J^i(\J_r-N^k\J_k)\),\NO\\
P^{ij}=\;&\lbar\J^i\Hat\G^{jk}\J_k+\lbar\J_k\Hat\G^{kj}\J^i,\qquad Z^{ab}=\frac 14\lbar\J_i\G\{\Hat\G^{ij},\G^{ab} \}\J_j,\NO\\
Y^{ij}=\;& -\frac12\g^{ij}\lbar\J_k\Hat\G^{kl}\J_l+\frac{ 1}{4N}\(\lbar\J_k[\Hat\G^{ik},\Hat\G^j](\J_r-N^l\J_l)-(\lbar\J_r-N^l\lbar\J_l)[\Hat\G^{ik},\Hat\G^j]\J_k\).
\eal
Notice that the Lagrangian $L$ does not contain any radial derivatives of the variables $N$, $N_i$, $a$, $\J_r$ and $\lbar\J_r$, and hence their conjugate momenta vanish identically. 

As we shall see momentarily, these non-dynamical Lagrange multipliers lead to a set of first class constraints, each associated with a local symmetry of the action \eqref{action}. In particular, the lapse and shift functions $N$ and $N_i$ reflect the diffeomorphism invariance of \eqref{action}, $a$ corresponds to local $U(1)$ gauge transformations, while $\J_r$ and $\lbar\J_r$ reflect the local supersymmetry invariance. Notice, however, that there is no Lagrange multiplier associated with local frame rotations, although the action \eqref{action} is clearly invariant under such transformations. The reason for this is that the spin connection here is not an independent field and so there is no torsion. In a first order formalism, where the spin connection is treated as an independent field, the Lagrange multiplier associated with local frame rotations would be the spin connection component $\o_{rab}$, while the phase space of the theory would include the generalized coordinate $\o_{iab}$ and its conjugate momentum. However, as can be seen in eq.~\eqref{spin-connection-components}, here $\o_{rab}$ is completely determined in terms of the vielbein $e_i^a$ on $\S_r$, as well as the lapse and shift functions, and hence it is not an independent Lagrange multiplier. Moreover, $\o_{iab}$ and its conjugate momenta are not independent phase space variables. 

Nevertheless, the fact that the torsion vanishes implies that the extrinsic curvature $K_{ij}$, which can be identified with a certain component of the bulk Christoffel symbol (see eq.~\eqref{Christoffel}), is symmetric, i.e. $K_{[ij]}=0$. It follows that canonical momenta \eqref{momenta} satisfy the {\em algebraic} constraint 
\be
\frac{\k^2}{\sqrt{-\g}}\(e^{aj}\p_a^{\;i}-e^{ai}\p_a^{\;j} \)=P^{[ij]}+e^{ai}e^{bj}Z_{ab},
\ee
or equivalently,
\be\label{frame-constraint}
e_{[a}^i\p_{b]i}=\frac14\(\p_\J^i\G_{ab}\J_{+i}-\lbar\J_{+i}\G_{ab}\p_{\lbar\J}^i\), 
\ee
and hence, on the reduced phase space where the fermion variables are set to zero the canonical variables $e_a^i$ and $\p_a^i$ can be replaced by the induced metric $\g_{ij}$ and its conjugate momentum $\p^{ij}$. As we shall see later, this ensures that on a background where all the fermion sources are set to zero, the dual field theory possesses a symmetric and relativistic stress tensor.

\paragraph{Radial Hamiltonian} The radial Hamiltonian $H$ corresponds to the Legendre transform of the Lagrangian \eqref{rLagrangian} with respect to all dynamical variables, namely
\be 
H=\int d^d x\(\dot e^a_{\;i}\p_a^{\;i}+\dot A_i\p^i+\p_\J^{i}\dot\J_{+i}+\dot{\lbar\J}_{+i}\p_{\lbar\J}^{i}\)-L.
\ee
Introducing the gauge invariant momentum
\be\label{gauge-momentum-shifted}
\cp^i=\p^i-\frac{\sqrt{-\g}}{\k^2}2c_1\;\Hat\e^{ijkl}F_{jk}A_l,
\ee
and the gauge-covariant derivative (see also \eqref{spinor-derivatives})
\be\label{gauge-der}
\cd_i\J_{+j}=\bb D_i\J_{+j}+i\frak g A_i\J_{+j},
\ee
we find that the Hamiltonian takes the form
\be\label{rHamiltonian}
H=\int d^d x\(N\ch+N_i\ch^i+a\cx+(\lbar\J_r-N^i\lbar\J_i)\cf+\lbar\cf(\J_r-N^i\J_i)\),
\ee
where
\bsub
\label{constraint-expressions}
\bal
\ch=&\frac{\k^2}{2\sqrt{-\g}}\Big[\Big(\frac{1}{d-1}e^a_i e^b_j-e^a_j e^b_i\Big)\p_a^i\p_b^j-\frac12\cp^i\cp_i
+\Big(\frac{1}{d-1}\g^{ij}\g^{kl}-\g^{ik}\g^{jl}\Big)e^a_{(i}\p_{aj)}(\lbar\J_{+k}\p_{\lbar\J l}+\p_{\J k}\J_{+l})\NO\\
&\hskip0.6in -\frac{1}{d-1}e^{a(i}\p_a^{j)}\(\p_\J^k(\Hat\G_{ki}-(d-2)\g_{ki})\Hat\G_j{}^l\J_{+l}+\lbar\J_{+k}\Hat\G^k{}_j(\Hat\G_{il}-(d-2)\g_{il})\p_{\lbar\J}^l\) \NO\\
&\hskip0.6in-\frac{1}{d-1}\p_\J^i\((\Hat\G_{ij}-(d-2)\g_{ij})\slashed{\cd}-\overleftarrow{\slashed{\cd}}(\Hat\G_{ij}-(d-2)\g_{ij})\)\p_{\lbar\J}^j\Big] \NO\\
&+\frac{d-1}{2\ell}\(\lbar\J_{+i}\p_{\lbar\J}^i+\p_\J^i\J_{+i}\)-\frac{i(d-1)c_2}{2\ell}\cp_k\lbar\J_{+i}\Hat\G^{ikj}\J_{+j}\NO\\
&+\(\frac{\k^2}{\sqrt{-\g}}\)^2\frac{ic_2}{2(d-1)\ell}\cp_k\p_\J^i\Big(\Hat\G_i{}^k{}_j-(d-1)(d-2)\Hat\G^k\g_{ij}-(d-2)(2\Hat\G_{[i}\d_{j]}^k+\Hat\G^k\Hat\G_{ij})\Big)\p_{\lbar\J}^j\NO\\
&+\frac{ic_2}{2\ell}F_{kl}\Big[\p_\J^i\(-2\Hat\G_i{}^{jkl}+3(d-3)\d_i^{[j}\Hat\G^{kl]}+2\g^{lj}(\Hat\G_{i}{}^k-(d-2)\d_{i}^k)\)\J_{+j}\NO\\
&\hskip0.8in+\lbar\J_{+i}\(-2\Hat\G^{ikl}{}_j+3(d-3)\Hat\G^{[ik}\d_j^{l]}+2\g^{ik}(\Hat\G^l{}_{j}-(d-2)\d^l_{j})\)\p_{\lbar\J}^j\Big]\label{ham} \\
&-\frac{\sqrt{-\g}}{2\k^2}\Big(R[\g]-2\L-F_{ij}F^{ij}+\lbar\J_{+i}\overleftarrow{\cd}_j\Hat\G^{ijk}\J_{+k}-\lbar\J_{+i}\Hat\G^{ijk}\cd_j\J_{+k}+2D^{i}(\lbar\J_{+[i}\Hat\G^{j}\J_{+j]})\Big),\NO\\
\rule{0.cm}{.8cm}\ch^i=& -D_j(e^{ai}\p_a^j+\p_\J^j\J_+^i+\lbar\J_+^i\p_{\lbar\J}^j)+\p_\J^j\cd^i\J_{+j}+\lbar\J_{+j}\overleftarrow{\cd }^i\p_{\lbar\J}^j+F^{ij}\cp_j,\\
\rule{0.cm}{.8cm}\cx=\;&-D_i\p^i -i\frak g\(\p_\J^i\J_{+i}-\lbar\J_{+i}\p_{\lbar\J}^i\)-\frac{\sqrt{-\g}}{2\k^2}c_1\;\Hat\e^{ijkl}F_{ij}F_{kl},\\
\rule{0.cm}{.8cm}\cf_+=&\frac{1}{2\ell}\Hat\G_i\p_{\lbar\J}^i+\frac{\k^2}{2\sqrt{-\g}}\(\frac{1}{d-1}\g_{ij}\g_{kl}-\g_{ik}\g_{jl}\)e^{a(i}\p^{j)}_a\Hat\G^k\p_{\lbar\J}^l-\frac{i(d-1)c_2}{2\ell}\cp^i\J_{+i}\NO\\
&+\frac{ic_2}{2\ell}F_{jk}\((d-2)\Hat\G^{jk}\Hat\G^i-(d-1)\Hat\G^{jki} \)\p_{\lbar\J i}+\frac{\sqrt{-\g}}{\k^2}\Hat\G^{ij}\cd_i \J_{+j},\label{cf-pos}\\
\rule{0.cm}{.8cm}\cf_-=& -\cd_i\p_{\lbar\J}^i+\frac12\p_a^i\G^a\J_{+i}-\frac{\k^2}{\sqrt{-\g}}\frac{ic_2}{2\ell}(\Hat\G_{ij}-(d-2)\g_{ij})\cp^i\p_{\lbar\J}^j\NO\\
&-\frac{\sqrt{-\g}}{\k^2}\frac{(d-1)}{2\ell}\(\Hat\G^i\J_{+i}+ic_2\Hat\G^{ijk}\J_{+i}F_{jk}\),\label{cf-neg}
\eal
\esub
and we have defined $\cf_\pm\equiv \G_\pm\cf$ so that $\cf=\cf_++\cf_-$. 

Since the momenta conjugate to the variables $N$, $N_i$, $a$ and $\J_r$ vanish identically, the form \eqref{rHamiltonian} of the Hamiltonian leads, via Hamilton's equations, to the first class constraints 
\be\label{constraints}
\ch=\ch^i=a=\cf_+=\cf_-=0.
\ee
We will see in the following that these constraints are directly related to the Ward identities of the dual field theory. An important observation is that the Hamiltonian constraint $\ch=0$, as well as the supersymmetry constraints $\cf_\pm=0$,  are quadratic in the canonical momenta, while all other constraints are linear. This reflects the fact that the symmetries associated with the Hamiltonian and supersymmetry constraints are spontaneously broken by the radial slice $\S_r$ and hence realized non linearly, while those associated with the remaining first class constraints are preserved. The non-linear constraints corresponding to the symmetries broken by the cutoff result in global anomalies in the dual quantum field theory. This is completely analogous to the way quantum anomalies appear in standard quantum field theory, through symmetry breaking by the ultraviolet regulator.

\paragraph{Hamilton-Jacobi formalism and flow equations} The Hamiltonian determines the radial evolution of the induced fields $e^a_i$, $\J_{+i}$ and $A_i$ through Hamilton's equations
\be 
\dot e^a_i=\frac{\d H}{\d \p_a^i},\qquad \dot A_i=\frac{\d H}{\d \p^i},\qquad \dot \J_{+i}=\frac{\d H}{\d \p_{\J}^i}.
\ee 
In the gauge \eqref{FG-gauge} only the Hamiltonian constraint contributes to these equations, which read
\bsub
\label{flow-eqs}
\bal
\dot e^a_i=&\frac{\k^2}{2\sqrt{-\g}}\Big[2\Big(\frac{1}{d-1}e^a_i e^b_j-e^a_j e^b_i\Big)\p_b^j
+\Big(\frac{1}{d-1}\g^{pq}\g^{kl}-\g^{pk}\g^{ql}\Big)e^a_{(p} \g_{q)i}(\lbar\J_{+k}\p_{\lbar\J l}+\p_{\J k}\J_{+l})\NO\\
&-\frac{1}{(d-1)}e^{a(p}\d^{q)}_i\(\p_\J^k(\Hat\G_{kp}-(d-2)\g_{kp})\Hat\G_q{}^l\J_{+l}+\lbar\J_{+k}\Hat\G^k{}_q(\Hat\G_{pl}-(d-2)\g_{pl})\p_{\lbar\J}^l\) \Big],\\
\rule{0.0cm}{0.8cm}
\dot A_i=&-\frac{\k^2}{2\sqrt{-\g}}\cp_i 
-\frac{i(d-1)c_2}{2\ell}\lbar\J_{+k}\Hat\G^{k}{}_i{}^{l}\J_{+l}\\
&+\(\frac{\k^2}{\sqrt{-\g}}\)^2\frac{ic_2}{2(d-1)\ell}\p_\J^k\Big(\Hat\G_{kil}-(d-1)(d-2)\Hat\G_i\g_{kl}-(d-2)(2\Hat\G_{[k}\g_{l]i}+\Hat\G_i\Hat\G_{kl})\Big)\p_{\lbar\J}^l,\NO\\
\rule{0.0cm}{0.8cm}
\dot \J_{+i}=&\frac{\k^2}{2\sqrt{-\g}}\frac{1}{d-1}\times\NO\\
&\Big[\Big(\g_{pq}\d^{l}_i-(d-1)\g_{ip}\g_q^{l}-(\Hat\G_{ip}-(d-2)\g_{ip})\Hat\G_q{}^l\Big)e^{a(p}\p_a^{q)}\J_{+l}-2(\Hat\G_{ip}-(d-2)\g_{ip})\slashed{\cd}\p_{\lbar\J}^p\Big] \NO\\
&+\(\frac{\k^2}{\sqrt{-\g}}\)^2\frac{ic_2}{2(d-1)\ell}\cp_k\Big(\Hat\G_i{}^k{}_j-(d-1)(d-2)\Hat\G^k\g_{ij}-(d-2)(2\Hat\G_{[i}\d_{j]}^k+\Hat\G^k\Hat\G_{ij})\Big)\p_{\lbar\J}^j,\NO\\
&+\frac{d-1}{2\ell}\J_{+i}+\frac{ic_2}{2\ell}F_{kl}\(-2\Hat\G_i{}^{jkl}+3(d-3)\d_i^{[j}\Hat\G^{kl]}+2\g^{lj}(\Hat\G_{i}{}^{k}-(d-2)\d^k_i)\)\J_{+j} .
\eal
\esub
Moreover, inverting the expression for $\p^{i}_{\lbar\J}$ in \eqref{momenta} we find that $\J_{-i}$ is given by 
\be\label{Psi-} 
\J_{-i}=\frac{\k^2}{\sqrt{-\g}}\frac{1}{d-1}(\Hat\G_{ij}-(d-2)\g_{ij})\p_{\lbar\J}^j.
\ee 

These equations allow us to determine the radial evolution of the induced fields without invoking the second order field equations, or equivalently the other half of the Hamilton equations, by using instead the Hamilton-Jacobi expressions  
\be\label{HJ-momenta}
\p_a^i=\frac{\d\bb S}{\d e^a_i},\qquad \p^i=\frac{\d\bb S}{\d A_i},\qquad \p_\J^i=-\frac{\d\bb S}{\d\J_{+i}}=\bb S \frac{\overleftarrow\d}{\d\J_{+i}},\qquad \p_{\lbar\J}^i=\frac{\d\bb S}{\d\lbar\J_{+i}},
\ee
where $\bb S[e,A,\J_+]$ is Hamilton's principal function. In  particular, inserting the expressions \eqref{HJ-momenta} for the canonical momenta in the first class constraints \eqref{constraints} leads to a set of functional differential equations for the functional $\bb S[e,A,\J_+]$. Given a solution $\bb S[e,A,\J_+]$ of the these Hamilton-Jacobi equations, the flow equations \eqref{flow-eqs} become {\em first} order equations for the induced fields. In the next section we will obtain the general asymptotic solution of the Hamilton-Jacobi equations. In combination with the relations \eqref{HJ-momenta} and the flow equations \eqref{flow-eqs} this solution allows us to determine the general asymptotic Fefferman-Graham expansions for the induced fields $e^a_i$, $\J_{+i}$ and $A_i$, without solving the second order field equations. Since Hamilton's principal functional $\bb S[e,A,\J_+]$ coincides with the on-shell action evaluated with a radial cut-off $r_o$, the asymptotic solution of the Hamilton-Jacobi equations also determines the divergences of the on-shell action, and hence the covariant boundary counterterms. Hence, the complete holographic dictionary can be obtained from a single object, namely the asymptotic solution of the Hamilton-Jacobi equations.

\section{Recursive solution of the Hamilton-Jacobi equations}
\label{algorithm}
\setcounter{equation}{0}

In this section we determine the asymptotic solution of the Hamilton-Jacobi equations, corresponding to the constraints \eqref{constraints}, with the canonical momenta expressed as gradients of Hamilton's principal function $\bb S[e,A,\J_+]$. As we will see, our algorithm requires solving only the Hamiltonian constraint $\ch=0$, with all other constraints being automatically satisfied, up to the corresponding order in the asymptotic expansion. The resulting asymptotic solution for Hamilton's principal function will allow us to obtain the asymptotic Fefferman-Graham expansions for the bulk fields, and will be the basis for constructing the holographic dictionary in the next section.

\subsection{Covariant expansion and recursive algorithm}

Since the conformal dimensions of the operators dual to the fields in the Lagrangian \eqref{action} are known and fixed, we can obtain the general asymptotic solution of the Hamilton-Jacobi equation $\ch=0$ by using the dilatation operator method \cite{Papadimitriou:2004ap}. From the leading asymptotic form of the fields in eq.~  
\eqref{leading-asymptotics} follows that the dilatation operator takes the form
\be\label{dilop}
\d_D=\int d^dx\;\(e^a_{\;i}\frac{\d}{\d e^a_{\;i}}+\frac12\lbar\J_{+l}\frac{\d}{\d\lbar\J_{+l}}+\frac12\frac{\overleftarrow\d}{\d\J_{+l}}\J_{+l}\).
\ee

This operator allows us to look for a solution $\bb S[e,A,\J_+]=\int d^d x\; \bb L$ of the Hamilton-Jacobi equation $\ch=0$ in the form of a formal expansion 
\be\label{exp}
\bb S= \bb S_{(0)}+\bb S_{(1)}+\bb S_{(2)}+\cdots,
\ee
where $\bb S\sub{n}=\int d^d x\; \bb L\sub{n}$ are eigenfunctions of the dilatation operator, i.e. $\d_D\bb S_{(n)}=(d-n)\bb S_{(n)}$. By construction, $\bb S_{(n)}$ are covariant functionals of the induced fields $e^a_i$, $A_i$ and $\J_{+i}$ and higher order eigenfunctions, i.e. larger $n$, are asymptotically subleading relative to lower order ones. The formal expansion \eqref{exp} is therefore a covariant asymptotic expansion. 

In order to utilize the expansion \eqref{exp} we observe that the Hamilton-Jacobi relations \eqref{HJ-momenta} imply that for a generic variation of the induced fields \cite{Papadimitriou:2011qb}
\be\label{variational-ID}
\p_a^i\d e_i^a+\p^i\d A_i+\d\lbar\J_{+i}\p^i_{\lbar\J}+\p^i_{\J}\d\J_{+i}=\d \bb L+\pa_iv^i,
\ee
where $\pa_iv^i$ indicates a generic total derivative term. Applying this identity to local scaling transformations generated by the dilatation operator \eqref{dilop} gives 
\be\label{dilatation-ID}\boxed{
	 e_i^a\p_a^i\sub{n}+\frac12\lbar\J_{+i}\p^i_{\lbar\J}\sub{n}+\frac12\p^i_{\J}\sub{n}\J_{+i}=(d-n) \bb L\sub{n},}
\ee
where we have used the fact that $\bb L\sub{n}$ are only defined up to a total derivative and hence $\pa_iv^i\sub{n}$ can be absorbed in $\bb L\sub{n}$. 

The final ingredient necessary to set up the recursive algorithm for determining the terms $\bb S\sub{n}$ in the expansion \eqref{exp} is the leading asymptotic behavior of the fields in eq.~  
\eqref{leading-asymptotics}. Together with the flow equations \eqref{flow-eqs}-\eqref{Psi-} and the expressions \eqref{HJ-momenta} for the canonical momenta, this leading asymptotic form of the fields determines the zero order solution $\bb S\sub{0}$ to be   
\be\label{S0}
	\bb S_{(0)}=\frac{1}{\k^2}\int d^dx\sqrt{-\g}\;\frac{(d-1)}{\ell}.
\ee
The corresponding canonical momenta are\footnote{The subscript in the canonical momenta indicates the dilatation weight of the corresponding potential $\bb S\sub{n}$ and not that of the momenta themselves. } 
\be\label{momenta-0}
\p^i_a\sub{0}=\frac{\sqrt{-\g}}{\k^2}\frac{(d-1)}{\ell}e^i_a,\qquad \p^i\sub{0}=0,\qquad \p^i_{\lbar\J}\sub{0}=\p^i_{\J}\sub{0}=0.
\ee

Inserting the expansion \eqref{exp} in \eqref{ham} and using the identity \eqref{dilatation-ID} and the leading solution \eqref{S0} of the Hamilton-Jacobi equation, the Hamiltonian constraint $\ch=0$ reduces to a tower of linear equations for $\bb L\sub{n}$, $n>0$, namely
\be\label{recursion}\boxed{
\frac{(d-n)}{\ell}\bb L\sub{n}=\bb R\sub{n},\quad n\geq 1,}
\ee
where the inhomogeneous terms up to order $n=4$ are given by 
\bsub
\label{sources}
\bal
\bb R\sub{1}=&\;0,\\
\bb R\sub{2} =&\;\frac{\sqrt{-\g}}{2\k^2}\Big(R[\g]+\lbar\J_{+i}\overleftarrow{\cd}_j\Hat\G^{ijk}\J_{+k}-\lbar\J_{+i}\Hat\G^{ijk}\cd_j\J_{+k}+2D^{i}(\lbar\J_{+[i}\Hat\G^{j}\J_{+j]})\Big),\\
\bb R\sub{3} =&\;0,\\
\bb R\sub{4} =&\;-\frac{\k^2}{2\sqrt{-\g}}\Big[\Big(\frac{1}{d-1}e^a_i e^b_j-e^a_j e^b_i\Big)\p_a^i\sub{2}\p_b^j\sub{2}\NO\\
&\hskip0.3in+\Big(\frac{1}{d-1}\g^{ij}\g^{kl}-\g^{ik}\g^{jl}\Big)e^a_{(i}\p_{aj)}\sub{2}(\lbar\J_{+k}\p_{\lbar\J l}\sub{2}+\p_{\J k}\sub{2}\J_{+l})\NO\\
&\hskip0.3in -\frac{1}{d-1}e^{a(i}\p_a^{j)}\sub{2}\(\p_\J^k\sub{2}(\Hat\G_{ki}-(d-2)\g_{ki})\Hat\G_j{}^l\J_{+l}+\lbar\J_{+k}\Hat\G^k{}_j(\Hat\G_{il}-(d-2)\g_{il})\p_{\lbar\J}^l\sub{2}\) \NO\\
&\hskip0.3in-\frac{1}{d-1}\p_\J^i\sub{2}\((\Hat\G_{ij}-(d-2)\g_{ij})\slashed{\cd}-\overleftarrow{\slashed{\cd}}(\Hat\G_{ij}-(d-2)\g_{ij})\)\p_{\lbar\J}^j\sub{2}\Big] \NO\\
&-\frac{ic_2}{2\ell}F_{kl}\Big[\p_\J^i\sub{2}\(-2\Hat\G_i{}^{jkl}+3(d-3)\d_i^{[j}\Hat\G^{kl]}+2\g^{lj}(\Hat\G_{i}{}^k-2(d-2)\d_{i}^k)\)\J_{+j}\NO\\
&\hskip0.2in+\lbar\J_{+i}\(-2\Hat\G^{ikl}{}_j+3(d-3)\Hat\G^{[ik}\d_j^{l]}+2\g^{ik}(\Hat\G^l{}_{j}-(d-2)\d^l_{j})\)\p_{\lbar\J}^j\sub{2}\Big]-\frac{\sqrt{-\g}}{2\k^2}F_{ij}F^{ij}.
\eal
\esub
In writing these expressions we have again ignored terms higher than quadratic in the gravitino. 

From \eqref{recursion} and \eqref{sources} follows that the first subleading term in the covariant expansion \eqref{exp} is
\be\label{S2}
	\bb S_{(2)}=\frac{\ell}{2(d-2)\k^2}\int d^dx\sqrt{-\g}\;\( R[\g]+\lbar\J_{+i}\overleftarrow{\cd}_j\Hat\G^{ijk}\J_{+k}-\lbar\J_{+i}\Hat\G^{ijk}\cd_j\J_{+k}\).
\ee
Varying this expression with respect to each of the induced fields gives the  canonical momenta
\bal\label{momenta-2}
\p^p_a\sub{2}=&\;\frac{\sqrt{-\g}}{\k^2}\frac{\ell}{(d-2)} \Bigg[\(\frac12\g^{pj}R-R^{pj}\)e_{ja}\NO\\
&+\frac12e^p_a\( \lbar\J_{+i}\overleftarrow{\cd}_j\Hat\G^{ijk}\J_{+k}-\lbar\J_{+i}\Hat\G^{ijk}\cd_j\J_{+k}\)-\frac12 e_a^i\( \lbar\J_{+i}\overleftarrow{\cd}_j\Hat\G^{pjk}\J_{+k}-\lbar\J_{+i}\Hat\G^{pjk}\cd_j\J_{+k}\)\NO\\
&-\frac12 e_a^j\( \lbar\J_{+i}\overleftarrow{\cd}_j\Hat\G^{ipk}\J_{+k}-\lbar\J_{+i}\Hat\G^{ipk}\cd_j\J_{+k}\)-\frac12 e_a^k\( \lbar\J_{+i}\overleftarrow{\cd}_j\Hat\G^{ijp}\J_{+k}-\lbar\J_{+i}\Hat\G^{ijp}\cd_j\J_{+k}\)\NO\\
&+\frac14D_j\(\lbar\J_{+i}(e_{qa}\Hat\G^{ijkpq}+4\Hat\G^{[i}\g^{k]p}e^j_a+2\Hat\G^j\g^{p[i}e^{k]}_a)\J_{+k}\)+e^p_aD^{[i}(\lbar\J_{+i}\Hat\G^{k]}\J_{+k})\NO\\
&+\frac12e^{[i}_aD^{k]}(\lbar\J_{+i}\Hat\G^{p}\J_{+k})+\frac12\g^{p[i}D^{k]}(\lbar\J_{+i}\Hat\G_a\J_{+k})\Bigg],\NO\\
\p_{\lbar\J}^i\sub{2}=&\;-\frac{\sqrt{-\g}}{\k^2}\frac{\ell}{(d-2)}\Hat\G^{ijk}\cd_j\J_{+k},\qquad \p_{\J}^i\sub{2}=\frac{\sqrt{-\g}}{\k^2}\frac{\ell}{(d-2)}\lbar\J_{+k}\overleftarrow\cd_j\Hat\G^{kji},
\eal  
where we have used the identities (see e.g. (7.96) in \cite{Freedman:2012zz})
\be
e^a_pe^b_q\d\o_{jab}=(D_{[j}\d e^a_{p]})e_{qa}-(D_{[p}\d e^a_{q]})e_{ja}+(D_{[q}\d e^a_{j]})e_{pa},
\ee
and 
\be
\{\Hat\G_{pq},\Hat\G^{ijk}\}=2\Hat\G^{ijk}{}_{pq}+4\Hat\G^i\d^{j}_{[q}\d^{k}_{p]}+4\Hat\G^k\d^{i}_{[q}\d^{j}_{p]}+4\Hat\G^j\d^{k}_{[q}\d^{i}_{p]},
\ee
in order to obtain the expression
\bal
e^a_pe^b_q\d\o_{jab}\lbar\J_{+i}\{\Hat\G^{pq},\Hat\G^{ijk}\}\J_{+k}&=\lbar\J_{+i}\Big[2(D_{j}\d e^a_{p})e_{qa}\Hat\G^{ijkpq}\NO\\
&\hskip-1.8in+\(8\Hat\G^{[i}\g^{k]p}e^j_a+4\Hat\G^j\g^{p[i}e^{k]}_a\)D_j\d e^a_p+\(8e^p_a\Hat\G^{[k}D^{i]}+4\Hat\G^pe^{[i}_aD^{k]}+4\Hat\G^le_{al}\g^{p[i}D^{k]}\)\d e^a_p\Big]\J_{+k}.
\eal

To proceed to the next order we insert the canonical momenta \eqref{momenta-2} in the expression for $\bb R\sub{4}$ in \eqref{sources}. After some algebra we obtain  
\bal\label{R4}
\bb R\sub{4} =&\;-\frac{\sqrt{-\g}}{2\k^2}\frac{\ell^2}{(d-2)^2}\Bigg\{\frac{d}{4(d-1)}R^2-R_{ij}R^{ij}+\frac{(d-2)^2}{\ell^2}F_{ij}F^{ij}\NO\\
&-\frac{1}{2(d-1)}R\( \lbar\J_{+i}\overleftarrow{\cd}_j\Hat\G^{ijk}\J_{+k}-\lbar\J_{+i}\Hat\G^{ijk}\cd_j\J_{+k}\)+\frac{d}{d-1}RD^{[i}(\lbar\J_{+i}\Hat\G^{j]}\J_{+j})\NO\\
&-\frac{(2d-3)}{d-1}R^k_l\( \lbar\J_{+i}\overleftarrow{\cd}_j\Hat\G^{ijl}\J_{+k}-\lbar\J_{+k}\Hat\G^{lji}\cd_j\J_{+i}\)+2R^{j[k}D_j(\lbar\J_{+i}\Hat\G^{i]}\J_{+k})+2R^{[i}_kD^{j]}(\lbar\J_{+i}\Hat\G^k\J_{+j})\NO\\
&-\frac{2}{d-1}R^{ik}\(\lbar\J_{+[i}\overleftarrow\cd_{j]}\Hat\G^j\J_{+k}-\lbar\J_{+[i}\Hat\G^j\cd_{j]}\J_{+k}\)-\frac{2}{d-1}R\(\lbar\J_{+[i}\overleftarrow\cd_{j]}\Hat\G^i\J_{+}^j-\lbar\J_{+[i}\Hat\G^i\cd_{j]}\J_{+}^j\)\NO\\
&+2R^i_k\(\lbar\J_{+[i}\overleftarrow\cd_{j]}\Hat\G^k\J_{+}^j-\lbar\J_{+[i}\Hat\G^k\cd_{j]}\J_{+}^j\)\NO\\
&\hskip0.3in+\frac{1}{d-1}\lbar\J_{+k}\overleftarrow\cd_l\Hat\G^{kli}\((\Hat\G_{ij}-(d-2)\g_{ij})\slashed{\cd}-\overleftarrow{\slashed{\cd}}(\Hat\G_{ij}-(d-2)\g_{ij})\)\Hat\G^{jpq}\cd_p\J_{+q}\Bigg\} \NO\\
&-\frac{\sqrt{-\g}}{2\k^2}\frac{ic_2}{(d-2)}F_{kl}\Big[\lbar\J_{+p}\overleftarrow\cd_q\Hat\G^{pqi}\(-2\Hat\G_i{}^{jkl}+3(d-3)\d_i^{[j}\Hat\G^{kl]}+2\g^{lj}(\Hat\G_{i}{}^k-(d-2)\d_{i}^k)\)\J_{+j}\NO\\
&\hskip0.6in-\lbar\J_{+i}\(-2\Hat\G^{ikl}{}_j+3(d-3)\Hat\G^{[ik}\d_j^{l]}+2\g^{ik}(\Hat\G^l{}_{j}-(d-2)\d^l_{j})\)\Hat\G^{jpq}\cd_p\J_{+q}\Big],
\eal
and from \eqref{recursion} follows that
\be
\bb L\sub{4}=\frac{\ell}{d-4}\bb R\sub{4}.
\ee

If the boundary dimension $d$ is not specified, the above procedure can be repeated indefinitely in order to determine the terms $\bb L\sub{n}$ to arbitrary order. However, the recursion procedure stops at the term of zero dilatation weight. In particular, we only need to determine the terms $\bb L\sub{n}$ for $n\leq d$. Since we are interested in the cases $d=3$ and $d=4$ here, the above results suffice. However, as is clear from the above expressions, there is a significant difference between the two cases, due to the fact that for $d=4$ one needs to go up to the term $\bb L\sub{4}$, which contains a pole when $d=4$. This pole indicates the presence of a conformal anomaly and should be removed through the dimensional regularization prescription \cite{Papadimitriou:2004ap,Papadimitriou:2011qb}   
\be\label{dimreg}
\frac{1}{d-4}\to r_o/\ell,
\ee
where $r_o$ is the radial cutoff in the Fefferman-Graham coordinates \eqref{ADM-decomposition}-\eqref{FG-gauge}. More specifically, one replaces $\bb L\sub{4}$ in the expansion \eqref{exp} according to the rule
\be
\bb L\sub{4}=\frac{\ell}{d-4}\bb R\sub{4}\to \wt{\bb L}\sub{4}\log(e^{-2r_o/\ell}) =-\frac{\ell}{2} \bb R\sub{4} \log (e^{-2r_o/\ell}).
\ee
The fact that the resulting asymptotic solution of the Hamilton-Jacobi equation depends explicitly on the radial cutoff is the holographic manifestation of the conformal anomaly \cite{Henningson:1998gx} and implies that the implicitly covariant and local expansion in \eqref{exp} is not a fully consistent asymptotic solution in this case. Alternatively, one could have started instead with a formal expansion that contains the logarithmically divergent term $\wt{\bb L}\sub{4}$, in which case the recursion procedure would not produce any poles at $d=4$. The two approaches are equivalent and an explicit comparison can be found in the recent review \cite{Papadimitriou:2016yit}.

The punchline of the above analysis is that the general asymptotic solutions of the Hamilton-Jacobi equation for $d=3$ and $d=4$ take the form
\be\label{HJ-sol}\boxed{
	\bb S=\left\{\begin{matrix}
		\bb S\sub{0}+\bb S\sub{2}+\bb S\sub{3}+\cdots, & d=3,\vspace{0.4cm}\\
		\bb S\sub{0}+\bb S\sub{2}+\wt{\bb S}\sub{4}\log (e^{-2r_o/\ell})+\bb S\sub{4}+\cdots, & d=4,
	\end{matrix}
	\right.}
\ee
where $\bb S\sub{d}$ is undetermined but finite (has dilatation weight zero), while $\bb S\sub{0}$, $\bb S\sub{2}$ and $\wt{\bb S}\sub{4}=-\frac{\ell}{2} \int d^4 x\;\bb R\sub{4}$ correspond to the divergent part of the solution $\bb S$ and are given in equations \eqref{S0}, \eqref{S2} and \eqref{R4}, respectively. The ellipses in \eqref{HJ-sol} stand for terms with negative dilation weight that asymptotically go to zero, and hence are not relevant for the subsequent analysis. As we show next, this asymptotic solution of the Hamilton-Jacobi equation can be used to determine the Fefferman-Graham asymptotic expansions of bulk fields, and to derive the holographic dictionary.

\subsection{Finite local counterterms and the supersymmetric renormalization scheme}

The first implication of the asymptotic solution \eqref{HJ-sol} of the Hamilton-Jacobi equation is that it determines the local and covariant boundary counterterms necessary to render the variational problem on the conformal boundary  well posed, and consequently to renormalize the on-shell action. 

Hamilton-Jacobi theory identifies Hamilton's principal function $\bb S$ with the on-shell action, and so the divergences of the on-shell action are in one-to-one correspondence with the divergences of the asymptotic solution \eqref{HJ-sol} of the Hamilton-Jacobi equation \cite{deBoer:1999tgo}. Hence, the counterterms can be defined as minus the divergent parts of the solution \eqref{HJ-sol}. The resulting renormalized action is invariant under boundary U(1) gauge transformations, diffeomorphisms, local Weyl transformations, as well as supersymmetry and superWeyl transformations, up to specific anomalies in the case $d=4$. We will discuss these symmetries and derive the corresponding anomalies in section \ref{dict}. In this subsection, however, we would like to specify the most general form of the boundary counterterms compatible with the above symmetries, including supersymmetry.       

In the case $d=4$, besides the divergent part of the solution \eqref{HJ-sol}, one may include in the boundary counterterms certain {\em finite, local}, and {\em covariant} terms that do not cancel any divergences, but may contribute to the value of the renormalized on-shell action, as well as to the value of certain one-point functions derived thereof. Such terms must preserve all local symmetries on the cutoff, since otherwise they would give rise to spurious anomalies, i.e. anomalies that are trivial cocycles and can be removed by the very same finite boundary terms. An example of a finite local boundary term that explicitly breaks local Weyl invariance (as well as supersymmetry) is the $R^2$ term, which leads to a total derivative contribution to the conformal anomaly. By construction, this contribution to the conformal anomaly is a trivial cocycle, that can be eliminated by removing the $R^2$ term from the action. Since we are interested in minimizing the possible contributions to the anomalies, in the following we will strictly insist that any candidate finite counterterm must be a local {\em superconformal invariant}, such that only non-trivial cocycles contribute to the anomalies.

Given the field content in the bulk, or equivalently the spectrum of local gauge invariant operators in the dual quantum field theory, the local superconformal invariants in any dimension can be classified. This amounts to a classification of the possible (non-trivial) contributions to the Weyl anomaly, since the Wess-Zumino consistency condition implies that the Weyl anomaly is itself a conformal invariant. For $\cn=1$ supersymmetric theories in $d=4$ this classification has been done using superspace techniques in \cite{Bonora:1984pn}. Explicit calculations of the Weyl anomaly have been carried out in superspace in \cite{McArthur:1983fk,Buchbinder:1986im} and in components in \cite{Anselmi:1997am}. The result that is relevant for us here is the general form of the bosonic part of the Weyl anomaly in eq.~(4.3) of \cite{Anselmi:1997am}, which is the sum of the only two non-trivial superconformal invariants corresponding to our field content.\footnote{The Pontryagin density and the four-form $F\wedge F$ are also finite, local and superconformal invariants, and so they can in principle be added as part of the specification of the renormalization scheme. However these terms are parity odd, and so we do not include them as finite counterterms here.} These superconformal invariants are  
\bsub
\label{invariants}
\bal
\ce=&\;R^{ijkl}R_{ijkl}-4R^{ij}R_{ij}+R^2,\\
\cw^2=&\;R^{ijkl}R_{ijkl}-2R^{ij}R_{ij}+\frac13R^2-\frac{8}{\ell^2}F^{ij}F_{ij}+\co(\J^2).
\eal
\esub
$\ce$ is the Euler density, which is a topological density and therefore does not receive fermionic corrections. The bosonic part of $\cw^2$ is a linear combination of the square of the Weyl tensor and $F_{ij}F^{ij}$. Notice that even though the square of the Weyl tensor and $F_{ij}F^{ij}$ are separately Weyl invariant, only the specific linear combination in $\cw^2$ preserves supersymmetry. This means that there is a {\em two}-parameter family of supersymmetric renormalization schemes in this case.\footnote{It may be useful to emphasize the distinction between a `regularization scheme' or `ultraviolet regulator', and a `renormalization scheme', both of which may break the symmetries of the theory in different ways. In holography the regularization scheme is the radial cutoff, which as we have seen above, breaks both supersymmetry and Weyl symmetry, leading to genuine anomalies  (i.e. non-trivial cocycles) for the corresponding symmetries, which we will derive below. The {\em renormalization} scheme, however, corresponds to the choice of finite local and covariant boundary terms, which, by construction, can only break symmetries through {\em trivial} contributions to the anomalies. Choosing a renormalization scheme that breaks the least symmetries --possibly none-- removes all trivial cocycles from the anomalies.} 

The term $\wt{\bb S}\sub{4}$ in the solution \eqref{HJ-sol} of the Hamilton-Jacobi equation is proportional to the linear combination 
\be
\cw^2-\ce=2R^{ij}R_{ij}-\frac23R^2-\frac{8}{\ell^2}F_{ij}F^{ij}+\co(\J^2)=\frac{2\k^2}{\sqrt{-\g}}\frac{2}{\ell^2}\bb R\sub{4},
\ee
which, as we will discuss in section \ref{dict}, corresponds to the holographic Weyl anomaly for $d=4$ and can be shown to be a superconformal invariant. Since the Euler density does not receive any contribution from the gravitino, it follows that the expression \eqref{R4} determines the fermionic part of the superconformal invariant $\cw^2$, up to quadratic order in the gravitino. To our knowledge, this is the first instance where this superconformal invariant, including the fermionic part, has been determined, at least in component language.

Combining the observations of this subsection with the general asymptotic solution of the Hamiton-Jacobi equation in \eqref{HJ-sol}, we conclude that the most general form of the local boundary counterterms compatible with supersymmetry (and parity), and which do not add trivial cocycles of the Wess-Zumino consistency conditions to the anomalies, is 
\be\label{ct}\boxed{
S\sbtx{ct}=\left\{\begin{matrix}
	-\int d^3x\;\(\bb L\sub{0}+\bb L\sub{2}\), & d=3,\vspace{0.2cm}\\ 
	-\int d^4x\;\(\bb L\sub{0}+\bb L\sub{2}+\wt{\bb L}\sub{4}\log (e^{-2r_o/\ell})+s_1\sqrt{-\g}\;\ce+s_2\sqrt{-\g}\;\cw^2\), & d=4,
	\end{matrix}
	\right.}
\ee
where $s_1$ and $s_2$ are arbitrary parameters, corresponding to a choice of renormalization scheme. In particular, the holographic renormalization scheme for $d=3$ is unique, while for $d=4$ there is a two-parameter family of supersymmetric schemes.  Once a choice of scheme has been made, the renormalized on-shell action is defined through the limit 
\be\label{Sren}
S\sbtx{ren}=\lim_{r_o\to\infty}(S+S\sbtx{GH}+S\sbtx{ct}).
\ee
The holographic dictionary identifies this quantity with the renormalized generating functional of the dual gauge-invariant operators, as we will review in section \ref{dict}.

\subsection{Fefferman-Graham asymptotic expansions from flow equations}

Besides the boundary counterterms, the asymptotic solution \eqref{HJ-sol} of the Hamilton-Jacobi equation also determines the general Fefferman-Graham expansions of the bulk fields. In particular, inserting the solution \eqref{HJ-sol} in the expressions \eqref{HJ-momenta} for the canonical momenta, and these in turn into the Hamilton equations \eqref{flow-eqs}, results in a set of first order flow equations that can be integrated to obtain the asymptotic expansions of the fields. Up to the order necessary to obtain both the normalizable and non-normalizable modes of the asymptotic expansions, the first order equations take the form
\bal\label{flow-eqs-FG} 
\dot e^a_i=&\; \frac1\ell e^a_i+\frac{\ell}{d-2}\(e^{a}_kR_{i}^k-\frac{1}{2(d-1)}Re^a_i\)\NO\\
&+\log e^{-2r/\ell}\Bigg[\frac{\ell^3}{4(d-2)^2}\Bigg(\frac{3}{d-1}e^a_i\Big(R_{pq}R^{pq}-\frac{d}{4(d-1)}R^2+\frac13\square R\Big)\NO\\
&+e^a_j\Big(\frac{d-2}{d-1}D^jD_iR+\frac{d}{d-1}RR^j_i-2\square R^j_i-4R_{pq}R^{pjq}{}_i\Big)\Bigg)+\ell\Big(e^a_j F^{jp}F_{ip}-\frac{3}{4(d-1)}e^a_iF^{pq}F_{pq}\Big)\Bigg]\NO\\
&+\Big(\frac{1}{d-1}e^a_i e^b_j-e^a_j e^b_i\Big)\k^2\Hat\p_{(4)b}^j(x)e^{-5r/\ell}+\cdots,\NO\\
\dot A_i=&\;\frac{\ell}{2}\log e^{-2r/\ell}D_jF^j{}_i+c_1\Hat\e_i{}^{jkl}F_{jk}A_l-\frac{\k^2}{2}\Hat\p_{(4)i}(x)e^{-2r/\ell}+\cdots, \NO\\
\dot \J_{+i}=&\;\frac{1}{2\ell}\J_{+i}+\frac{\ell}{2(d-2)}\Big(R^j_i-\frac{1}{2(d-1)}R\d^j_i\Big)\J_{+j}\NO\\
&+\frac{\ell}{2(d-1)(d-2)}(\Hat\G_{ip}-(d-2)\g_{ip})\Big[2\Hat\G^q\Hat\G^{pkj}\cd_q\cd_k\J_{+j}+\Big(R^{pq}-\frac12\g^{pq}R\Big)\Hat\G_q{}^j\J_{+j}\Big]\NO\\
&+\frac{ic_2}{2\ell}F_{kl}\(-2\Hat\G_i{}^{jkl}+3(d-3)\d_i^{[j}\Hat\G^{kl]}+2\g^{lj}(\Hat\G_{i}{}^{k}-(d-2)\d^k_i)\)\J_{+j}+\cdots,\NO\\
\J_{-i}=&\;-\frac{\ell}{(d-1)(d-2)}(\Hat\G_{ij}-(d-2)\g_{ij})\Hat\G^{jkl}\cd_k\J_{+l}\NO\\
&+\frac{1}{d-1}\log(e^{-2r/\ell})(\Hat\G_{ij}-(d-2)\g_{ij})\Bigg\{\frac{\ell^3}{16}\Big[\frac{1}{6} (D_kR)\Hat\G^{jkl}\J_{+l}+\frac13R\Hat\G^{jkl}\cd_k\J_{+l}\NO\\
&-\frac{1}{3}(D^{[j}R)\Hat\G^{k]}\J_{+k}+\frac{5}{3}\((D_k R^p_l)\Hat\G^{ikl}\J_{+p}+2R^{[p}_l\Hat\G^{j]kl}\cd_k\J_{+p}\)+2(D^{[j}R^{k]}_l)\Hat\G^l\J_{+k}\NO\\
&+\frac{2}{3}(D_kR^{l[j})\Hat\G^{k]}\J_{+l}+\frac43R^{l[j}\Hat\G^{k]}\cd_k\J_{+l}+\frac23(D_kR)\Hat\G^{[j}\J_{+}^{k]}+\frac43R\Hat\G^{[j}\cd_k\J_{+}^{k]}\NO\\
&-2(D_kR^{[j}_l)\Hat\G^l\J_+^{k]}-4R^{[j}_l\Hat\G^l\cd_k\J_+^{k]}-\frac{2}{3}\Hat\G^{jkl}(\Hat\G_{ls}-2\g_{ls})\cd_k\slashed{\cd}\Hat\G^{spq}\cd_p\J_{+q}\Big] \NO\\
&-\frac{i\ell^2 }{4\sqrt{3}}\Big[(D_sF_{kl})\Hat\G^{jsp}\(-2\Hat\G_p{}^{qkl}+3\d_p^{[q}\Hat\G^{kl]}+2\g^{lq}(\Hat\G_{p}{}^k-2\d_{p}^k)\)\J_{+q}\NO\\
&\hskip0.6in+2F_{kl}\(-2\Hat\G^{jkl}{}_s+3\Hat\G^{[jk}\d_s^{l]}+2\g^{jk}(\Hat\G^l{}_{s}-2\d^l_{s})\)\Hat\G^{spq}\cd_p\J_{+q}\Big]\Bigg\}\NO\\
&+\frac{\k^2}{d-1}(\Hat\G_{ij}-(d-2)\g_{ij})\Hat\p_{\lbar\J(4)}^j(x)e^{-\frac{9r}{2\ell}}+\cdots, 
\eal 
where the ellipses stand for asymptotically subleading terms that only affect the asymptotic expansions beyond the normalizable mode. We have written these equations explicitly for the case $d=4$, but kept the coefficients generic so that one can also apply them to the simpler case $d=3$. Notice that the unspecified quantities  
$\Hat\p_{(4)b}^j(x)$, $\Hat\p_{(4)i}(x)$, $\Hat\p_{\lbar\J(4)}^j(x)$ correspond to the contribution to the flow equations from the undetermined finite part $\bb S\sub{4}$ (or $\bb S\sub{3}$ in the case $d=3$) in the solution \eqref{HJ-sol} of the Hamilton-Jacobi equation. As we will see in section \ref{dict}, these quantities are directly related to the corresponding local operators in the dual field theory.    

The first order equations \eqref{flow-eqs-FG} can be solved either recursively (see \cite{Papadimitriou:2016yit} for a step by step example), or by making an ansatz for the asymptotic form of the fields. The resulting asymptotic solutions for $d=4$ and $d=3$ are as follows.    

\paragraph{Fefferman-Graham expansions for $d=4$:} The asymptotic solutions take the form
\bal\label{FG4d}
e^a_i=&\; e^{r/\ell}e_i^a\sub{0}(x)+e^{-r/\ell}e_i^a\sub{2}(x)+e^{-3r/\ell}\(\log e^{-2r/\ell}\wt e_i^a\sub{4}(x)+e_i^a\sub{4}(x)\)+\cdots,\NO\\
A_i=&\;A_{(0)i}(x)+e^{-2r/\ell}\(\log e^{-2r/\ell}\wt A_{(4)i}(x)+ A_{(4)i}(x)\)+\cdots,\NO\\
\J_{+i}=&\;e^{\frac{r}{2\ell}}\J_{(0)+i}(x)+e^{-\frac{3r}{2\ell}}\J_{(2)+i}(x)+\cdots,\NO\\
\J_{-i}=&\;e^{-\frac{r}{2\ell}}\J_{(2)-i}(x)+e^{-\frac{5r}{2\ell}}\(\log e^{-2r/\ell} \wt \J_{(4)-i}(x)+\J_{(4)-i}(x)\)+\cdots,
\eal
where $e_i^a\sub{0}(x)$, $A_{(0)i}(x)$ and $\J_{(0)+i}(x)$ are arbitrary, while the remaining coefficients are given by
\bal\label{FG4d-coeffs}
e_i^a\sub{2}=&\;-\frac{\ell^2}{4}e^{a}_k\sub{0}\(R^k_i[g\sub{0}]-\frac{1}{6}R[g\sub{0}]\d^k_i\),\NO\\
\wt e_i^a\sub{4}=&\;-\frac{\ell}{4}\Bigg[\frac{\ell^3}{16}\Bigg(e^a_i\sub{0}\Big(R_{pq}[g\sub{0}]R^{pq}[g\sub{0}]-\frac13R^2[g\sub{0}]+\frac13\square\sub{0} R[g\sub{0}]\Big)\NO\\
&+e^a_j\sub{0}\Big(\frac{2}{3}D_{(0)}^jD_{(0)i}R[g\sub{0}]+\frac{4}{3}R[g\sub{0}]R^j_i[g\sub{0}]-2\square\sub{0} R^j_i[g\sub{0}]-4R_{pq}[g\sub{0}]R^{pjq}{}_i[g\sub{0}]\Big)\Bigg)\NO\\
&+\ell\Big(e^a_j\sub{0} F^{jp}_{(0)}F_{(0)ip}-\frac{1}{4}e^a_i\sub{0}F^{pq}_{(0)}F_{(0)pq}\Big)\Bigg],\NO\\
e_i^a\sub{4}=&\;-\frac{\ell}{4}\Big(\frac{1}{3}e^a_i\sub{0} e^b_j\sub{0}-e^a_j\sub{0} e^b_i\sub{0}\Big)\k^2\Hat\p_{(4)b}^j-\frac12\wt e_i^a\sub{4}\NO\\
&-\frac{\ell}{4}\Big[e^a_k\sub{0}\Big(R^{l[k}[g\sub{0}]g_{(2)i]l}-R^{kp}{}_i{}^q[g\sub{0}]g_{(2)pq}+D_{(0)}^{(k}D_{(0)}^lg_{(2)li)}-\frac12\square_{(0)}g^k_{(2)i}-\frac12D_{(0)}^kD_{(0)i}g\sub{2}\NO\\
&-\frac16\d^k_i(-R^{pq}[g\sub{0}]g_{(2)pq}+D_{(0)}^pD_{(0)}^qg_{(2)pq}-\square\sub{0}g\sub{2})\Big)+e^a_k\sub{2}\Big(R^k_i[g\sub{0}]-\frac{1}{6}R[g\sub{0}]\d^k_i\Big)\Big],\NO\\
\wt A_{(4)i}=&\;-\frac{\ell^2}{4}D_{(0)k}F_{(0)}{}^k{}_i,\NO\\
A_{(4)i}=&\;\frac{\k^2\ell}{4}\Hat\p_{(4)i}+\frac{\ell^2}{4}D_{(0)k}F_{(0)}{}^k{}_i+\frac{\ell^2}{3\sqrt{3}}\Hat\e_{(0)i}{}^{jkl}F_{(0)jk}A_{(0)l},\NO\\
\J_{(2)+i}=&\;-\frac{\ell^2}{8}\Big(R^j_i[g\sub{0}]-\frac{1}{6}R[g\sub{0}]\d^j_i\Big)\J_{(0)+j}\NO\\
&-\frac{\ell^2}{24}(\Hat\G_{(0)ip}-2g_{(0)ip})\Big[2\Hat\G_{(0)}^q\Hat\G^{pkj}\cd_{(0)q}\cd_{(0)k}\J_{(0)+j}+\Big(R^{pq}[g\sub{0}]-\frac12g_{(0)}^{pq}R[g\sub{0}]\Big)\Hat\G_{(0)q}{}^j\J_{(0)+j}\Big]\NO\\
&-\frac{i\ell}{8\sqrt{3}}F_{(0)kl}\(-2\Hat\G_{(0)i}{}^{jkl}+3\d_i^{[j}\Hat\G_{(0)}^{kl]}+2g_{(0)}^{lj}(\Hat\G_{(0)i}{}^{k}-2\d^k_i)\)\J_{(0)+j},\NO\\
\J_{(2)-i}=&\;-\frac{\ell}{6}(\Hat\G_{(0)ij}-2g_{(0)ij})\Hat\G_{(0)}^{jkl}\cd_{(0)k}\J_{(0)+l},\NO\\
\wt \J_{(4)-i}=&\;\frac{1}{3}(\Hat\G_{(0)ij}-2g_{(0)ij})\Bigg\{\frac{\ell^3}{16}\Big[\frac{1}{6} (D_{(0)k}R[g\sub{0}])\Hat\G_{(0)}^{jkl}\J_{(0)+l}+\frac13R[g\sub{0}]\Hat\G_{(0)}^{jkl}\cd_{(0)k}\J_{(0)+l}\NO\\
&-\frac{1}{3}(D_{(0)}^{[j}R[g\sub{0}])\Hat\G_{(0)}^{k]}\J_{(0)+k}+\frac{5}{3}\((D_{(0)k} R[g\sub{0}]^p_l)\Hat\G_{(0)}^{ikl}\J_{(0)+p}+2R^{[p}_l[g\sub{0}]\Hat\G_{(0)}^{j]kl}\cd_{(0)k}\J_{(0)+p}\)\NO\\
&+2(D_{(0)}^{[j}R^{k]}_l[g\sub{0}])\Hat\G_{(0)}^l\J_{(0)+k}+\frac{2}{3}(D_{(0)k}R^{l[j}[g\sub{0}])\Hat\G_{(0)}^{k]}\J_{(0)+l}+\frac43R^{l[j}[g\sub{0}]\Hat\G_{(0)}^{k]}\cd_{(0)k}\J_{(0)+l}\NO\\
&+\frac23(D_{(0)k}R[g\sub{0}])\Hat\G_{(0)}^{[j}\J_{(0)+}^{k]}+\frac43R[g\sub{0}]\Hat\G_{(0)}^{[j}\cd_{(0)k}\J_{(0)+}^{k]}-2(D_{(0)k}R^{[j}_l[g\sub{0}])\Hat\G_{(0)}^l\J_{(0)+}^{k]}\NO\\
&-4R^{[j}_l[g\sub{0}]\Hat\G_{(0)}^l\cd_{(0)k}\J_{(0)+}^{k]}-\frac{2}{3}\Hat\G_{(0)}^{jkl}(\Hat\G_{(0)ls}-2g_{(0)ls})\cd_{(0)k}\slashed{\cd}_{(0)}\Hat\G_{(0)}^{spq}\cd_{(0)p}\J_{(0)+q}\Big]\NO\\
&-\frac{i\ell^2 }{4\sqrt{3}}\Big[(D_{(0)s}F_{(0)kl})\Hat\G_{(0)}^{jsp}\(-2\Hat\G_{(0)p}{}^{qkl}+3\d_p^{[q}\Hat\G_{(0)}^{kl]}+2g_{(0)}^{lq}(\Hat\G_{(0)p}{}^k-2\d_{p}^k)\)\J_{(0)+q}\NO\\
&\hskip0.6in+2F_{(0)kl}\(-2\Hat\G_{(0)}^{jkl}{}_s+3\Hat\G_{(0)}^{[jk}\d_s^{l]}+2g_{(0)}^{jk}(\Hat\G_{(0)}^l{}_{s}-2\d^l_{s})\)\Hat\G_{(0)}^{spq}\cd_{(0)p}\J_{(0)+q}\Big]\Bigg\},\NO\\
\J_{(4)-i}=&\;\frac{\k^2}{3}(\Hat\G_{(0)ij}-2g_{(0)ij})\Hat\p_{\lbar\J(4)}^j+\frac{\ell}{6}\Big\{\Big[e_{ia}\sub{2}\Hat\G_{(0)}^{akl}+2e^a_j\sub{2}\Big(e^{[k}_a\sub{0}\Hat\G_{(0)i}{}^{l]j}+2\d^{[k}_ie^{l]}_a\sub{0}\Hat\G_{(0)}^j\Big)\Big]\cd_{(0)k}\J_{(0)+l}\NO\\
&+(\Hat\G_{(0)i}{}^{kl}-4\d^{[k}_i\Hat\G_{(0)}^{l]})\Big[\cd_{(0)k}\J_{(2)+l}\NO\\
&+\frac14\Big((D_{(0)[k}e^a_{p]}\sub{2})e_{qa}\sub{0}-(D_{(0)[p}e^a_{q]}\sub{2})e_{ka}\sub{0}+(D_{(0)[q}e^a_{k]}\sub{2})e_{pa}\sub{0}\Big)\Hat\G_{(0)}^{pq}\J_{(0)+l}\Big]\Big\}.
\eal
In these expressions $g_{(0)ij}=e_i^a\sub{0}e_{ja}\sub{0}$ is the boundary metric and the notations $g^i_{(2)j}$ and $g_{(2)}$ stand respectively for $g_{(0)}^{ik}g_{(2)kj}$ and $g_{(0)}^{ij}g_{(2)ij}$. Notice that the normalizable modes $e_i^a\sub{4}$, $A_{(4)i}$ and $\J_{(4)-i}$ are related respectively to the undetermined canonical momenta $\Hat\p_{(4)b}^j(x)$, $\Hat\p_{(4)i}(x)$, $\Hat\p_{\lbar\J(4)}^j(x)$ obtained from the finite part $\bb S\sub{4}$ in \eqref{HJ-sol}. Since the latter are directly related to the dual operators, the expressions for $e_i^a\sub{4}$, $A_{(4)i}$ and $\J_{(4)-i}$ in \eqref{FG4d-coeffs} provide the explicit relation between the dual operators and the coefficients of the Fefferman-Graham expansions. Finally, although the solution \eqref{HJ-sol} allows one to determine the Fefferman-Graham expansions of the bosonic fields up to quadratic order in the gravitino, we only give explicitly these expansions to zero order in the gravitino in \eqref{FG4d-coeffs}. These expansions generalize earlier results for the case of flat boundary metric and vanishing gauge field \cite{Argurio:2014uca}.

\paragraph{Fefferman-Graham expansions for $d=3$:} The asymptotic expansions for $d=3$ do not contain any logarithmic terms and therefore take simpler form 
\bal\label{FG3d}
e^a_i=&\; e^{r/\ell}e_i^a\sub{0}(x)+e^{-r/\ell}e_i^a\sub{2}(x)+e^{-2r/\ell}e_i^a\sub{3}(x)+\cdots,\NO\\
A_i=&\;A_{(0)i}(x)+e^{-r/\ell} A_{(3)i}(x)+\cdots,\NO\\
\J_{+i}=&\;e^{\frac{r}{2\ell}}\J_{(0)+i}(x)+e^{-\frac{3r}{2\ell}}\J_{(2)+i}(x)+\cdots,\NO\\
\J_{-i}=&\;e^{-\frac{r}{2\ell}}\J_{(2)-i}(x)+e^{-\frac{3r}{2\ell}}\J_{(3)-i}(x)+\cdots,
\eal\label{FG3d-coeffs}
where $e_i^a\sub{0}(x)$, $A_{(0)i}(x)$ and $\J_{(0)+i}(x)$ are again arbitrary and the remaining coefficients are 
\bal
e_i^a\sub{2}=&\;-\frac{\ell^2}{2}e^{a}_k\sub{0}\(R^k_i[g\sub{0}]-\frac{1}{4}R[g\sub{0}]\d^k_i\),\NO\\
e_i^a\sub{3}=&\;-\frac{\ell}{3}\Big(\frac{1}{2}e^a_i\sub{0} e^b_j\sub{0}-e^a_j\sub{0} e^b_i\sub{0}\Big)\k^2\Hat\p_{(3)b}^j,\NO\\
A_{(3)i}=&\;\frac{\k^2\ell}{2}\Hat\p_{(3)i},\NO\\
\J_{(2)+i}=&\;-\frac{\ell^2}{4}\Big(R^j_i[g\sub{0}]-\frac{1}{4}R[g\sub{0}]\d^j_i\Big)\J_{(0)+j}\NO\\
&-\frac{\ell^2}{8}(\Hat\G_{(0)ip}-g_{(0)ip})\Big[2\Hat\G_{(0)}^q\Hat\G^{pkj}\cd_{(0)q}\cd_{(0)k}\J_{(0)+j}+\Big(R^{pq}[g\sub{0}]-\frac12g_{(0)}^{pq}R[g\sub{0}]\Big)\Hat\G_{(0)q}{}^j\J_{(0)+j}\Big]\NO\\
&-\frac{i\ell}{8}F_{(0)kl}\(-2\Hat\G_{(0)i}{}^{jkl}+2g_{(0)}^{lj}(\Hat\G_{(0)i}{}^{k}-\d^k_i)\)\J_{(0)+j},\NO\\
\J_{(2)-i}=&\;-\frac{\ell}{2}(\Hat\G_{(0)ij}-g_{(0)ij})\Hat\G_{(0)}^{jkl}\cd_{(0)k}\J_{(0)+l},\NO\\
\J_{(3)-i}=&\;\frac{\k^2}{2}(\Hat\G_{(0)ij}-g_{(0)ij})\Hat\p_{\lbar\J(3)}^j.
\eal
Notice that in this case the relation between the normalizable modes $e_i^a\sub{3}$, $A_{(3)i}$ and $\J_{(3)-i}$ and the undetermined momenta $\Hat\p_{(3)b}^j(x)$, $\Hat\p_{(3)i}(x)$, $\Hat\p_{\lbar\J(3)}^j(x)$ obtained from the finite part $\bb S\sub{3}$ is very simple. These asymptotic expansions are in agreement with those obtained in \cite{Amsel:2009rr} in a similar setup.

\section{Local operators, Ward identities and superconformal anomalies}
\label{dict}
\setcounter{equation}{0}

In this section we use the boundary counterterms \eqref{ct} in order to construct the holographic dictionary for the theory \eqref{action}. In particular, we define the local operators dual to the bulk fields and derive the Ward identities these operators satisfy, including all bosonic and fermionic anomalies. These anomalies  play a central role in our analysis of supersymmetric backgrounds in section \ref{backgrounds}.    

The operators dual to the bulk fields correspond to the renormalized canonical momenta, namely 
\bal
\label{ops}
&\ct^i_a =-\lim_{r_o\to\infty}\(\frac{e^{(d+1)r_o/\ell}}{\sqrt{-\g}}\(\p^i_a+\frac{\d S\sbtx{ct}}{\d e^a_i}\)\),
&&\cj^i =\lim_{r_o\to\infty}\(\frac{e^{dr_o/\ell}}{\sqrt{-\g}}\(\p^i+\frac{\d S\sbtx{ct}}{\d A_i}\)\),\NO\\
&\cs^i =\lim_{r_o\to\infty}\(\frac{e^{(d+1/2)r_o/\ell}}{\sqrt{-\g}}\(\p^i_{\lbar\J}+\frac{\d S\sbtx{ct}}{\d \lbar\J_i}\)\),
&&\lbar\cs^i =\lim_{r_o\to\infty}\(\frac{e^{(d+1/2)r_o/\ell}}{\sqrt{-\g}}\(\p^i_{\J}+ S\sbtx{ct}\frac{\overleftarrow\d}{\d \lbar\J_i}\)\).
\eal
These relations are the holographic version of the local renormalization group definition of local operators \cite{Osborn:1991gm}, which is why we omit the bracket notation $\<\cdot\>$. Each of these operators is a function of arbitrary sources and so correlation functions can be computed by further differentiating these operators with respect to the sources. Notice that for $d=4$, these operators depend on the value of the parameters $s_1$ and $s_2$ that parameterize the supersymmetric renormalization schemes. Specifically, the definition \eqref{ops} of the local operators implies that they are identified respectively with the quantities $\Hat\p_{(3)b}^j$, $\Hat\p_{(3)i}$ and $\Hat\p_{\lbar\J(3)}^j$ appearing in the Fefferman-Graham expansions for $d=3$, while for $d=4$ they correspond to $\Hat\p_{(4)b}^j$, $\Hat\p_{(4)i}$ and $\Hat\p_{\lbar\J(4)}^j$, up to terms proportional to the parameters $s_1$ and $s_2$ coming from the finite local counterterms parameterizing the renormalization scheme.   

The Ward identities that the local operators \eqref{ops} satisfy can be derived by inserting the covariant expansion \eqref{exp} in the constraints
\eqref{constraint-expressions} and \eqref{frame-constraint} and isolating the terms of dilatation weight zero. This leads to the identities 
\bsub
\label{WIDs-momentum}
\bal
&D_j(-e^{a}_i\p_a^j\sub{d}-\p_\J^j\sub{d}\J_{+i}-\lbar\J_{+i}\p_{\lbar\J}^j\sub{d})+\p_\J^j\sub{d}\cd_i\J_{+j}+\lbar\J_{+j}\overleftarrow{\cd }_i\p_{\lbar\J}^j\sub{d}+F_{ij}\p^j_{(d)}=\sqrt{-\g}\;\ca_{Mi},\hskip-0.1cm\\
&D_i\p^i\sub{d} +i\frak g\(\p_\J^i\sub{d}\J_{+i}-\lbar\J_{+i}\p_{\lbar\J}^i\sub{d}\)=\sqrt{-\g}\;\ca_R,\\
& \cd_i\p_{\lbar\J}^i\sub{d}-\frac12\p_a^i\sub{d}\G^a\J_{+i}-\frac{ic_2}{2(d-2)}\p^i_{(d)}(\Hat\G_{ij}-(d-2)\g_{ij})\Hat\G^{jpq}\cd_p\J_{+q}=\sqrt{-\g}\;\ca_S,\\
&-e_i^a\p_a^i\sub{d}-\frac12\lbar\J_{+i}\p^i_{\lbar\J}\sub{d}-\frac12\p^i_{\J}\sub{d}\J_{+i}=-\ell\; \bb R\sub{d}=\sqrt{-\g}\;\ca_W,\\
&\Hat\G_i\p_{\lbar\J}^i\sub{d}-\frac{i(d-1)c_2}{2}\p^i\sub{d}\J_{+i}=\sqrt{-\g}\;\ca_{sW},\\
&e^{i[a}\p^{b]}_i\sub{d}-\frac14(\p_\J^i\sub{d}\G^{ab}\J_{+i}-\lbar\J_{+i}\G^{ab}\p_{\lbar\J}^i\sub{d})=0, 
\eal
\esub
where the expressions on the r.h.s. vanish identically for $d=3$, while for $d=4$ are local in terms of the induced fields and take the form\footnote{Although these anomalies apply specifically to the case $d=4$, for technical reasons we keep $d$ generic.} 
\bsub
\label{anomalies}
\bal
\ca_{Mi} &=\frac{2c_1}{\k^2}\;\Hat\e^{jklp}F_{ij}F_{kl}A_p,\\
\ca_R &=-\frac{c_1}{2\k^2}\;\Hat\e^{ijkl}F_{ij}F_{kl},\\
\ca_{sW} &= \frac{\ell}{2\k^2}\Big[\frac{\ell^2}{(d-2)^2}\(R_{ij}-\frac{1}{2(d-1)}R\g_{ij}\)\Hat\G^i\Hat\G^{jkl}\cd_k\J_{+l}-\frac{2i(d-1)c_1c_2}{\ell}\;\Hat\e^{ijkl}F_{jk}A_l\J_{+i}\NO\\
&\hskip0.5in+\frac{ic_2}{d-2}F_{jk}\((d-2)\Hat\G^{jk}\Hat\G^i-(d-1)\Hat\G^{jki} \)\Hat\G_i{}^{pq}\cd_p\J_{+q}\Big],\\
\ca_{S} &=-\frac{1}{\k^2}\frac{ic_1c_2}{(d-2)}\;\Hat\e^{iskl}F_{sk}A_l(\Hat\G_{ij}-(d-2)\g_{ij})\Hat\G^{jpq}\cd_p\J_{+q},\\
\ca_W &=\frac{\ell^3}{2\k^2(d-2)^2}\Bigg\{\frac{d}{4(d-1)}R^2-R_{ij}R^{ij}+\frac{(d-2)^2}{\ell^2}F_{ij}F^{ij}\NO\\
&-\frac{1}{2(d-1)}R\( \lbar\J_{+i}\overleftarrow{\cd}_j\Hat\G^{ijk}\J_{+k}-\lbar\J_{+i}\Hat\G^{ijk}\cd_j\J_{+k}\)+\frac{d}{d-1}RD^{[i}(\lbar\J_{+i}\Hat\G^{j]}\J_{+j})\NO\\
&-\frac{(2d-3)}{d-1}R^k_l\( \lbar\J_{+i}\overleftarrow{\cd}_j\Hat\G^{ijl}\J_{+k}-\lbar\J_{+k}\Hat\G^{lji}\cd_j\J_{+i}\)+2R^{j[k}D_j(\lbar\J_{+i}\Hat\G^{i]}\J_{+k})+2R^{[i}_kD^{j]}(\lbar\J_{+i}\Hat\G^k\J_{+j})\NO\\
&-\frac{2}{d-1}R^{ik}\(\lbar\J_{+[i}\overleftarrow\cd_{j]}\Hat\G^j\J_{+k}-\lbar\J_{+[i}\Hat\G^j\cd_{j]}\J_{+k}\)-\frac{2}{d-1}R\(\lbar\J_{+[i}\overleftarrow\cd_{j]}\Hat\G^i\J_{+}^j-\lbar\J_{+[i}\Hat\G^i\cd_{j]}\J_{+}^j\)\NO\\
&+2R^i_k\(\lbar\J_{+[i}\overleftarrow\cd_{j]}\Hat\G^k\J_{+}^j-\lbar\J_{+[i}\Hat\G^k\cd_{j]}\J_{+}^j\)\NO\\
&\hskip0.3in+\frac{1}{d-1}\lbar\J_{+k}\overleftarrow\cd_l\Hat\G^{kli}\((\Hat\G_{ij}-(d-2)\g_{ij})\slashed{\cd}-\overleftarrow{\slashed{\cd}}(\Hat\G_{ij}-(d-2)\g_{ij})\)\Hat\G^{jpq}\cd_p\J_{+q}\Bigg\} \NO\\
&+\frac{ic_2\ell}{2\k^2(d-2)}F_{kl}\Big[\lbar\J_{+p}\overleftarrow\cd_q\Hat\G^{pqi}\(-2\Hat\G_i{}^{jkl}+3(d-3)\d_i^{[j}\Hat\G^{kl]}+2\g^{lj}(\Hat\G_{i}{}^k-2(d-2)\d_{i}^k)\)\J_{+j}\NO\\
&\hskip0.6in-\lbar\J_{+i}\(-2\Hat\G^{ikl}{}_j+3(d-3)\Hat\G^{[ik}\d_j^{l]}+2\g^{ik}(\Hat\G^l{}_{j}-(d-2)\d^l_{j})\)\Hat\G^{jpq}\cd_p\J_{+q}\Big].
\eal
\esub

Combining the definitions \eqref{ops} of the dual operators with the identities 
\eqref{WIDs-momentum} (taking into account the fact that the finite local counterterms in \eqref{ct} parameterizing the supersymmetric renormalization scheme by construction obey the identities \eqref{WIDs-momentum} without the anomaly terms)
and removing the radial cutoff results in the Ward identities
\bsub
\label{WIDs}
\bal
&\hskip-0.1cm D_{(0)j}(e^{a}_{(0)i}\ct_a^j-\lbar\cs^j\J_{(0)+i}-\lbar\J_{(0)+i}\cs^j)+\lbar\cs^j\cd_{(0)i}\J_{(0)+j}+\lbar\J_{(0)+j}\overleftarrow{\cd }_{\hskip-0.08cm(0)i}\cs^j+F_{(0)ij}\cj^j=\ca_{M(0)i},\label{WI-diff} \hskip-0.0cm\\
&\hskip-0.1cm D_{(0)i}\cj^i +i\frak g(\lbar\cs^i\J_{(0)+i}-\lbar\J_{(0)+i}\cs^i)=\ca_{R(0)},\label{WI-R}\\
&\hskip-0.1cm \cd_{(0)i}\cs^i+\frac12\ct^i_a\G^a\J_{(0)+i}-\frac{ic_2}{2(d-2)}\cj^i(\Hat\G_{(0)ij}-(d-2)g_{(0)ij})\Hat\G^{jpq}_{(0)}\cd_{(0)p}\J_{(0)+q}=\ca_{S(0)},\\
&\hskip-0.1cm e_{(0)i}^a\ct_a^i-\frac12\lbar\J_{(0)+i}\cs^i-\frac12\lbar\cs^i\J_{(0)+i}=\ca_{W(0)},\label{WI-trace}\\
&\hskip-0.1cm \Hat\G_{(0)i}\cs^i-\frac{i(d-1)c_2}{2}\cj^i\J_{(0)+i}=\ca_{sW(0)},\\
&\hskip-0.1cm e^{i[a}_{(0)}\ct^{b]}_i+\frac14(\lbar\cs^i\G^{ab}\J_{(0)+i}-\lbar\J_{(0)+i}\G^{ab}\cs^i)=0,\label{WI-Lorentz}
\eal
\esub
where the local terms on the r.h.s. of these identities are identical to the expressions on the radial cutoff in \eqref{anomalies}, except that the induced fields are appropriately replaced with the corresponding sources on the boundary of AdS. Analogous Ward identities, but with flat metric $g_{(0)ij}$, were obtained for the $\cn=1$ theory dual to the Klebanov-Strassler background in \cite{Bertolini:2015hua}. It cannot be overemphasized that the Ward identities \eqref{WIDs} and the related anomalies \eqref{anomalies} apply to generic $\cn=2$ superconformal quantum field theories in three and generic $\cn=1$ superconformal quantum field theories four dimensions, except that the form of the anomalies \eqref{anomalies} in the four dimensional case holds for theories with $a=c$ only, and the value of these anomaly coefficients may be different in a generic theory. Crucially, the form of the Ward identities \eqref{WIDs} applies to any such theory, irrespectively of whether it admits a holographic dual. In particular, the alternative derivation of the Ward identities discussed in the following subsection is entirely field theoretic and allows one to extrapolate the results of section \eqref{backgrounds} beyond holographic theories.

\subsection{Boundary supersymmetry and current anomalies}

The Ward identities \eqref{WIDs} reflect the local bulk symmetries surviving on the conformal boundary, including those broken by anomalies arising from the regularization and renormalization procedure. An alternative derivation of these identities, therefore, may be achieved by identifying the bulk symmetries acting on the boundary sources and considering how the renormalized generating functional transforms. From a bulk perspective, such symmetries correspond to all local bulk transformations preserving the Fefferman-Graham gauge \eqref{FG-gauge}. In the case of bulk diffeomorphisms these are known as Penrose-Brown-Henneaux (PBH) transformations \cite{Penrose:1987uia,Brown:1986nw,Imbimbo:1999bj}. In appendix \ref{sPBH} we derive the most general bulk symmetry transformations of the supergravity action \eqref{action} preserving the Fefferman-Graham gauge. We will refer to these as generalized Penrose-Brown-Henneaux transformations.   

One of the main results of the analysis in appendix B is the identification of the local symmetries on the conformal boundary and their action on the field theory sources. Specifically, we find that the local symmetry transformations on the boundary are generated by the arbitrary bosonic parameters $\s(x)$, $\x_o^i(x)$, $\l_o^{ab}(x)$, as well as the fermionic variables $\e_{o+}(x)$ and $\e_{o-}(x)$, all of which are functions of the boundary coordinates only. Removing the radial cutoff in the asymptotic relations 
\eqref{PBH-sources-cutoff} we determine that these transformations act on the field theory sources as
\bsub
\label{PBH-sources}
\bal
\d_{\s,\x_o,\l_o,\th_o,\e_{o+},\e_{o-}} e_i^a\sub{0} &= \frac{\s}{\ell} e_i^a\sub{0}+ \x^j_o\pa_j e_i^a\sub{0}+(\pa_i \x^j_o) e_j^a\sub{0}-\l_o^a{}_b e_i^b\sub{0}\NO\\
&\hskip0.5cm+\frac12(\lbar\e_{o+}\G^a\J_{(0)+i}-\lbar\J_{(0)+i}\G^a\e_{o+}),\\
\d_{\s,\x_o,\l_o,\th_o,\e_{o+},\e_{o-}} A_{(0)i} &= \x^j_o\pa_j A_{(0)i}+(\pa_i \x^j_o)A_{(0)j}+\pa_i\th_o\NO\\
&\hskip0.5cm +ic_3\(\lbar\J_{(0)+i}\e_{o-}+\lbar\J_{(2)-i}\e_{o+}-\lbar\e_{o+}\J_{(2)-i}-\lbar\e_{o-}\J_{(0)+i}\),\\
\d_{\s,\x_o,\l_o,\th_o,\e_{o+},\e_{o-}}\J_{(0)+i} &=\frac{\s}{2\ell}\J_{(0)+i}+\x^j_o\pa_j\J_{(0)+i}+(\pa_i\x^j_o)\J_{(0)+j}-\frac14\l_o^{ab}\G_{ab}\J_{(0)+i}-i\frak g\th_o\J_{(0)+i}\NO\\
&\hskip0.5cm +\cd_{(0)i}\e_{o+}-\frac{1}{\ell}\Hat\G_{(0)i}\e_{o-},
\eal
\esub
where from \eqref{Psi-} and \eqref{momenta-2}
\be
\J_{(2)-i}=-\frac{\ell}{(d-1)(d-2)}(\Hat\G_{(0)ij}-(d-2)g_{(0)ij})\Hat\G^{jkl}_{(0)}\cd_{(0)k}\J_{(0)+l}.
\ee 
Notice that, although these transformations are derived holographically here, they should apply to any $\cn=2$ superconformal theory in $d=3$ and any $\cn=1$ superconformal theory $d=4$. 

Given the general variation of the renormalized on-shell action 
\be\label{Sren-var}
\d S\sbtx{ren}=\int d^dx\sqrt{-g\sub{0}}\(-\ct^i_a\d e^a_i\sub{0}+\cj^i\d A_{(0)i}+\lbar\cs^i\d \J_{(0)+i}+\d \lbar\J_{(0)+i}\cs^i\),
\ee
the Noether procedure for the transformations \eqref{PBH-sources} leads to an alternative derivation of the Ward identities \eqref{WIDs}. In particular, inserting the transformations \eqref{PBH-sources} for the sources in \eqref{Sren-var} and parameterizing the anomalies such that 
\bal\label{Sren-anomalies}
\d_{\s,\x_o,\l_o,\th_o,\e_{o+},\e_{o-}} S\sbtx{ren}=&\;\int d^dx\sqrt{-g\sub{0}}\Big(-\frac{\s}{\ell}\ca_{W(0)}-\th_o\ca_{R(0)}\NO\\
&\hskip2.5cm-\lbar\e_{o+}\ca_{S(0)}-\lbar\ca_{S(0)}\e_{o+}+\frac1\ell\lbar\e_{o-}\ca_{sW(0)}+\frac1\ell\lbar\ca_{sW(0)}\e_{o-}\Big),
\eal
reproduces the Ward identities \eqref{WIDs}. However, the anomalies must be computed independently in this case, either holographically as above, or through field theory arguments. Note that the term $\ca_{M(0)i}$ in \eqref{WI-diff} cancels out in the variation of the generating function \eqref{Sren-anomalies}, implying that the theory is invariant under local diffeomorphisms.

Even though the field theory sources transform as tensors under the local symmetries, the corresponding operators do not, due to the anomalies. The simplest way to obtain the transformation of the currents under the local symmetries is by introducing a suitable Poisson bracket on the symplectic space of sources and local operators as discussed in appendix \ref{sPBH}. In particular, the general transformation of the current densities is given in \eqref{PBH-momenta}, with the anomalous part of the transformation shown in terms of the anomaly densities in \eqref{current-anomalies}. For our purpose of studying supersymmetric backgrounds in section \ref{backgrounds} it suffices to determine explicitly the transformations of the supercurrent under boundary supersymmetry and superWeyl transformations, generated respectively by $\e_{o+}(x)$ and $\e_{o-}(x)$. The general transformation identities \eqref{PBH-momenta} for the currents and the explicit form of the anomalies in \eqref{anomalies} determine  
\bsub
\label{supercurrent-anomalies}
\bal
\d_{\e_{o+}}\cs^i =&\;-\frac12\ct_a^i\G^a\e_{o+}\NO\\
&\hskip0.15cm+\frac{ic_2}{2(d-2)}\Hat\G_{(0)}^{ijk}\(\Hat\G_{(0)kl}-(d-2)g_{(0)kl}\)\cd_{(0)j}\[\Big(\cj^l-\frac{2c_1}{\k^2}\Hat\e^{lpqs}F_{(0)pq}A_{(0)s}\Big)\e_{o+}\],\\
\rule{0cm}{1.0cm}
\d_{\e_{o-}}\cs^i =&\;-\frac{i(d-1)c_2}{2\ell}\Big(\cj^i-\frac{2c_1}{\k^2}\Hat\e^{ipqs}F_{(0)pq}A_{(0)s}\Big)\e_{o-}\NO\\
&\hskip0.15cm-\frac{\ell^2}{2(d-2)^2\k^2}\Hat\G_{(0)}^{ijk}\Hat\G_{(0)}^l\cd_{(0)j}\[\Big(R_{kl}[g_{(0)}]-\frac{1}{2(d-1)}R[g\sub{0}]g_{(0)kl}\Big)\e_{o-}\]\NO\\
&\hskip0.15cm-\frac{ic_2}{2(d-2)\k^2}\Hat\G_{(0)}^{ij}{}_k\Big((d-2)\Hat\G_{(0)}^k\Hat\G_{(0)}^{pq}-(d-1)\Hat\G_{(0)}^{kpq}\Big)\cd_{(0)j}(F_{(0)pq}\e_{o-}).
\eal
\esub
Notice that, besides the tensorial part involving the currents, these transformations contain additional local terms due to the anomalies. This result will be central to our analysis of supersymmetric backgrounds in the next section. Another observation that will play an important role in the subsequent analysis  is that the anomalous supersymmetry transformations  
\eqref{supercurrent-anomalies} of the supercurrent can be obtained in a third way,  namely from the bulk supersymmetry transformation of the gravitino. Finally, it should be stressed that the anomalous transformations \eqref{supercurrent-anomalies} of the supercurrent are independent of the choice of supersymmetric renormalization scheme, i.e. of the choice of the parameters $s_1$ and $s_2$ in \eqref{ct}.

\section{Fermionic anomalies and supersymmetry breaking}
\label{backgrounds}
\setcounter{equation}{0}

The anomalous transformation \eqref{supercurrent-anomalies} of the supercurrent under local supersymmetry and superWeyl transformations is our main result. In this section we explore some of the consequences of this result for $\cn=1$ superconformal theories in four dimensions. In particular, we show that even if a field theory background admits a Killing spinor and all anomalies are numerically vanishing, supersymmetry can still be broken by the anomalous transformation of the supercurrent.    

\subsection{Field theory backgrounds admitting conformal Killing spinors}

In order to explore the consequences of the anomalous supercurrent transformation we consider a simple class of field theory backgrounds admitting Killing spinors that were originally obtained in \cite{Klare:2012gn,Dumitrescu:2012ha} and were the subject of the recent analysis of \cite{Genolini:2016ecx}. 

Setting the AdS$_5$ radius to $\ell=1$, this class of rigid four-dimensional supersymmetric field theory backgrounds takes the form\footnote{Notice that the gauge field in \eqref{susy-background} is the one obtained from background {\em conformal supergravity} \cite{Sohnius:1981tp} and not that of new minimal supergravity used in \cite{Dumitrescu:2012ha}. It is the former that is induced on the conformal boundary by the bulk graviphoton in holography \cite{Klare:2012gn,Cassani:2012ri}. } 
\bal\label{susy-background}
ds^2_{(0)} &=-\tx dt^2+(\tx d\j+a)^2+4e^w \tx dz\tx d\bar z,\NO\\
A_{(0)} & =-\frac{1}{\sqrt 3}\Big(-\frac 18 u\tx dt+\frac 14 u(\tx d\j+a)+\frac{\tx  i}{4}(\pa_{\bar z}w\tx d\bar z -\pa_z w\tx dz )+\g^\prime \tx dt+\g \tx d\j+\tx d\l \Big),
\eal 
where $w(z,\bar z)$ is a local function on the compact Riemann surface $\S_2$ parameterized by the complex coordinates $(z,\bar z)$, $a=a_z(z,\bar z)\tx dz+a_{\bar z}(z,\bar z)\tx d\bar z$ is a local one-form on $\S_2$ satisfying the condition  
\be\label{one-form-condition} 
\tx da=\tx i ue^w\tx dz\wedge \tx d\bar z,
\ee 
for a globally defined function $u(z,\bar z)$ on $\S_2$, and $\pa_\j$ is a nowhere vanishing Killing vector. The real constants $\g$ and $\g'$, as well as the  function $\l=\l(z,\bar z)$ on $\S_2$ are locally pure gauge, but their values are determined by the requirement that $A_{(0)}$ be a globally defined one-form. In particular, global considerations require that $\g'=0$ and $\g$ be proportional to the first Chern class of $\S_2$. For an extensive discussion of the global properties the backgrounds \eqref{susy-background} and explicit expressions for these parameters in terms of topological invariants we refer the reader to \cite{Genolini:2016ecx}.    

In the following it will be convenient to slightly reparameterize the background \eqref{susy-background} in order to solve explicitly the constraint \eqref{one-form-condition}. In particular, writing the components of the one-form $a$ in terms of a local function $\m(z,\bar z)$ as    
\be
a_z=-\frac{\tx i}{2}\pa_z\m,\qquad a_{\bar z}=\frac{\tx i}{2} \pa_{\bar z}\m,
\ee
the condition \eqref{one-form-condition} is automatically satisfied provided 
$u(z,\bar z)$ is expressed in terms of $w(z,\bar z)$ and $\m(z,\bar z)$ through the relation
\be
u=e^{-w}\pa_z\pa_{\bar z}\m.
\ee 
This allows us to parameterize the supersymmetric backgrounds \eqref{susy-background} in terms of the two unconstrained functions $w(z,\bar z)$ and $\m(z,\bar z)$ as
\bal\label{susy-background-mu}
ds^2_{(0)} =&\; -\tx dt^2+\Big(\tx d\j+\frac{\tx i}{2}\pa_{\bar z}\m \tx d\bar z-\frac{\tx i}{2}\pa_z\m \tx dz\Big)^2+4e^w \tx dz\tx d\bar z,\NO\\
A_{(0)} =&\; -\frac{1}{\sqrt 3}\Big[-\frac 18 e^{-w}\pa_z\pa_{\bar z}\m\tx dt+\frac 14 e^{-w}\pa_z\pa_{\bar z}\m\Big(\tx d\j+\frac{\tx i}{2}\pa_{\bar z}\m \tx d\bar z-\frac{\tx i}{2}\pa_z\m \tx dz\Big)\NO\\
&\hskip1.4cm+\frac{\tx i}{4}(\pa_{\bar z}w\tx d\bar z -\pa_z w\tx dz )+\g^\prime \tx dt+\g \tx d\j+\tx d\l \Big].
\eal

\paragraph{Conformal Killing spinors and vectors} A Killing vector $\ck^i$ of a generic bosonic background specified by the sources $e_{(0)i}^a$ and $A_{(0)i}$ corresponds to a diffeomorphism that leaves the sources invariant, up to local frame rotations and $U(1)$ gauge transformations. The general local transformations of the sources \eqref{PBH-sources} imply that in terms of the metric and the gauge invariant fieldstrength the conditions for $\ck^i$ to be a Killing vector are 
\be
\cl_{\ck}g_{(0)ij}=0,\qquad \cl_{\ck}F_{(0)ij}=0,
\ee
where $\cl_K$ denotes the Lie derivative with respect to $\ck^i$. Similarly, a {\em conformal} Killing vector corresponds to a combined diffeomorphism and Weyl transformation that leaves the sources invariant up to the same transformations. In this case \eqref{PBH-sources} give 
\be
\cl_{\ck}g_{(0)ij}+2\frac{\s_\ck}{\ell} g_{(0)ij}=0,\qquad \cl_{\ck}F_{(0)ij}=0,
\ee
and so the Weyl factor $\s$ is determined in terms of the vector $\ck^i$ as
\be
\s_\ck=-\frac\ell d D_{(0)i}\ck^i.
\ee

Killing spinors and conformal Killing spinors on a bosonic background are defined analogously. Namely, from the transformations \eqref{PBH-sources} follows that a conformal Killing spinor $\z_+$ satisfies\footnote{Killing spinors correspond to special supersymmetry transformations and so the radiality of such spinors is unambiguously determined by the generalized PBH transformations \eqref{PBH-sources}. Our spinor conventions therefore differ from other common choices in the literature, and in particular from those in \cite{Genolini:2016ecx}.} 
\be\label{Killing}
\cd_{(0)i}\z_+=\frac1\ell\Hat\G_{(0)i}\z_-,\qquad \lbar\z_+\overleftarrow\cd_{(0)i}=-\frac1\ell\lbar\z_-\Hat\G_{(0)i}, 
\ee
where
\be
\z_-=\frac{\ell}{2}\Hat\G_{(0)}^j\cd_{(0)_j}\z_+,\qquad \lbar\z_-=-\frac\ell 2 \lbar\z_+\overleftarrow\cd_{(0)j}\Hat\G_{(0)}^j.
\ee
Killing spinors are the subclass of such spinors having in addition $\z_-=0$. It is a straightforward exercise to show that for any conformal Killing spinor $\z_+$, the spinor bilinear 
\be\label{spinor-bilinear}
\ck^i_\z=-\tx i\lbar\z_+\Hat\G_{(0)}^i\z_+,
\ee
is a conformal Killing vector with Weyl factor
\be
\s_\ck=\tx i(\lbar\z_+\z_--\lbar\z_-\z_+).
\ee
In particular, if $\z_+$ is a Killing spinor, then $\ck^i_\z$ is a Killing vector.

In order to determine the Killing spinors that the backgrounds \eqref{susy-background-mu} admit, it is necessary to first make a choice of vielbein for the background metric, and to specify a basis for the gamma matrices. In the following we will choose the vielbein components as 
\be 
e^0_{(0)}=\tx dt,\quad e^1_{(0)}=\tx d\j+\frac{\tx i}{2}\pa_{\bar z}\m \tx d\bar z-\frac{\tx i}{2}\pa_z\m \tx dz,\quad e^2_{(0)}=e^{\frac w2}(\tx dz+\tx d\bar z),\quad  e^3_{(0)}=-\tx ie^{\frac w2}(\tx dz-\tx d\bar z),
\ee 
and the gamma matrix basis
\bal\label{gamma-basis}
\G^{\dot t}=\left(\begin{matrix}
	0 & \s_0 \\
	-\s_0 & 0
	\end{matrix}\right),\qquad 
	\G^{\dot \j}=\left(\begin{matrix}
		0 & \s_1 \\
		\s_1 & 0
	\end{matrix}\right),\qquad 
	\G^{\dot z}=\left(\begin{matrix}
		0 & \s_2 \\
		\s_2 & 0
	\end{matrix}\right),\qquad 
	\G^{\dot{\bar z}}=\left(\begin{matrix}
		0 & \s_3 \\
		\s_3 & 0
	\end{matrix}\right),
\eal
where $\s_a$ are the Pauli matrices. To avoid potential confusion with the complex conjugate coordinate $\bar z$, in this section we denote frame indices with a dot rather than an overbar, as in the rest of the paper. With the above choice of vielbein and gamma matrices, one finds that the spinor $\z_-$ of any conformal Killing spinor of the background \eqref{susy-background-mu} is given by
\be
\z_-=\frac{\tx i}{8}u\G^{\dot t}\z_+.
\ee

Solving the conformal Killing spinor condition on a generic background of the form \eqref{susy-background-mu} with $u\neq 0$ we find that there is a unique solution given by 
\be\label{CKS}
\z_+=\frac{1}{\sqrt{2}}e^{\tx i\g' t+\tx i\g\j+\tx i\l}\left(\begin{matrix}
	0\\ 0\\ 1\\ -1
	\end{matrix}\right),\qquad \z_-=\frac{\tx i}{8\sqrt{2}}ue^{\tx i\g' t+\tx i\g\j+\tx i\l}\left(\begin{matrix}
	1\\ -1\\ 0\\ 0
\end{matrix}\right).
\ee
Note that these spinors are globally well defined provided $\g'=0$ \cite{Genolini:2016ecx}. With this normalization the corresponding conformal Killing vector \eqref{spinor-bilinear} takes the form
\be
K_\z=-\pa_t+\pa_\j.
\ee
In the special case $u=0$, however, there are two independent solutions to the conformal Killing spinor conditions, namely 
\be
\z_+^{(1)}=\frac{1}{\sqrt{2}}e^{\tx i\g' t+\tx i\g\j+\tx i\l}\left(\begin{matrix}
	0\\ 0\\ 1\\ -1
\end{matrix}\right),\qquad \z_+^{(2)}=\frac{1}{\sqrt{2}}e^{\tx i\g' t+\tx i\g\j+\tx i\l}\left(\begin{matrix}
1\\ -1\\ 0 \\ 0
\end{matrix}\right),\qquad \z_-^{(1),(2)}=0,
\ee
both of which are in fact Killing spinors since $\z_-=0$. The corresponding Killing vectors are 
\be
K_\z^{(1)}=-\pa_t+\pa_\j,\qquad K_\z^{(2)}=-\pa_t-\pa_\j.
\ee
In the following we will focus on the generic case admitting only the conformal Killing spinor \eqref{CKS}.

\subsection{Supercurrent anomalies and supersymmetric vacua}

Restricting the local supersymmetry and superWeyl transformations of the supercurrent in \eqref{supercurrent-anomalies} to the conformal Killing spinor \eqref{CKS} results in the transformation of the supercurrent under {\em rigid} supersymmetry, namely
\bal\label{supercurrent-var}
\d_{\z}\cs^i =&\;-\frac12\ct^{ij}\Hat\G_{(0)j}\z_{+}+\frac{ic_2}{2(d-2)}\Hat\G_{(0)}^{ijk}\(\Hat\G_{(0)kl}-(d-2)g_{(0)kl}\)\z_+ D_{(0)j}\Big(\cj^l-\frac{2c_1}{\k^2}\Hat\e^{lpqs}F_{(0)pq}A_{(0)s}\Big)\NO\\
&\hskip0.15cm+\frac{ic_2}{\ell}\(\Hat\G^i_{(0)l}-(d-1)\d^i_l\)\z_-\Big(\cj^l-\frac{2c_1}{\k^2}\Hat\e^{lpqs}F_{(0)pq}A_{(0)s}\Big)\NO\\
&\hskip0.15cm-\frac{\ell^2}{2(d-2)^2\k^2}\Hat\G_{(0)}^{ijk}\Hat\G_{(0)}^l\cd_{(0)j}\[\Big(R_{kl}[g_{(0)}]-\frac{1}{2(d-1)}R[g\sub{0}]g_{(0)kl}\Big)\z_{-}\]\NO\\
&\hskip0.15cm-\frac{ic_2}{2(d-2)\k^2}\Hat\G_{(0)}^{ij}{}_k\Big((d-2)\Hat\G_{(0)}^k\Hat\G_{(0)}^{pq}-(d-1)\Hat\G_{(0)}^{kpq}\Big)\cd_{(0)j}(F_{(0)pq}\z_{-}),
\eal
where we have used the fact that on a bosonic background the Ward identity \eqref{WI-Lorentz} implies that the stress tensor $\ct^{ij}$ is symmetric, as well as the gamma matrix identity 
\be
\Hat\G_{(0)}^{ijk}\(\Hat\G_{(0)kl}-(d-2)g_{(0)kl}\)\Hat\G_{(0)j}=2(d-2)\Hat\G_{(0)}^i{}_l-(d-1)(d-2)\d^i_l,
\ee
and the conformal Killing spinor relation \eqref{Killing}. Although all anomalies in the superconformal Ward identities \eqref{WIDs} numerically vanish on the class of supersymmetric backgrounds \eqref{susy-background-mu}, the anomalous term in the supercurrent transformation \eqref{supercurrent-var} under rigid supersymmetry is {\em not} vanishing for a generic background of the form \eqref{susy-background-mu}.

The fact that the supercurrent transforms anomalously under rigid supersymmetry on the backgrounds \eqref{susy-background-mu} has important implications for the supersymmetric vacua of theories defined on such backgrounds. As we remarked earlier, the transformations \eqref{supercurrent-anomalies} of the supercurrent under local supersymmetry and superWeyl transformations can alternatively be derived from the bulk supersymmetry transformation of the gravitino. It follows that for supersymmetric vacua, corresponding to supergravity solutions satisfying the bulk BPS equations, the expectation value of the supersymmetry variation of the supercurrent vanishes, namely 
\be\label{susy-vacua}\boxed{
	\<\d_{\z}\cs^i\>_{\text{susy}}=0.}
\ee
This is commonly used to argue that the linear combination of the stress tensor and the supercurrent
\be\label{Q-exact}
\hskip-0.3cm-\frac12\ct^{ij}\Hat\G_{(0)j}\z_{+}+\frac{ic_2}{2(d-2)}\Hat\G_{(0)}^{ijk}\(\Hat\G_{(0)kl}-(d-2)g_{(0)kl}\)\z_+ D_{(0)j}\cj^l+\frac{ic_2}{\ell}\(\Hat\G^i_{(0)l}-(d-1)\d^i_l\)\z_-\cj^l,\hskip-.1cm
\ee
is $Q$-exact. However, the anomaly in the supercurrent transformation under rigid supersymmetry invalidates this argument and hence, the operator \eqref{Q-exact} is $Q$-exact only on the subclass of backgrounds of the form \eqref{susy-background-mu} where the supercurrent anomaly is numerically zero. 
A sufficient but not strictly necessary condition for the supercurrent anomaly to vanish is 
\be
R_{ij}[g_{(0)}]=0,\qquad F_{(0)ij}=0,
\ee
but clearly this is a too restrictive condition. By a careful analysis of the supercurrent anomaly in \eqref{supercurrent-var} one can determine the most general subclass of backgrounds of the form \eqref{susy-background-mu} for which the supercurrent anomaly vanishes, but we will not address this question here.

\subsection{The $w$ and $\m$ deformations}

The fact that the operator \eqref{Q-exact} is not in general $Q$-exact due to the  supercurrent anomaly under rigid supersymmetry implies that the argument first presented in \cite{Closset:2013vra,Closset:2014uda,Assel:2014paa}, showing that the supersymmetric partition function on backgrounds of the form  \eqref{susy-background-mu} is independent of the functions $w$ and $\m$, needs revisiting. In this subsection we reexamine this argument using the anomalous supercurrent transformation \eqref{supercurrent-var}, and we provide a resolution to an apparent paradox in the recent analysis of \cite{Genolini:2016ecx}. We would like to emphasize that the subsequent analysis relies only on the superconformal Ward identities and the anomalous transformation of the supercurrent and, hence, it is applicable to any $\cn=1$ superconformal field theory on backgrounds of the form \eqref{susy-background-mu}, irrespectively of whether they admit a holographic dual.  

The general variation \eqref{Sren-var} of the generating functional implies that under a bosonic variation  
\be\label{Sren-var-bosonic}
\d S\sbtx{ren}=\int d^dx\sqrt{-g\sub{0}}\Big(-\frac12\ct^{ij}\d g_{(0)ij}+\cj^i\d A_{(0)i}\Big).
\ee
Since the supersymmetric backgrounds \eqref{susy-background-mu} are parameterized by the two unconstrained functions $w(z,\bar z)$ and $\m(z,\bar z)$, we would like to examine the dependence of the supersymmetric partition function on these functions. As we now show, the variation of the partition function under infinitesimal deformations of the functions $w(z,\bar z)$ and $\m(z,\bar z)$ can be expressed as a linear combination of different components of the {\em homogeneous} part of the supercurrent transformation \eqref{supercurrent-var}, i.e. the part proportional to the stress tensor and the $R$-symmetry current given in \eqref{Q-exact}. This observation implies that the transformation of the supersymmetric partition function under $w$ and $\m$ deformations, for which the total variation of the supercurrent vanishes due to the identity \eqref{susy-vacua}, is determined entirely by the supercurrent anomaly!

\paragraph{The $w$ deformation} An infinitesimal deformation of the function $w$, keeping $\m$ fixed, corresponds to the following transformations of the bosonic sources:  
\bal
& \d_w g_{(0)z\bar z}= 2e^w\d w,\qquad \d_w A_{(0)t}=-\frac{1}{8\sqrt 3}e^{-w}\pa_z\pa_{\bar z}\m\d w,\qquad \d_w A_{(0)\j}=\frac{1}{4\sqrt 3} e^{-w}\pa_z\pa_{\bar z}\m\d w,\\
&\d_w A_{(0)z}=-\frac{\tx i}{8\sqrt{3}}e^{-w}\pa_z\pa_{\bar z}\m\;\pa_z\m\;\d w+\frac{\tx  i}{4\sqrt{3}}\pa_z\d w,\qquad \d_w A_{(0)\bar z}=\frac{\tx i}{8\sqrt{3}}e^{-w}\pa_z\pa_{\bar z}\m\;\pa_{\bar z}\m\;\d w-\frac{\tx i}{4\sqrt{3}}\pa_{\bar z}\d w,\NO
\eal
with the variation of all other components vanishing.

\paragraph{The $\m$ deformation} Similarly, an infinitesimal $\m$ deformation, keeping $w$ fixed, corresponds to the source variations 
\bal
&\hskip-0.2cm\d_\m g_{(0)\j z}=-\frac{\tx i}{2}\pa_{z}\d\m,\quad \d_\m g_{(0)\j \bar z}=\frac{\tx i}{2}\pa_{\bar z}\d\m,\quad \d_\m g_{(0)z z}=-\frac12\pa_z\m\pa_z\d\m,\quad \d_\m g_{(0)\bar z\bar z}=-\frac12\pa_{\bar z}\m\pa_{\bar z}\d\m,\\
&\hskip-0.2cm\d_\m g_{(0) z\bar z}=\frac14(\pa_{\bar z}\m\pa_z\d\m+\pa_{\bar z}\d\m\pa_z\m),\quad \d_\m A_{(0)t}=\frac{1}{8\sqrt{3}} e^{-w}\pa_z\pa_{\bar z}\d\m,\quad \d_\m A_{(0)\j}=-\frac{1}{4\sqrt{3}} e^{-w}\pa_z\pa_{\bar z}\d\m,\NO\\
&\hskip-0.2cm\d_\m A_{(0)z}=\frac{\tx i}{8\sqrt{3}} e^{-w}\(\pa_z\pa_{\bar z}\d\m\;\pa_z\m+\pa_z\pa_{\bar z}\m\;\pa_z\d\m\),\quad \d_\m A_{(0)\bar z}=-\frac{\tx i}{8\sqrt{3}} e^{-w}\(\pa_z\pa_{\bar z}\d\m\;\pa_{\bar z}\m+\pa_z\pa_{\bar z}\m\;\pa_{\bar z}\d\m\).\NO
\eal

We next insert these transformations of the sources in the identity \eqref{Sren-var-bosonic} and, by dropping various total derivative terms in the complex coordinates $z$ and $\bar z$, we express these variations in terms of certain components of the homogeneous transformation of the supercurrent under rigid supersymmetry. Specifically, for the $w$ deformation we obtain 
\bal
\d_w S\sbtx{ren}=&\;\int d^4x\sqrt{-g\sub{0}}\;\d w\Big(-2e^w\ct^{z\bar z}-\frac{1}{8\sqrt 3}e^{-w}\pa_z\pa_{\bar z}\m\cj^t+\frac{1}{4\sqrt 3} e^{-w}\pa_z\pa_{\bar z}\m\cj^\j\NO\\
&\;\hskip.28in -\frac{\tx i}{8\sqrt{3}}e^{-w}\pa_z\pa_{\bar z}\m(\cj^z\pa_z\m-\cj^{\bar z}\pa_{\bar z}\m)-\frac{\tx i}{4\sqrt{3}}e^{-w}\[\pa_z(e^w\cj^z)-\pa_{\bar z}(e^w\cj^{\bar z})\]\Big)\NO\\
\rule{0.0cm}{0.8cm}
=&\;\int d^4x\sqrt{-g\sub{0}}\;\d w\Big[-\tx i\sqrt{2}e^{w/2}\(\left.\d_\z^{\text{hom}}\cs^z\right|_1+\left.\d_\z^{\text{hom}}\cs^z\right|_2\)-\frac{\tx i}{4\sqrt{3}}e^{-w}\Big(\pa_z(e^w\cj^z)+\pa_{\bar z}(e^w\cj^{\bar z})\Big)\Big]\NO\\
\rule{0.0cm}{0.8cm}
=&\;\int d^4x\sqrt{-g\sub{0}}\;\d w\;\tx i\sqrt{2}e^{w/2}\(\left.\d_\z^{\text{anom}}\cs^z\right|_1+\left.\d_\z^{\text{anom}}\cs^z\right|_2\)\\
\rule{0.0cm}{0.8cm}
=&\;\frac{1}{2^{6}3\k^2}\int d^4x\sqrt{-g\sub{0}}\;\d w\Big(-u^2R_{2d}-\frac12\square_{2d}u^2+\frac{19}{32}u^4+\frac{8}{9}(\g+2\g')(2uR_{2d}+2\square_{2d}u-u^3)\Big).\NO
\eal
The subscript $1$ or $2$ in the variations of the supercurrent denotes the relevant spinorial component. Moreover, from the second to the third equality we have used the $R$-symmetry Ward identity $D_{(0)i}\cj^i=0$, since the $R$-symmetry anomaly vanishes on the backgrounds \eqref{susy-background-mu}, as well as the identity \eqref{susy-vacua} for the supersymmetric partition function. In the last step we have evaluated the indicated components of the supercurrent anomaly in \eqref{supercurrent-var} on the backgrounds \eqref{susy-background-mu}. Remarkably, the final result agrees completely with that obtained in eq.~(4.35) of \cite{Genolini:2016ecx}, provided the ``scheme'' parameters $\varsigma$ and $\varsigma'$ there are set to zero.\footnote{\label{orientation}The only difference is the sign of the term proportional to $\g+2\g'$, which can be traced to different choices for the orientation of the bulk manifold. In \cite{Genolini:2016ecx} the signature of the bulk metric is $(-,+,+,+,+)$, while we use $(+,-,+,+,+)$. The orientation affects the sign of the Chern-Simons term, which is the origin of the term proportional to $\g+2\g'$.} Since these parameters do not multiply superconformal invariants, they contribute trivial cocycles to the anomalies, and hence, should be set to zero. As we discussed in section \ref{algorithm}, the supersymmetric schemes correspond instead to the parameters $s_1$ and $s_2$ in \eqref{ct} and neither of these affects the supercurrent anomaly, which is scheme independent.   

Similarly, the $\m$ variation of the partition function can also be related to a certain linear combination of the supercurrent variation under rigid superymmetry, namely 
\bal
\hskip-0.3cm
\d_\m S\sbtx{ren}=&\int d^4x\sqrt{-g\sub{0}}\Big[-\frac{\tx i}{2}\Big(\ct^{\bar z\j}-\frac{\tx i}{2}\ct^{\bar zz}\pa_z\m+\frac{\tx i}{2}\ct^{\bar z\bar z}\pa_{\bar z}\m-\frac{\tx i}{8\sqrt{3}}e^{-w}\pa_z(\cj^t-2\cj^\j)\NO\\
&-\frac{1}{8\sqrt{3}}e^{-w}(\pa_{z}\cj^{\bar z})\pa_{\bar z}\m+\frac{1}{8\sqrt{3}}e^{-w}\cj^{\bar z}\pa_z\pa_{\bar z}\m+\frac{1}{8\sqrt{3}} e^{-w}\pa_z(\cj^z\pa_z\m)\Big)\pa_{\bar z}\d\m+\text{h.c.}\Big]\NO\\
\rule{0.0cm}{0.8cm}
=&\int d^4x\sqrt{-g\sub{0}}\Big\{-\sqrt{2}\Big[\frac{\tx i}{2}\(\left.\d_\z^{\text{hom}}\cs^{\bar z}\right|_1-\left.\d_\z^{\text{hom}}\cs^{\bar z}\right|_2\)+\frac14e^{-\frac w2}\(\left.\d_\z^{\text{hom}}\cs^t\right|_1+\left.\d_\z^{\text{hom}}\cs^t\right|_2\)\Big]\pa_{\bar z}\d\m+\text{h.c.}\Big\}\NO\\
\rule{0.0cm}{0.8cm}
=&\int d^4x\sqrt{-g\sub{0}}\Big\{\sqrt{2}\Big[\frac{\tx i}{2}\(\left.\d_\z^{\text{anom}}\cs^{\bar z}\right|_1-\left.\d_\z^{\text{anom}}\cs^{\bar z}\right|_2\)+\frac14e^{-\frac w2}\(\left.\d_\z^{\text{anom}}\cs^t\right|_1+\left.\d_\z^{\text{anom}}\cs^t\right|_2\)\Big]\pa_{\bar z}\d\m+\text{h.c.}\Big\}\NO\\
\rule{0.0cm}{0.8cm}
=&\frac{-1}{2^{10}3^2\k^2}\int d^4x\sqrt{-g\sub{0}}\Big[e^{-w}\pa_z\Big(24uR_{2d}-19u^3+\frac{32}{3}(\g+2\g')(3u^2-4R_{2d})\Big)\pa_{\bar z}\d\m+\text{h.c.}\Big]\NO\\
\rule{0.0cm}{0.8cm}
=&\;\frac{1}{2^93^2\k^2}\int d^4x\sqrt{-g\sub{0}}(e^{-w}\pa_z\pa_{\bar z}\d\m)\Big(24uR_{2d}-19u^3+\frac{32}{3}(\g+2\g')(3u^2-4R_{2d})\Big). 
\eal
Once again, this result agrees with the corresponding expression in eq.~(4.36) of \cite{Genolini:2016ecx}, after setting the parameters $\varsigma$ and $\varsigma'$ there to zero. 

We would like to emphasize that the above argument for the dependence of the supersymmetric partition function on $w$ and $\m$ is fundamentally different from that in \cite{Genolini:2016ecx}. The derivation in \cite{Genolini:2016ecx} relies exclusively on the bosonic operators and their sources. In particular, the variation of the supersymmetric partition function is obtained by explicitly evaluating the bulk on-shell action on supersymmetric supergravity solutions and it is therefore inherently a holographic derivation. The argument we provide above, however, does not require the evaluation of the partition function at any stage. Instead, by including the supercurrent in the analysis, we express the $w$ and $\m$ variation of the supersymmetric partition function in terms of the supercurrent transformation under rigid supersymmetry. This relation is a direct consequence of the superconformal Ward identities and hence, our argument is in fact a {\em field theory} one! As we stressed a number of times, in our entire analysis we only rely on holography in order to derive the form of the superconformal Ward identities. These identities, however, apply more generally to SCFTs, with or without a holographic dual.      

An immediate conclusion that follows from the above derivation is that, contrary to earlier claims, the supersymmetric field theory partition function is in fact {\em not} invariant under $w$ and $\m$ deformations around a generic background of the form \eqref{susy-background-mu}. Moreover, holographic renormalization reproduces precisely this field theory property. In particular, the anomalous transformation of the supersymmetric partition function under $w$ and $\m$ deformations is scheme independent on both the field theory and holographic sides (provided the scheme preserves supersymmetry), and, since it is a direct consequence of a non-trivial supercurrent anomaly, it cannot possibly be removed by a local covariant boundary counterterm. This anomalous transformation is physical and so there is no need to add non-covariant counterterms as is done in \cite{Genolini:2016ecx}, which explicitly break other symmetries of the theory! As we now explain, these observations also imply that the BPS relation between the conserved charges on supersymmetric vacua is also anomalous!

\subsection{Conserved charges and the BPS relation}

The conserved charges are a consequence of the Ward identities \eqref{WIDs} \cite{Papadimitriou:2005ii}. The presence of anomalies in the Ward identities renders the argument leading to the conserved charges less trivial in general \cite{thermo2}, but for the supersymmetric backgrounds \eqref{susy-background-mu} we are interested in here these anomalies numerically vanish.     

\paragraph{Electric charge} The $R$-symmetry Ward identity \eqref{WI-R} implies that the quantity  
\be\label{charge-e}
\cq\sbtx{e}^{\o}=\frac{1}{\sqrt{3}}\int d\s_i\;\Big(\<\cj^i\>+\o\frac{c_1}{\k^2}\Hat\e^{ipqs}F_{(0)pq}A_{(0)s}\Big),
\ee
where the integral is over a constant time slice of the geometry \eqref{susy-background-mu}, is conserved for any background field $A_{(0)i}$ if $\o=1$, or for any value of $\o$ if the $R$-symmetry anomaly is numerically zero, as is the case for the backgrounds \eqref{susy-background-mu}. Notice that we have introduced the notation $\<\cdot\>$ here in order to emphasize that the charges are associated with a given state of the theory and so involve the expectation value of the operators in a specific state. Moreover, the overall factor $1/\sqrt{3}$ in the definition of the electric charge is included in order to compensate the normalization of the background gauge field in \eqref{susy-background-mu} and get a canonical BPS relation between the charges later on.

The transformations \eqref{PBH-momenta} of the currents under the local symmetries imply that the integrand in \eqref{charge-e} is gauge invariant only when $\o=-2$. However, the electric charge is invariant under small gauge transformations for any value of $\o$ due to the Bianchi identity $D_{(0)[i}F_{(0)jk]}=0$. Notice that for $\o=-2$ the charge \eqref{charge-e} is the Maxwell charge, while for $\o=1$ it is the Page charge \cite{Page:1984qv,Marolf:2000cb,Benini:2007gx}:
\be\boxed{
 \cq\sbtx{e}^{\o=-2}=\cq\sbtx{e}^{\tx{Maxwell}},\qquad \cq\sbtx{e}^{\o=1}=\cq\sbtx{e}^{\tx{Page}}.}
\ee

\paragraph{Conformal Killing charges} The conserved charges associated with conformal Killing vectors of a bosonic background follow from the diffeomorphism Ward identity \eqref{WI-diff}, namely  
\be
D_{(0)j}\ct^j_i+F_{(0)ij}\Big(\cj^j+\o\frac{c_1}{\k^2}\Hat\e^{jpqs}F_{(0)pq}A_{(0)s}\Big)=\frac{(\o+2)c_1}{\k^2}\Hat\e^{jklp}F_{(0)ij}F_{(0)kl}A_{(0)p}.
\ee
Contracting this identity with a conformal Killing vector $\ck^i$ and utilizing the trace \eqref{WI-trace} and $R$-symmetry \eqref{WI-R} Ward identities, one finds that the quantity 
\be\label{Killing-charge}
\cq^{\o}[\ck]=-\int d\s_i\; \[\<\ct^i_j\>-\Big(\<\cj^i\>+\o\frac{c_1}{\k^2}\Hat\e^{ipqs}F_{(0)pq}A_{(0)s}\Big)A_{(0)j}\]\ck^j,
\ee
is conserved provided both the Weyl anomaly $\ca_{W(0)}$ and $\ca_{M(0)i}$ vanish numerically, which is the case for the supersymmetric backgrounds \eqref{susy-background-mu}. Note that these charges are not gauge invariant except when $\o=-2$, but they can be made gauge invariant for any $\o$ by a minor modification \cite{thermo2}.

\paragraph{Supercharges} Finally, the supercharges associated with a conformal Killing spinor $\z_+$ are 
\be
\lbar\cq[\z_+]=\int d\s_i\; \lbar\z_+\<\cs^i\>,\qquad \cq[\z_+]=\int d\s_i\; \<\lbar\cs^i\>\z_+.
\ee
These charges are conserved provided the supersymmetry and superWeyl anomalies vanish, and they are of course zero in a bosonic background. 

\paragraph{BPS relation} The transformation of the supercurrent under rigid supersymmetry in \eqref{supercurrent-var} leads to the BPS relation between the conserved charges in supersymmetric vacua. In particular, contracting the identity \eqref{supercurrent-var} from the left with $\tx i\lbar\z_+$ and taking the expectation value in a supersymmetric vacuum we obtain  
\bal\label{supercurrent-var-BPS}
0=\tx i\lbar\z_+\<\d_{\z}\cs^i\>\sbtx{susy} =&\;\frac{1}{2}\<\ct_j^i\>\sbtx{susy}\ck_\z^j+\frac{(d-1)c_2}{2\ell}\Big(\<\cj^i\>\sbtx{susy}-\frac{2c_1}{\k^2}\Hat\e^{ipqs}F_{(0)pq}A_{(0)s}\Big)(\lbar\z_+\z_{-}+\lbar\z_-\z_{+})\NO\\
&\hskip0.15cm+\frac{\tx i\ell}{2(d-2)\k^2}\lbar\z_-\(\Hat\G_{(0)}^ig_{(0)}^{kl}-g_{(0)}^{i(k}\Hat\G_{(0)}^{l)}\)\z_{-}\Big(R_{kl}[g_{(0)}]-\frac{1}{2(d-1)}R[g\sub{0}]g_{(0)kl}\Big)\NO\\
&\hskip0.15cm-\frac{(d-1)c_2}{2\k^2\ell}\lbar\z_-\Hat\G_{(0)}^{ijk}\z_{-}F_{(0)jk}+D_{(0)j}\cv^{ij},
\eal 
where 
\bal
\cv^{ij}=&\;-\frac{c_2}{2(d-2)}\lbar\z_+\Hat\G_{(0)}^{ijk}\(\Hat\G_{(0)kl}-(d-2)g_{(0)kl}\)\z_{+}\Big(\cj^l-\frac{2c_1}{\k^2}\Hat\e^{lpqs}F_{(0)pq}A_{(0)s}\Big)\NO\\
&\hskip0.15cm-\frac{\tx i\ell^2}{2(d-2)^2\k^2}\lbar\z_+\Hat\G_{(0)}^{ijk}\Hat\G_{(0)}^l\z_{-}\Big(R_{kl}[g_{(0)}]-\frac{1}{2(d-1)}R[g\sub{0}]g_{(0)kl}\Big)\NO\\
&\hskip0.15cm+\frac{c_2}{2(d-2)\k^2}\lbar\z_+\Hat\G_{(0)}^{ij}{}_k\Big((d-2)\Hat\G_{(0)}^k\Hat\G_{(0)}^{pq}-(d-1)\Hat\G_{(0)}^{kpq}\Big)\z_{-}F_{(0)pq},
\eal
and we have used the Killing spinor property \eqref{Killing}. Moreover, from the conformal Killing spinor solution \eqref{CKS} we determine
\be
A_{(0)i}\ck_\z^i=-\frac{1}{\sqrt{3}}\(\frac38 u+\g-\g'\),
\ee
and 
\be
(\lbar\z_+\z_{-}+\lbar\z_-\z_{+})=\frac u4=\frac23 (-\sqrt{3}A_{(0)i})\ck_\z^i-\frac23(\g-\g'),
\ee
which allow us to rewrite \eqref{supercurrent-var-BPS} in the form
\bal
0=&\;\<\ct_j^i\>\sbtx{susy}\ck_\z^j-\Big(\<\cj^i\>\sbtx{susy}+\o\frac{c_1}{\k^2}\Hat\e^{ipqs}F_{(0)pq}A_{(0)s}\Big)\Big(A_{(0)i}\ck_\z^i+\frac{1}{\sqrt{3}}(\g-\g')\Big)\NO\\
&\hskip0.15cm+\frac{\tx i\ell}{(d-2)\k^2}\lbar\z_-\(\Hat\G_{(0)}^ig_{(0)}^{kl}-g_{(0)}^{i(k}\Hat\G_{(0)}^{l)}\)\z_{-}\Big(R_{kl}[g_{(0)}]-\frac{1}{2(d-1)}R[g\sub{0}]g_{(0)kl}\Big)\NO\\
&\hskip0.15cm-\frac{(d-1)c_2}{\k^2\ell}\lbar\z_-\Hat\G_{(0)}^{ijk}\z_{-}F_{(0)jk}+\frac{(\o+2)\ell}{3\k^2}\d_{d,4}\Hat\e^{ipqs}F_{(0)pq}A_{(0)s}(\lbar\z_+\z_{-}+\lbar\z_-\z_{+})+2D_{(0)j}\cv^{ij}.
\eal 
Integrating this identity over a constant time slice of the geometries \eqref{susy-background-mu} we obtain the relation 
\be\label{BPS-id}
\cq^{\o}[\ck_\z]+(\g-\g')\cq\sbtx{e}^{\o}=\cq\sbtx{anomaly}^{\o},
\ee
where,
\bal
\cq\sbtx{anomaly}^{\o}=&\;-\int d\s_i\;\Big[-\frac{\tx i\ell}{(d-2)\k^2}\lbar\z_-\(\Hat\G_{(0)}^ig_{(0)}^{kl}-g_{(0)}^{i(k}\Hat\G_{(0)}^{l)}\)\z_{-}\Big(R_{kl}[g_{(0)}]-\frac{1}{2(d-1)}R[g\sub{0}]g_{(0)kl}\Big)\NO\\
&\hskip1.cm+\frac{(d-1)c_2}{\k^2\ell}\lbar\z_-\Hat\G_{(0)}^{ijk}\z_{-}F_{(0)jk}-\frac{(\o+2)\ell}{3\k^2}\d_{d,4}\Hat\e^{ipqs}F_{(0)pq}A_{(0)s}(\lbar\z_+\z_{-}+\lbar\z_-\z_{+})\Big].
\eal
Notice that, as the conformal Killing charges \eqref{Killing-charge}, the anomaly charge is in general gauge invariant only for $\o=-2$. However, the identity \eqref{BPS-id} is gauge invariant and holds for any $\o$.
Identifying the mass and angular momentum respectively with the conserved charges
\be
M^{\o}=\cq^{\o}[-\pa_t],\qquad J^{\o}=\cq^{\o}[\pa_\j],
\ee
leads to the BPS condition
\be\label{BPS}\boxed{
M^{\o}+J^{\o}+(\g-\g') \cq\sbtx{e}^{\o}=\cq\sbtx{anomaly}^{\o}.}
\ee
In particular, the supercurrent anomaly gives rise to an anomaly in the BPS condition! This observation provides a resolution to another apparent paradox discussed in \cite{Genolini:2016ecx}, as well as the earlier works \cite{Assel:2015nca,Martelli:2015kuk,Genolini:2016sxe,DiPietro:2016ond,Brunner:2016nyk}, concerning the supersymmetric Casimir energy, which we will discuss momentarily. Interestingly,  
the anomalous BPS condition \eqref{BPS} seems to be compatible with the recent conjecture \cite{Ooguri:2016pdq} for non-supersymmetric AdS vacua.\footnote{I thank Gavin Hartnett for suggesting this connection.}

\paragraph{BPS states and Casimir charges} In order to compute the supersymmetric Casimir charges it is necessary to determine the expectation values of the stress tensor and the $R$-current in general supersymmetric vacua. These expectation values must simultaneously satisfy the supersymmetry condition \eqref{susy-vacua}, as well as the trace, diffeomorphism and $R$-symmetry Ward identities. These conditions together account for $6+1+4+1=12$ constraints, while the stress tensor and the $R$-current together have $10+4=14$ components. It follows that the general form of the one-point functions for supersymmetric vacua must be parameterized in terms of two arbitrary functions. 

The supersymmetry condition \eqref{susy-vacua} and the Ward identities can be solved systematically to obtain the following general solution for the supersymmetric one-point functions:
\bal\label{BPS-stress-tensor}
\<\ct^{tt}\>\sbtx{BPS} =&\;\frac{1}{\k^2}\Upsilon(z,\lbar z)-\tx i\pa_{z}\m\<\ct^{\j z}\>\sbtx{BPS}+\tx i\pa_{\lbar z}\m\<\ct^{\j\lbar z}\>\sbtx{BPS}-\frac14(\pa_{z}\m)^2\<\ct^{zz}\>\sbtx{BPS}-\frac14(\pa_{\lbar z}\m)^2\<\ct^{\lbar z\lbar z}\>\sbtx{BPS}\NO\\
&+(4e^w+\pa_z\m\pa_{\lbar z}\m)\<\ct^{z\lbar z}\>\sbtx{BPS},\NO\\
\<\ct^{t\j}\>\sbtx{BPS} =&\;-\frac{1}{\k^2}\Upsilon(z,\lbar z)-\frac{1}{3\times 2^{11}\k^2}\Big[2^8u\F(z,\lbar z)+3u^4+16 u^2R_{2d}+64\square_{2d}u^2\NO\\
&-2e^{-w}\pa_{\lbar z}\Big(\(e^{-w}(\pa_z\m)\pa_{\lbar z}u^2+14u^3-24uR_{2d}\)\pa_z\m\Big)\NO\\
&-2e^{-w}\pa_{z}\Big(\(e^{-w}(\pa_{\lbar z}\m)\pa_{z}u^2+14u^3-24uR_{2d}\)\pa_{\lbar z}\m\Big)\Big],\NO\\
\<\ct^{tz}\>\sbtx{BPS} =&\;\frac{\tx i}{3^2\times 2^6\k^2}e^{-w}\pa_{\lbar z}\Big(3uR_{2d}-u^3+48\F(z,\lbar z)\Big),\NO\\
\<\ct^{t\bar z}\>\sbtx{BPS} =&\;-\frac{\tx i}{3^2\times 2^6\k^2}e^{-w}\pa_{ z}\Big(3uR_{2d}-u^3+48\F(z,\lbar z)\Big),\NO\\
\<\ct^{\j\j}\>\sbtx{BPS} =&\;\frac{1}{\k^2}\Upsilon(z,\lbar z)+\frac12\pa_{\lbar z}\m\pa_z\m\;\<\ct^{z\bar z}\>\sbtx{BPS},\NO\\
\<\ct^{\j z}\>\sbtx{BPS} =&\;-\frac{\tx i}{2}\pa_{\lbar z}\m\; \<\ct^{z\lbar z}\>\sbtx{BPS}-\frac{\tx i}{3\times 2^9\k^2}e^{-w}\pa_{\lbar z}\Big(\frac{17}{3}u^3-8uR_{2d}+e^{-w}(\pa_{z}\m)\pa_{\lbar z}u^2+64\F(z,\lbar z)\Big),\NO\\
\<\ct^{\j \bar z}\>\sbtx{BPS} =&\;\frac{\tx i}{2}\pa_{z}\m\; \<\ct^{z\lbar z}\>\sbtx{BPS}+\frac{\tx i}{3\times 2^9\k^2}e^{-w}\pa_{z}\Big(\frac{17}{3}u^3-8uR_{2d}+e^{-w}(\pa_{\lbar z}\m)\pa_{z}u^2+64\F(z,\lbar z)\Big),\NO\\
\<\ct^{zz}\>\sbtx{BPS} =&\;-\frac{1}{3\times 2^8\k^2}e^{-w}\pa_{\lbar z}(e^{-w}\pa_{\lbar z} u^2),\NO\\
\<\ct^{\bar z\bar z}\>\sbtx{BPS} =&\;-\frac{1}{3\times 2^8\k^2}e^{-w}\pa_{ z}(e^{-w}\pa_{ z} u^2),\NO\\
\<\ct^{z\bar z}\>\sbtx{BPS} =&\;\frac{1}{3\times 2^8\k^2}e^{-w}\Big(\square_{2d}u^2+2u^2R_{2d}-\frac{19}{16}u^4-16u\F(z,\lbar z)\Big),
\eal
and
\be\label{BPS-current}
\<\cj^i\>\sbtx{BPS} =\left\{\begin{matrix}
	\frac{2c_1}{\k^2}\Hat\e^{ipqs}F_{(0)pq}A_{(0)s}+\frac{1}{\sqrt{3}\;\k^2}\F(z,\lbar z), & i=t, \\&\\
	\hskip-2.5cm\frac{2c_1}{\k^2}\Hat\e^{ipqs}F_{(0)pq}A_{(0)s}, & i\neq t.
	\end{matrix}\right.
\ee
As anticipated, these supersymmetric expectation values are parameterized by the two 
arbitrary functions $\Upsilon(z,\lbar z)$ and $\F(z,\lbar z)$.

This solution for the supersymmetric one-point functions can be generalized further by replacing 
\be
\<\ct^i_j\>\sbtx{BPS}\to\<\ct^i_j\>\sbtx{BPS} +\n\<\ct^i_j\>\sbtx{loc},\qquad \<\cj^i\>\sbtx{BPS}\to\<\cj^i\>\sbtx{BPS} +\n\<\cj^i\>\sbtx{loc},
\ee
where $\n$ is an arbitrary constant and 
\bal\label{loc-sol}
\<\ct^{i}_j\>\sbtx{loc} =&\;\frac{\ell^3}{24\k^2}\Big(D_{(0)}^iD_{(0)j}R[g_{(0)}]-3\square_{(0)}R^i_j[g_{(0)}]+\frac12\square_{(0)}R\d^i_j+6R^i_k[g\sub{0}]R^k_j[g\sub{0}]-2R[g\sub{0}]R^i_j[g\sub{0}]\NO\\
&-\frac32R^k_l[g\sub{0}]R^l_k[g\sub{0}]\d^i_j+\frac12 R^2[g\sub{0}]\d^i_j-6W^i{}_{kjl}[g_{(0)}]R^{kl}[g_{(0)}]\Big)\NO\\
&+\frac{\ell}{\k^2}\Big(F^i_{(0)k}F_{(0)j}{}^k-\frac14F_{(0)kl}F_{(0)}^{kl}\d^i_j\Big),\NO\\
\<\cj^i\>\sbtx{loc} =&\;-\frac{\ell}{\k^2}D_{(0)j}F^{ji}_{(0)},
\eal
where $W_{(0)ijkl}$ is the Weyl tensor of the metric $g_{(0)ij}$, which for backgrounds of the form \eqref{susy-background-mu} satisfies the identity 
\be
W_{(0)}^{iklm}W_{(0)jklm}-\frac{2}{\ell^2}F_{(0)}^{kl}F_{(0)kl}\d^i_j=0.
\ee 
The expressions \eqref{loc-sol} correspond respectively to the derivative of 
\be
\int d^4x\sqrt{-g\sub{0}}\;\cw_{(0)}^2,
\ee  
where $\cw_{(0)}^2$ is the supersymmetric Weyl squared density defined in eq.~\eqref{invariants}, with respect to the metric $g_{(0)ij}$ and the gauge field $A_{(0)i}$. Notice that although $\cw_{(0)}^2$ vanishes on the backgrounds \eqref{susy-background-mu}, its derivatives $\<\ct^{i}_j\>\sbtx{loc}$ and $\<\cj^i\>\sbtx{loc}$ are not generically zero. However, it can be shown that they only contribute total derivative terms to the one-point functions of the currents, and so do not affect the value of the conserved charges. This fact also implies that the conserved charges are independent of the choice of supersymmetric renormalization scheme, parameterized by the two parameters $s_1$ and $s_2$ in \eqref{ct}.

The general solution \eqref{BPS-stress-tensor} and \eqref{BPS-current} for the supersymmetric one-point functions allows us to compute explicitly the conserved charges in terms of the functions $\Upsilon(z,\lbar z)$ and $\F(z,\lbar z)$. We find
\bal\label{BPS-charges}
M^{\o}=&\;-\frac{1}{3\times 2^{9}\k^2}\int\tx{vol}_3\Big(3\times 2^9 \Upsilon+2^6u\F+\frac{11}{6}u^4\Big)\NO\\
&+\frac{\g'}{3^2\k^2}\int\tx{vol}_3\Big(3\F-(\o+2)\Big(\frac13\g R_{2d}-\frac{1}{24}(u^3-4uR_{2d})\Big)\Big)\NO\\
&+\frac{(\o+2)}{3^2\times 2^3\k^2}\int\tx{vol}_3\;\Big(u\Big(\frac13\g R_{2d}-\frac{1}{24}(u^3-4uR_{2d})\Big)-\frac{1}{24}(2\g u^3+u^2 R_{2d})\Big),\NO\\ \NO\\
J^{\o}=&\;-\frac{1}{3\times 2^{8}\k^2}\int\tx{vol}_3\Big(\frac{23}{24}u^4-6u^2R_{2d}-2^5u\F-3\times 2^8\Upsilon\Big)\NO\\
&-\frac{\g}{3^2\k^2}\int\tx{vol}_3\Big(3\F-(\o+2)\Big(\frac13\g R_{2d}-\frac{1}{24}(u^3-4uR_{2d})\Big)\Big)\NO\\
&+\frac{(\o+2)}{3^2\times 2^2\k^2}\int\tx{vol}_3\;\Big(u\Big(\frac13\g R_{2d}-\frac{1}{24}(u^3-4uR_{2d})\Big)-\frac{1}{24}(2\g u^3+u^2 R_{2d})\Big),\NO\\\NO\\
\cq\sbtx{e}^{\o}=&\;\frac{1}{3^2\k^2}\int \tx{vol}_3\; \Big(3\F-(\o+2)\Big(\frac13\g R_{2d}-\frac{1}{24}(u^3-4uR_{2d})\Big)\Big),\NO\\\NO\\
\cq\sbtx{anomaly}^{\o}=&\;\frac{1}{ 2^{9}\k^2}\int\tx{vol}_3\Big(4u^2R_{2d}-\frac{5}{4}u^4\Big)\NO\\
&+\frac{(\o+2)}{3\times 2^3\k^2}\int\tx{vol}_3\;\Big(u\Big(\frac13\g R_{2d}-\frac{1}{24}(u^3-4uR_{2d})\Big)-\frac{1}{24}(2\g u^3+u^2 R_{2d})\Big).
\eal
As expected, these charges satisfy the BPS relation \eqref{BPS} for any value of the parameter $\o$ and for arbitrary functions $\Upsilon(z,\lbar z)$ and $\F(z,\lbar z)$. Moreover, they apply to any BPS state, including states corresponding to supersymmetric black holes in the bulk. 

We would now like to evaluate the conserved charges \eqref{BPS-charges} explicitly in the global supersymmetric vacuum of the theory. This vacuum corresponds to a smooth horizonless geometry in the bulk, which allows one to determine the functions $\Upsilon\sbtx{vac}(z,\lbar z)$ and $\F\sbtx{vac}(z,\lbar z)$ explicitly. As was pointed out in \cite{Genolini:2016ecx}, for smooth horizonless bulk solutions the Page $R$-charge vanishes identically, i.e. $\cq\sbtx{e}^{\o=1}=0$, which implies that the time component of 
\be\label{vac-current-condition}
\<\cj^i\>\sbtx{vac}+\frac{c_1}{\k^2}\Hat\e^{ipqs}F_{(0)pq}A_{(0)s},
\ee
is a total derivative. This condition, together with the general supersymmetric solution \eqref{BPS-current}, determines the function $\F(z,\lbar z)$ to be 
\be\label{Phivac}\boxed{
\F\sbtx{vac}(z,\lbar z)=\frac13\g R_{2d}-\frac{1}{24}(u^3-4uR_{2d}).}
\ee

In order to determine the function $\Upsilon(z,\lbar z)$ in the global supersymmetric vacuum we need an additional condition. In \cite{Genolini:2016ecx} this additional condition was that the Ricci potential be a globally defined one-form in the bulk. However, physically this condition is equivalent to requiring that the entropy vanishes, which is not a priori obvious how to express as a condition on the one-point functions of the stress tensor and the $R$-current. To circumvent this, below we determine the function $\Upsilon\sbtx{vac}(z,\lbar z)$ by requiring the value of the free energy to agree with that computed in \cite{Genolini:2016ecx} (without the ``new counterterms'') through the global Ricci potential condition. We then manage to express the value of $\Upsilon\sbtx{vac}(z,\lbar z)$ as a condition on the one-point functions of the stress tensor and the $R$-current. The resulting condition provides an expression for the supersymmetric entropy in terms of the local one-point functions. For the sake of presentation, however, we find it convenient to start with the zero entropy condition in terms of the one-point functions, and justify it a posteriori by the fact that it leads to the correct value of $\Upsilon\sbtx{vac}(z,\lbar z)$.  

In order to write down an expression for the entropy, or to be able to relate the free energy to the conserved charges, we need to identify the timelike Killing vector that would become null on the Killing horizon. The form of the graviphoton in \eqref{susy-background-mu} implies that for generic $u(z,\lbar z)$ the null generator of the horizon is uniquely determined to be  
\be\label{smooth-Killing}
\c=\pa_t+\frac12\pa_\j,
\ee
so that
\be
A_{(0)i}\c^i=-\frac{1}{2\sqrt{3}}(\g+2\g'),
\ee
is constant. Given this timelike Killing vector, we conjecture that the entropy of any supersymmetric state can be expressed in terms of the expectation values of the stress tensor and the $R$-current as 
\be\label{BPS-entropy}\boxed{
S\sbtx{BPS}\propto\int_{\cc_{\c_\perp}}\hskip-0.35cm d\s_i\Big[\Big(\<\ct^i_j\>+\Big(\<\cj^i\>+\frac{c_1}{\k^2}\Hat\e^{ipqs}F_{(0)pq}A_{(0)s}\Big)A_{(0)j}\Big)\c^j-\frac{(\g^2+2\g'^2)}{9\sqrt{3}\k^2}R_{2d}\c^i_\perp\Big],}
\ee
where 
\be
\c_\perp^i=-\frac{1}{\sqrt{3}}(1,2),\qquad \c_{\perp\;i}\c_\perp^i=1,\qquad \c_{\perp i}\c^i=0,
\ee
is the unit vector orthogonal to $\c$. Notice that the integral is done over the surface orthogonal to $\c_\perp$ and the relative sign between the stress tensor and the $R$-current is opposite that appearing in the conserved charges \eqref{Killing-charge}. Obtaining a first principles derivation of this expression is highly desirable, but here we will be content with the fact that it gives the correct expression for $\Upsilon\sbtx{vac}(z,\lbar z)$.

In particular, demanding that 
\be
\left. S\sbtx{BPS}\right|\sbtx{vac}=0,
\ee
determines
\begin{align}
\boxed{
\begin{aligned}\label{Yvac}
\Upsilon\sbtx{vac}(z,\lbar z)=&\;-\frac{1}{3^2\times 2^9}\Big(3\times 2^6 u\F\sbtx{vac}(z,\lbar z)+\frac{59}{12} u^4+4u^2R_{2d}+32\square_{2d}u^2\Big)\\
&\;-\frac{1}{27}(\g+2\g')\Big(\F\sbtx{vac}(z,\lbar z)-\frac13(\g+2\g')R_{2d}\Big)-\frac{2}{3^4}(\g^2+2\g'^2)R_{2d}.
\end{aligned}}
\end{align}
In the next subsection we will see that this expression for $\Upsilon\sbtx{vac}(z,\lbar z)$ leads to the free energy obtained in  \cite{Genolini:2016ecx} (without the ``new counterterms'').

Having determined the functions $\Upsilon\sbtx{vac}(z,\lbar z)$ and $\F\sbtx{vac}(z,\lbar z)$ for the supersymmetric vacuum, we can evaluate explicitly the charges \eqref{BPS-charges}, which in this case are identified with the Casimir charges:
\bal\label{Casimir-charges}
M^{\o}=&\;\frac{1}{3^2\times 2^{9}\k^2}\int\tx{vol}_3\Big(4u^2R_{2d}-\frac{7}{12}u^4\Big)+\frac{1}{27\k^2}\int\tx{vol}_3\Big((\g+2\g')\F\sbtx{vac}+\frac13\g(\g-4\g')R_{2d}\Big)\NO\\
&+\frac{(1-\o)\g'}{3^2\k^2}\int\tx{vol}_3\;\F\sbtx{vac}+\frac{(\o+2)}{3^2\times 2^3\k^2}\int\tx{vol}_3\;\Big(u\F\sbtx{vac}-\frac{1}{24}(2\g u^3+u^2 R_{2d})\Big),\NO\\ \NO\\
J^{\o}=&\;\frac{1}{3^2\times 2^{4}\k^2}\int\tx{vol}_3\Big(u^2R_{2d}-\frac{1}{3}u^4\Big)-\frac{1}{27\k^2}\int\tx{vol}_3\Big((\g+2\g')\F\sbtx{vac}+\frac13\g(\g-4\g')R_{2d}\Big)\NO\\
&-\frac{(1-\o)\g}{3^2\k^2}\int\tx{vol}_3\;\F\sbtx{vac}+\frac{(\o+2)}{3^2\times 2^2\k^2}\int\tx{vol}_3\;\Big(u\F\sbtx{vac}-\frac{1}{24}(2\g u^3+u^2 R_{2d})\Big),\NO\\\NO\\
\cq\sbtx{e}^{\o}=&\;\frac{(1-\o)}{3^2\k^2}\int \tx{vol}_3\; \F\sbtx{vac},\NO\\\NO\\
\cq\sbtx{anomaly}^{\o}=&\;\frac{1}{ 2^{9}\k^2}\int\tx{vol}_3\Big(4u^2R_{2d}-\frac{5}{4}u^4\Big)+\frac{(\o+2)}{3\times 2^3\k^2}\int\tx{vol}_3\;\Big(u\F\sbtx{vac}-\frac{1}{24}(2\g u^3+u^2 R_{2d})\Big).
\eal

\subsection{Supersymmetric partition function}

We conclude with some remarks regarding the value of the supersymmetric partition function. On general grounds, the Euclidean partition function $I$ is proportional to the Gibbs free energy, namely 
\be
I=\b(M-\b^{-1}S-\O J-\F\sbtx{e}\cq\sbtx{e}),
\ee
where $\b$ is the perimeter of the Euclidean time circle, $\O$ is the angular velocity, and $\F\sbtx{e}$ is the electric potential. This relation applies to both supersymmetric and non-supersymmetric vacua and is renormalization scheme independent. In particular, the scheme dependence of the partition function compensates that of the mass $M$, while the remaining charges are scheme independent. This relation applies to any field theory, independently of whether it admits a holographic dual, but a general holographic proof was provided in \cite{Papadimitriou:2005ii} by combining the Noether current approach to entropy \cite{Iyer:1994ys} and the holographic definition of the conserved charges.

In order to evaluate the supersymmetric partition function we use the Casimir charges \eqref{Casimir-charges} and set $S=0$. Moreover, from the Killing vector \eqref{smooth-Killing} follows that the angular momentum and electric potential are given respectively by  
\be
\O=\frac12,\qquad \F\sbtx{e}=-\sqrt{3}A_{(0)i}\ck^i=\frac12(\g+2\g').
\ee
Using these expressions we obtain
\bal
I\sbtx{vac}=&\;\b\Big(M^\o-\frac12 J^\o-\frac{1}{2}(\g+2\g')\cq\sbtx{e}^\o\Big)\\
=&\;\frac{\b}{3^2\times 2^{11}\k^2}\int\tx{vol}_3\Big(19u^4-48u^2R_{2d}-\frac{128}{3}(\g+2\g')(u^3-4uR_{2d})\Big)+\frac{\b(\g-\g')\g}{27\k^2}\int\tx{vol}_3\; R_{2d}.\NO
\eal
This result is independent of the value of $\o$ and of the choice of supersymmetric renormalization scheme. Moreover, it agrees with the free energy obtained in \cite{Genolini:2016ecx} (without the ``new counterterms'' and with opposite sign for the term proportional to ($\g+2\g'$), due to our different orientation conventions, as discussed in footnote \ref{orientation}).

\section*{Acknowledgments}

I would like to thank Francesco Aprile, Matteo Bertolini, Nima Doroud, Gavin Hartnett, Danielle Musso, Flavio Porri, Himanshu Raj and Kostas Skenderis for useful discussions. I am particularly thankful to Davide Cassani, Zohar Komargodski and Dario Martelli for very useful comments on the manuscript.

\appendix

\renewcommand{\thesection}{\Alph{section}}
\renewcommand{\theequation}{\Alph{section}.\arabic{equation}}

\section*{Appendices}
\setcounter{section}{0}

\section{Conventions and radial ADM decomposition}
\label{conventions}
\setcounter{equation}{0}

In this appendix we define our conventions and collect several identities that are extensively used in this paper. 

\subsection{Indices and orientation}

The following table summarizes the different sets of indices we use to denote bulk and boundary coordinates, as well as the corresponding frame bundle indices.
\begin{center}
\begin{tabular}{|l|l|l|}
	\hline &&\\
	$\m,\n,\r,\ldots$ & $1,\ldots,d+1$ & bulk coordinate idices \\&&\\
	$\a,\b,\g,\ldots$ & $\bar 1,\ldots,\lbar{d+1}$ & bulk frame indices \\&&\\
	$i,j,k,\ldots$ & $1,\ldots,d$ & boundary coordinate indices \\&&\\
	$a,b,c,\ldots$ & $\bar 1,\ldots,\bar d$ & boundary frame indices\\&&\\
	\hline
\end{tabular}
\end{center}
An overline, as in $\bar r$ or $\bar t$, will be used to denote the frame indices associated with the corresponding coordinates, respectively $r$ and $t$ in this example. We take the Minkowski metric $\h_{\a\b}$ on the frame bundle to be
$\h=\diag(1,-1,1,\ldots,1)$, where $\h_{\bar t\bar t }=-1$. Accordingly, we choose the orientation of the bulk manifold such that the Levi-Civita symbol $\ve_{\m\n\r\ldots}=\pm 1$ satisfies $\ve_{r, t,\ldots}=1$. The corresponding Levi-Civita tensor is defined as usual by $\e_{\m\n\r\ldots}=\sqrt{-g}\;\ve_{\m\n\r\ldots}$.

\subsection{Gamma matrix conventions and identities}

The Gamma matrices $\G^\a$ with flat frame bundle indices satisfy the 
Clifford algebra
\be
\{\G^\a,\G^\b\}=2\h^{\a\b}.
\ee
The Gamma matrices with coordinate indices are defined using the inverse vielbein $E^\m_\a$ as
\be
\G^\m=E^{\m}_\a\G^\a,
\ee
and satisfy
\be\label{Clifford-bulk}
\{\G^\m,\G^\n\}=2g^{\m\n}.
\ee
We do not need the explicit matrix representation of the Clifford algebra here, except for the calculation in section \ref{backgrounds}, where we specify the gamma matrix representation used in \eqref{gamma-basis}. As in \cite{Freedman:2012zz}, we will assume that the representation is Hermitian so that   
\be\label{Hermiticity-bulk} 
\G^{\a\dagger}=\G^{\bar t}\G^\a\G^{\bar t},\qquad \G^{\m\dagger}=\G^{\bar t}\G^\m\G^{\bar t}.
\ee  

We also adopt the standard notation for  
totally antisymmetric products of Gamma matrices
\be
\G^{\m_1\m_2\ldots\m_n}\equiv\G^{[\m_1}\G^{\m_2}\cdots\G^{\m_n]},
\ee
where the indices are antisymmetrized with weight one. Several identities these products satisfy in $D$-dimensions can be found e.g. in section 3 of \cite{Freedman:2012zz}. Some of the identities that we utilize here are 
\bsub
\label{Gamma-IDs} 
\bal
\G^{\m\n\r}&=\frac12\{\G^\m,\G^{\n\r}\},\\
\G^{\m\n\r\s}&=\frac12[\G^\m,\G^{\n\r\s}],\\
\G^{\m\n}\G_{\r\s}&=\G^{\m\n}{}_{\r\s}+4\G^{[\m}{}_{[\s}\d^{\n]}{}_{\r]}+2\d^{[\m}{}_{[\s}\d^{\n]}{}_{\r]},\\
\G_{\m}\G^{\n_1\ldots\n_p}&=\G_\m{}^{\n_1\ldots\n_p}+p\d ^{[\n_1}_{\m}\G^{\n_2\ldots\n_p]},\\
\G^{\n_1\ldots\n_p}\G_{\m}&=\G^{\n_1\ldots\n_p}{}_{\m}+p\G^{[\n_1\ldots\n_{p-1}}\d ^{\n_p]}_{\m},\\
\G^{\m\n\r}\G_{\s\t}&=\G^{\m\n\r}{}_{\s\t}+6\G^{[\m\n}{}_{[\t}\d ^{\r]}{}_{\s]}+6\G^{[\m}\d^\n{}_{[\t}\d^{\r]}{}_{\s]},\\
\G^{\m\n\r\s}\G_{\t\l}&=\G^{\m\n\r\s}{}_{\t\l}+8\G^{[\m\n\r}{}_{[\l}\d^{\s]}{}_{\t]}+12\G^{[\m\n}\d^\r{}_{[\l}\d^{\s]}{}_{\t]},\\
\G^{\m\n\r}\G_{\s\t\l}&=\G^{\m\n\r}{}_{\s\t\l}+9\G^{[\m\n}{}_{[\t\l}\d^{\r]}{}_{\s]}+18\G^{[\m}{}_{[\l}\d^\n{}_{\t}\d^{\r]}{}_{\s]}+6\d^{[\m}{}_{[\l}\d^\n{}_{\t}\d^{\r]}{}_{\s]},\\
\G^{\m_1\ldots\m_r\n_1\ldots\n_s}\G_{\n_s\ldots\n_1}&=\frac{(D-r)!}{(D-r-s)!}\G^{\m_1\ldots\m_r},\\
\G^{\m\r}\G_{\r\n}&=(D-2)\G^{\m}{}_\n+(D-1)\d^\m_\n,\\
\G^{\m\n\r}\G_{\r\s}&=(D-3)\G^{\m\n}{}_\s+2(D-2)\G^{[\m}\d^{\n]}{}_\s,\\
\G^{\m\n\l}\G_{\l\r\s}&=(D-4)\G^{\m\n}{}_{\r\s}+4(D-3)\G^{[\m}{}_{[\s}\d^{\n]}{}_{\r]}+2(D-2)\d^{[\m}{}_{[\s}\d^{\n]}{}_{\r]},\\
\G_{\m\r}\G^{\r\s\t}\G_{\t\n}&=(D-4)^2\G_\m{}^\s{}_\n+(D-4)(D-3)\(\G_\m\d_\n^\s-\G^\s g_{\m\n}\)\NO\\
&\hskip0.5cm+(D-3)(D-2)\d_\m^\s\G_\n-(D-3)\G^\s\G_{\m\n},\\
\G_\r \G^{\m_1\m_2\ldots \m_p}\G^\r&=(-1)^p(D-2p)\G^{\m_1\m_2\ldots \m_p}.
\eal 
\esub

\subsection{Radial ADM decomposition}

The radial ADM decomposition of the dynamical variables consists in 
choosing a suitable radial coordinate $r$ emanating from the conformal boundary and describing the bulk space as a foliation by constant $r$-slices, which we denote by $\S_r$. Accordingly, all fields with tensor indices are decomposed along the radial and transverse directions as
\bal\label{ADM-decomposition}
&ds^2=g_{\m\n}dx^\m dx^\n=(N^2+N^i N_i)dr^2+2N_i dr dx^i+\g_{ij}dx^idx^j,\NO\\
&A=a dr+A_i dx^i,\qquad \J=\J_r dr+\J_i dx^i,
\eal
where $\g_{ij}$, $A_i$ and $\J_i$ are dynamical induced fields on the slices $\S_r$, while $N$, $N_i$, $a$ and $\J_r$, are non-dynamical Lagrange multipliers conjugate to first class constraints. 

\paragraph{Fefferman-Graham gauge} Since $N$, $N_i$, $a$ and $\J_r$, are non-dynamical Lagrange multipliers, once the corresponding first class constraints have been derived, we can set them to any fixed value that is convenient. A choice that is particularly useful for reading off the holographic dictionary is the Fefferman-Graham (FG) gauge
\be \label{FG-gauge}
N=1,\qquad N^i=0,\qquad a=0,\qquad \J_r=0.
\ee
Whenever we gauge-fix the Lagrange multipliers in this paper we adopt this choice of gauge.

\paragraph{Vielbein and gamma matrices}

The vielbeins compatible with the ADM metric \eqref{ADM-decomposition} can be expressed in terms of a $(d+1)$-vector $n^\a$ and a $(d+1)\times d$ matrix $e^\a_j$ as
\be\label{bulk-vielbein}
E^\a=\left(N n^\a+N^j e_j^\a\right)dr+e_j^\a dx^j,
\ee
so that
\be
g_{\m\n}=E_\m^\a E_\n^\b\h_{\a\b},\qquad \g_{ij}=e_i^\a
e_j^\b\h_{\a\b},\qquad n_\a e_i^\a=0,\qquad \h_{\a\b}n^\a n^\b
=1.
\ee
The inverse vielbeins are then given by 
\be
E_\a^r=\frac{1}{N}n_\a,\qquad E_\a^i=e_\a^i-\frac{N^i}{N}n_\a,
\ee
and satisfy the orthogonality relation 
\be
e_\a^i e_i^\b+n_\a n^\b=\d_\a^\b.
\ee
In order to simplify the calculations we sometimes choose a convenient vielbein frame so that
\be\label{default-frame}
n_\a=(1,0),\quad e_\a^i=(0,e_a^i),\quad e_i^\a=(0,e_i^a),
\ee
and $e_i^a$ becomes the vielbein on the slice $\S_r$.

\paragraph{Gamma matrix decomposition and radiality}

The decomposition of the vielbein allows us to decompose the gamma matrices in radial and transverse components as follows:  
\be
\G^r=\G^\a E_\a^r=\frac1N n_\a\G^\a\equiv \frac1N \G,\qquad \G^i=\G^\a E_\a{}^i=\Hat\G^i-\frac{N^i}{N}\G,\qquad \Hat\G^i\equiv \G^\a e_\a^i.
\ee
These relations imply that 
\bal\label{Gamma-id}
\G^{ri_1i_2\ldots i_n} & =E_{\a_1}^{\;r}E_{\a_2}^{\;i_1}\cdots E_{\a_{n+1}}^{\;i_n}\G^{\a_1\a_2\ldots \a_{n+1}}\NO\\
& =\frac 1N n_{\a_1} \(e_{\a_2}^{\;i_1}-\frac{N^{i_1}}{N}n_{\a_2}\)\ldots \(e_{\a_{n+1}}^{\;i_n}-\frac{N^{i_n}}{N}n_{\a_{n+1}}\)\G^{\a_1\a_2\ldots \a_{n+1}}\NO\\
&=\frac 1N \G\;\Hat \G^{i_1i_2\ldots i_n}.
\eal
Moreover, the Hermiticity property \eqref{Hermiticity-bulk} translates into the identities  
\be\label{Hermiticity-slice} 
\G^{\dagger}=\G^{\lbar t}\G\;\G^{\bar t},\qquad \Hat\G^{i\dagger}=\G^{\lbar t}\Hat\G^i\G^{\bar t},
\ee  
while the Clifford algebra \eqref{Clifford-bulk} becomes
\be 
\{\Hat\G^i,\Hat\G^j\}=2\g^{ij},\quad \{\Hat\G^i,\G\}=0.
\ee 

The fact that $\G$ anticommutes with all gamma matrices $\Hat\G^i$ allows us to introduce the {\em radiality} projectors and the corresponding radially projected spinors \cite{Freedman:2016yue} 
\be\label{radiality}
\G_\pm\equiv \frac12\(1\pm\G\),\qquad \j_\pm\equiv \G_\pm \j.
\ee
Note that the radiality projectors are independent of the spacetime dimension, but coincide with the usual chirality projectors for odd $D=d+1$. Moreover, splitting the spinors according to their radiality is necessary not only for formulating a consistent Hamiltonian description of the fermion dynamics \cite{Kalkkinen:2000uk}, but also in order to construct the asymptotic Fefferman-Graham expansions, since fermions of different radiality have different asymptotic behavior.   

\paragraph{Levi-Civita tensor} The Levi-Civita tensor in the $(d+1)$-dimensional bulk is given by
\be
\e^{\m_1\cdots\m_{d+1}}=E_{\a_1}^{\m_1}\cdots E_{\a_{d+1}}^{\m_{d+1}}\e^{\a_1\cdots\a_{d+1}}.
\ee
In the coordinate system \eqref{ADM-decomposition} it takes the form 
\be
\e^{ri_1\cdots i_d}=\frac 1N n_{\a_0}\(e_{\a_1}^{i_1}-\frac{N^{i_1}}{N}n_{\a_1}\)\cdots\(e_{\a_d}^{i_d}-\frac{N^{i_d}}{N}n_{\a_d}\)\e^{\a_0\cdots \a_d}=\frac 1N\Hat\e^{i_1\cdots i_d},
\ee
where $\Hat\e^{i_1\cdots i_d}$ is the Levi-Civita tensor on $\S_r$ and 
\be
\Hat\e^{\a_1\cdots\a_d}\equiv n_{\a_0}\e^{\a_0\cdots\a_d}.
\ee

The Levi-Civita tensor is related with the antisymmetric products of the gamma matrices through the identity 
\be
\G^{\m_1\m_2\ldots \m_D}\propto \e^{\m_1\m_2\ldots \m_D}.
\ee
In particular, for $D=d+1$ with $d$ even (see (3.31) in \cite{Freedman:2012zz})
\be
\Hat\G^{i_1i_2\ldots i_d}=i^{d/2+1}\Hat\e^{i_1i_2\ldots i_d}\G.
\ee
It follows that
\be
\G^{ri_1i_2\ldots i_d}=
\frac1N\G\Hat\G^{i_1i_2\ldots i_d}
=\frac{i^{d/2+1}}{N}\G^2\Hat\e^{i_1i_2\ldots i_d}
=\frac{i^{d/2+1}}{N}\Hat\e^{i_1i_2\ldots i_d}=i^{d/2+1}\e^{ri_1i_2\ldots i_d},
\ee
and hence 
\be\label{5D-5Gs}
\G^{\m_1\m_2\ldots \m_D}= i^{d/2+1}\e^{\m_1\m_2\ldots \m_D}, \quad D=d+1\,\, \text{odd}.
\ee

\paragraph{Radial decomposition of the Christoffel and spin connections} 

We further need the ADM decomposition of the Christoffel symbol and of the spin connection. A straightforward calculation using the metric \eqref{ADM-decomposition} determines the bulk Christoffel symbol:
\bea\label{Christoffel}
&&\G^r_{rr}=N^{-1}(\dot{N}+N^i\pa_i N-N^iN^jK_{ij}),\NO\\
&&\G^r_{ri}=N^{-1}\left(\pa_iN-N^jK_{ij}\right),\NO\\
&&\G^r_{ij}=-N^{-1}K_{ij},\NO\\
&&\G^i_{rr}=-N^{-1}N^i\dot{N}-ND^iN-N^{-1}N^iN^j\pa_jN+\dot{N}
^i+N^jD_jN^i+2NN^jK^i_j+N^{-1}N^iN^kN^lK_{kl},\NO\\
&&\G^i_{rj}=-N^{-1}N^i\pa_jN+D_jN^i+N^{-1}N^iN^kK_{kj}+NK^i_j,\NO\\
&&\G^k_{ij}=\G^k_{ij}[\g]+N^{-1}N^kK_{ij},
\eea
where 
\be 
K_{ij}=\frac{1}{2N}(\dot\g_{ij}-D_i N_j-D_j N_i),
\ee 
is the extrinsic curvature of the radial slice $\S_r$ and $K\equiv \g^{ij}K_{ij}$.

The torsion free spin connection is expressed in terms of the vielbein as 
\be\label{spin-connection}
\o_{\m\a\b} =E_{\n\a}\pa_\m E_\b^\n+\G^{\r}_{\m\n}E_{\r\a}E_\b^\n.
\ee
Using the above decompositions of the vielbein and of the Christoffel symbol we determine the components of the spin connection (see also (88) and (89) in \cite{Kalkkinen:2000uk}):
\bsub
\label{spin-connection-components}
\bal
\o_{r\a\b} =&\,E_{\n\a}\dot E_\b{}^\n+\G^{\r}_{r\n}E_{\r\a}E_\b{}^\n\NO\\
=&\,
E_{r\a}\pa_r E_\b{}^r+E_{i\a}\pa_r
E_\b{}^i+\G^{r}_{rr}E_{r\a}E_\b{}^r+\G^{r}_{ri}E_{r\a}E_\b{}^i+\G^{i}_{rr}E_{i\a
} E_\b{}^r+\G^{i}_{rj}E_{i\a}E_\b{}^j\NO\\
=&\,n_{[\a}\dot n_{\b]}+e_{i[\a}\dot e_{\b]}{}^i+2n_{[\a}e_{\b]}{}^i\left(\pa_i
N-N^jK_{ji}\right)
-D_iN_je_{[\a}{}^ie_{\b]}{}^j,\\\NO\\
\o_{i\a\b} =&\,E_{\n\a}\pa_i E_\b{}^\n+\G^{\r}_{i\n}E_{\r\a}E_\b{}^\n\NO\\
=&\,E_{r\a}\pa_i E_\b{}^r+E_{j\a}\pa_i
E_\b{}^j+\G^{r}_{ir}E_{r\a}E_\b{}^r+\G^{r}_{ij}E_{r\a}E_\b{}^j+\G^{j}_{ir}E_{j\a
} E_\b{}^r+\G^{k}_{ij}E_{k\a}E_\b{}^j\NO\\
=&\,n_\a\pa_i n_\b+e_{j\a}\pa_i
e_\b{}^j+\G^k_{ij}[\g]e_{k\a}e_\b{}^j+2K^j_ie_{j[\a}n_{\b]}.
\eal
\esub
Choosing the convenient frame \eqref{default-frame}, these identities allow us to write 
\bsub
\bal 
& \o_{iab}=e_{ja}\pa_i e_b^{\;j}+\G^k_{\;ij}[\g]e_{ka}e_b^{\;j}\equiv\Hat\o_{iab},\\
& \o_{i\a\b}\G^{\a\b}=\Hat\o_{iab}\G^{ab}+2K_{ji}e_\a^{\;j}n_\b
\G^{\a\b}=\Hat\o_{iab}\G^{ab}+2 K_{ji}\Hat\G^j\G,\\
& \o_{r\a\b}\G^{\a\b} =e_{i\a}\dot e_\b^{\;i}\G^{\a\b}+2\G\Hat\G^i\(\pa_i N-N^j K_{ji}
\)-\Hat\G^{ij}D_i N_j.
\eal
\esub
It follows that the components of the covariant derivative acting on the gravitino take the form  
\bsub
\bal
& \nabla_i \J_j=\bb D_i \J_j+\frac 12K_{li}\Hat \G^l\G\J_j+\frac 1N K_{ij}(\J_r-N^k\J_k) ,\\
& \nabla_i \J_r=\bb D_i \J_r+\frac 12 K_{ji}\Hat \G^j\G\J_r-\G^j_{ir}\J_j-\G^r_{ir}\J_r,\\
& \nabla_r\J_i=\dot \J_i+\frac 14\( e_{ai}\dot e_b^{i}\G^{ab}+2\G\Hat\G^j (\pa_j
N-N^l K_{lj})-\Hat\G^{jl}D_j N_l \)\J_i-\G^j_{ir}\J_j-\G^r_{ir}\J_r,
\eal
\esub
where
\bsub\label{spinor-derivatives} 
\bal
& \bb D_i \J_j=\pa_i \J_j +\frac
14\Hat\o_{iab}\G^{ab}\J_j-\G^k_{ij}[\g]\J_k,\\
& \bb D_i \J_r=\pa_i \J_r +\frac 14\Hat\o_{iab}\G^{ab}\J_r,
\eal
\esub
are the covariant derivatives of, respectively, a vector and a scalar spinor on the radial slice $\S_r$.

\section{Generalized PBH transformations}
\label{sPBH}
\setcounter{equation}{0}

Penrose-Brown-Henneaux (PBH) transformations \cite{Penrose:1987uia,Brown:1986nw,Imbimbo:1999bj} are bulk diffeomorphisms that preserve the Fefferman-Graham gauge  \eqref{FG-gauge}. In addition to diffeomorphisms, the supergravity action \eqref{action} is invariant under local supersymmetry and $U(1)$ gauge transformations. Moreover, local $SO(1,d)$ frame rotations not only leave the action \eqref{action} invariant, but also automatically preserve the FG gauge \eqref{FG-gauge}. However, as we discussed in appendix \ref{conventions}, it is particularly useful to make a specific choice of frame defined by the conditions \eqref{default-frame}. In combination with the FG gauge \eqref{FG-gauge}, these conditions amount to requiring that  
\be \label{FG-strong}
E_r^{\bar r}=1,\qquad E_r^a=0,\qquad E_i^{\bar r}=0,\qquad a=0,\qquad \J_r=0.
\ee
We will refer to this gauge fixing condition as the {\em strong} Fefferman-Graham gauge. It is equivalent to the FG gauge \eqref{FG-gauge}, but also partially gauge fixes the frame rotations. In the remaining of this appendix we will determine the most general local bulk transformations that preserve the gauge conditions \eqref{FG-strong}. Such transformations generalize the PBH diffeomorphisms and have a central role in constructing the holographic dictionary, and in particular, in determining how supersymmetry acts on the boundary.

We start by considering the transformation of all bulk fields under infinitesimal diffeomorphisms $\x^\m$, frame rotations $\l^{\a\b}=-\l^{\b\a}$, $U(1)$ gauge transformations $\th$, as well as local supersymmetry transformations $\e$. The transformation of the fields under local supersymmetry transformations is given in \eqref{SUSY-trans}. Under bulk diffeomorphisms 
\be\label{diff-trans}
\d_\x E_\m^\a =\x^\n\pa_\n E_\m^\a+(\pa_\m \x^\n) E_\n^\a,\quad
\d_\x A_\m  =\x^\n\pa_\n A_\m+(\pa_\m \x^\n) A_\n,\quad
\d_\x\J_\m =\x^\n\pa_\n\J_\m+(\pa_\m\x^\n)\J_\n,
\ee
while under local frame rotations 
\be\label{frame-trans}
\d_\l E_\m^\a =-\l^\a{}_\b E_\m^\b ,\qquad
\d_\l A_\m =0,\qquad
\d_\l\J_\m =-\frac14\l^{\a\b}\G_{\a\b}\J_\m,
\ee
and under $U(1)$ gauge transformations
\be\label{gauge-trans}
\d_\th E_\m^\a =0 ,\qquad
\d_\th A_\m =\pa_\m\th,\qquad
\d_\th\J_\m =-i\frak g\th\J_\m.
\ee
Requiring that the total transformation of all the fields preserves the conditions \eqref{FG-strong} leads to the following set of differential equations for the local parameters: 
\bsub
\label{PBHeqs}
\bal
&\dot\x^r=0,\\
&\dot\x^i e^a_i-\l^a{}_{\bar r} =0,\\
&(\pa_i\x^r)-\l^{\bar r}{}_b e^b_i+\frac12\(\lbar\e\G\J_i-\lbar\J_i\G\e\)=0,\\
&\dot \x^i A_i+\dot\th=0,\\
&\dot \x^i \J_i+\dot\e+\frac14e_{ia}\dot e_b^{\;i}\Hat\G^{ab}\e+ic_4\(\G\Hat\G^{ij}F_{ij}-2(d-2)\Hat\G^i \dot A_i\)\e-\frac{1}{2\ell}\G\e=0.
\eal
\esub
The last equation can be simplified by decomposing the gravitino and the spinor parameter $\e$ using the radiality projectors \eqref{radiality}, namely
\be
\J_{\pm i}\equiv \G_\pm\J_i,\qquad \e_\pm\equiv\G_\pm\e.
\ee
This leads to the two spinor equations
\bsub
\bal
&\dot \x^i \J_{+i}+\dot\e_++\frac14e_{ia}\dot e_b^{\;i}\Hat\G^{ab}\e_++ic_4\(\Hat\G^{ij}F_{ij}\e_+-2(d-2)\Hat\G^i \dot A_i\e_-\)-\frac{1}{2\ell}\e_+=0,\\
&\dot \x^i \J_{-i}+\dot\e_-+\frac14e_{ia}\dot e_b^{\;i}\Hat\G^{ab}\e_--ic_4\(\Hat\G^{ij}F_{ij}\e_-+2(d-2)\Hat\G^i \dot A_i\e_+\)+\frac{1}{2\ell}\e_-=0.
\eal 
\esub 
Moreover, combining the second and third equations in \eqref{PBHeqs} using $\l^{\bar r}{}_b=-\l^a{}_{\bar r}\h_{ab}$, gives
\be
\pa_i\x^r+\g_{ij}\dot\x^j+\frac12\(\lbar\e\G\J_i-\lbar\J_i\G\e\)=0.
\ee

The above equations can be solved to determine the general form of the local transformations preserving the strong FG gauge conditions \eqref{FG-strong}. The general solution contains a number of integration functions and takes the form (see also section C of \cite{Amsel:2009rr})
\bsub
\label{PBHsols}
\bal
\x^r&=\s(x),\\
\x^i & =\x_o^i(x)-\int ^r dr'\g^{ij}(r',x)\(\pa_j\s+\frac12\(\lbar\e_-\J_{+i}-\lbar\e_+\J_{-i}-\lbar\J_{-i}\e_++\lbar\J_{+i}\e_-\)\),\\
\l^a{}_{\bar r}&=-e^a_i\g^{ij}\(\pa_j\s+\frac12\(\lbar\e_-\J_{+i}-\lbar\e_+\J_{-i}-\lbar\J_{-i}\e_++\lbar\J_{+i}\e_-\)\),\\
\l^a{}_b &=\l_o^a{}_b(x)+\cdots,\\
\th &=\th_o(x)-\int ^r dr'\g^{ij}(r',x)\(\pa_j\s+\frac12\(\lbar\e_-\J_{+i}-\lbar\e_+\J_{-i}-\lbar\J_{-i}\e_++\lbar\J_{+i}\e_-\)\)A_i(r',x),\\
\e_+ &=\X_+\e_{o+}(x)+\X_+i\sqrt{\frac{d-2}{2(d-1)}}\int^r dr' \X_+^{-1}(r',x)\(\Hat\G^i\dot A_i\e_-+\co(\J^2)\),\\
\e_- &=\X_-\e_{o-}(x)+\X_-i\sqrt{\frac{d-2}{2(d-1)}}\int^r dr' \X_-^{-1}(r',x)\(\Hat\G^i\dot A_i\e_++\co(\J^2)\),
\eal
\esub
where
\be\label{Xi}
\X_\pm=\exp\int^rdr'\(\pm\frac{1}{2\ell}-\frac14e_{ia}\dot e_b^{\;i}\Hat\G^{ab}\mp\frac{i}{\sqrt{8(d-2)(d-1)}}\Hat\G^{ij}F_{ij}+\co(\J^2)\),
\ee
and $\s(x)$, $\x_o^i(x)$, $\l_o^a{}_b(x)$, $\th_o(x)$, and $\e_{o\pm}(x)$ are arbitrary functions of the transverse coordinates only. 

Note that in determining the solution for $\l^a{}_{\bar r}$ and $\e_\pm$ we used the leading asymptotic form of the fields for a general asymptotically locally AdS$_{d+1}$ background, namely 
\be\label{leading-asymptotics}
e^a_i\sim e^{r/\ell}e_{(0)i}^a(x),\qquad A_i\sim A_{(0)i}(x),\qquad \J_{\pm i}\sim e^{\pm r/2\ell}\J_{(0)\pm i}(x).
\ee
In particular, although the gauge fixing conditions \eqref{FG-strong} allow $\l^a{}_{\bar r}(r,x)$ to be arbitrary, preserving the asymptotic form of the vielbein $e_i^a$ requires that $\l^a{}_{\bar r}(r,x)\sim \l_o^a{}_{\bar r}(x)$, up to subleading terms that remain arbitrary, but do not affect any physical observable. Moreover, the expressions for $\e_\pm$ look formal since $\e_\pm$ enter on the r.h.s. of the last two equations in \eqref{PBHsols} as well. However, this dependence on the r.h.s. is asymptotically subleading, and so are the last two terms in the expression for $\X_\pm$ in \eqref{Xi}. The expressions for $\e_\pm$, together with that for $\X_\pm$ in \eqref{Xi}, therefore, allow one to recursively determine the asymptotic solution of the spinors $\e_\pm$. To leading order asymptotically the general solution \eqref{PBHsols} takes the form
\be\label{PBH-asympt-sols}
\x^r=\s(x),\qquad \x^i\sim\x_o^i(x),\qquad \l^a{}_b \sim\l_o^a{}_b(x),\qquad
\th \sim \th_o(x),\qquad
\e_\pm \sim e^{\pm r/2\ell}\e_{o\pm}(x),
\ee
with
\be
\l^a{}_{\bar r}\sim-e^{-r/\ell}e^a_{(0)i}g_{(0)}^{ij}\Big(\pa_j\s(x)+\frac12\(\lbar\e_{o-}\J_{(0)+i}-\lbar\e_{o+}\J_{(0)-i}-\lbar\J_{(0)-i}\e_{o+}+\lbar\J_{(0)+i}\e_{o-}\)\Big).
\ee
Notice that each local transformation is sourced by an arbitrary function of the transverse coordinates. As is discussed in the main part of the paper, these parameters correspond to the local symmetries of the dual field theory in a background of arbitrary sources. To determine how the local sources transform under these local symmetries we next consider how the generalized PBH transformations act on the induced fields on $\S_r$, in the limit that $r\to\infty$.

Under local bulk transformations that preserve the strong FG gauge \eqref{FG-strong} the induced fields on the radial slices $\S_r$ transform as 
\bsub
\label{PBH-induced}
\bal
\d_{\x,\l,\th,\e} e_i^a &=\s\dot e_i^a+ \x^j\pa_j e_i^a+(\pa_i \x^j) e_j^a-\l^a{}_b e_i^b+\frac12(\lbar\e\Hat\G^a\J_i-\lbar\J_i\Hat\G^a\e),\\
\d_{\x,\l,\th,\e} A_i &=\s\dot A_i+\x^j\pa_j A_i+(\pa_i \x^j)A_j+\pa_i\th+ic_3\(\lbar\J_i\e-\lbar\e\J_i\),\\
\d_{\x,\l,\th,\e}\J_{i} &=\s\dot\J_i+\x^j\pa_j\J_{i}+(\pa_i\x^j)\J_{j}-\frac14\l^{\a\b}\G_{\a\b}\J_{i}
+\cd_i\e+\frac12 K_{ji}\Hat\G^j\G\e-\frac{1}{2\ell}\Hat\G_i\e\NO\\
&\hskip0.5cm+ic_4\([\Hat\G_i{}^{jk}-2(d-2)\Hat\G^k \d^j_i]F_{jk}-2[\Hat\G_i{}^{j}-(d-2)\d_i^j]\G\dot A_j\)\e-i\frak g\th\J_i,
\eal
\esub
where 
\be
\cd_i\equiv\bb D_i+i\frak g A_i.
\ee
Using the leading asymptotic form of the induced fields in eq.~\eqref{leading-asymptotics}, and of the local symmetry parameters in eq.~\eqref{PBH-asympt-sols}, we find that to leading order asymptotically  
\bsub
\label{PBH-sources-cutoff}
\bal
\d_{\x,\l,\th,\e} e_i^a &\sim \frac{\s}{\ell} e_i^a+ \x^j_o\pa_j e_i^a+(\pa_i \x^j_o) e_j^a-\l_o^a{}_b e_i^b+\frac12(\lbar\e_+\G^a\J_{+i}-\lbar\J_{+i}\G^a\e_+),\\
\d_{\x,\l,\th,\e} A_i &\sim \x^j_o\pa_j A_i+(\pa_i \x^j_o)A_j+\pa_i\th_o+ic_3\(\lbar\J_{+i}\e_-+\lbar\J_{-i}\e_+-\lbar\e_+\J_{-i}-\lbar\e_-\J_{+i}\),\\
\d_{\x,\l,\th,\e}\J_{+i} &\sim\frac{\s}{2\ell}\J_{+i}+\x^j_o\pa_j\J_{+i}+(\pa_i\x^j_o)\J_{+j}-\frac14\l_o^{ab}\G_{ab}\J_{+i}
+\cd_i\e_+-\frac{1}{\ell}\Hat\G_i\e_--i\frak g\th_o\J_{+i}.
\eal
\esub
Taking the limit $r\to\infty$, these expressions result in the transformations of the field theory sources given in eq.~\eqref{PBH-sources}. Finally, projecting the gravitino transformation in \eqref{PBH-induced} with the $P_-$ radiality projector gives 
\bal\label{psi-minus-trans}
\d_{\x,\l,\th,\e}\J_{-i}&\sim -\frac{\s}{2\ell}\J_{-i}+\x^j_o\pa_j\J_{-i}+(\pa_i\x^j_o)\J_{-j}-\frac14\l_o^{ab}\Hat\G_{ab}\J_{-i}\NO\\
&\hskip0.5cm+\cd_i\e_-+\frac12 K_{(2)ji}\Hat\G^j\e_++ic_4(\Hat\G_i{}^{jk}-2(d-2)\Hat\G^k \d^j_i)F_{jk}\e_+-i\frak g\th_o\J_{-i},
\eal
where 
\be
K_{(2)ij}=\frac{\ell}{d-2}\(R_{ij}[\g]-\frac{1}{2(d-1)}R[\g]\g_{ij}\)+\co(\J^2).
\ee
From eq.~\eqref{Psi-} follows that this corresponds to the leading asymptotic transformation of the canonical momentum $\p_{\lbar \J}^i$.

\subsection{Ward identities as generators of generalized PBH transformations}

In section \ref{dict} we argued that the generalized PBH transformations are directly related with the Ward identities \eqref{WIDs}, which were obtained from the first class constraints \eqref{constraints}. Specifically, varying the renormalized on-shell action \eqref{Sren} with respect to arbitrary PBH transformations of the form \eqref{PBH-sources} provides an alternative derivation of the Ward identities. As we now demonstrate, the connection between the Ward identities and generalized PBH transformations admits an elegant formulation on the symplectic space of sources and local operators, which can be employed in order to obtain the transformations of the local operators under PBH transformations. 

Inserting the Fefferman-Graham expansions in the symplectic form associated with the bulk action \eqref{action} \cite{Crnkovic:1986ex,Lee1990}, it can be shown that the field theory sources and the corresponding operators defined through \eqref{ops} parameterize a symplectic manifold equipped with the symplectic 2-form 
\be\label{symplectic-form}
\O=\int d^dx\(\d e_i^a\sub{0}\wedge \d\P^i_a+\d A\sub{0}_i\wedge\d\P^i+\d\lbar\J_{(0)+i}\wedge\d\P_{\lbar\J}^i+\d\P_{\J}^i\wedge\d\J_{(0)+i}\),
\ee
where 
\be
\P^i_a=-\sqrt{-g\sub{0}}\;\ct^i_a,\qquad \P^i_a=\sqrt{-g\sub{0}}\;\cj^i,\qquad \P_{\lbar\J}^i=\sqrt{-g\sub{0}}\;\cs^i,\qquad \P_{\J}^i=\sqrt{-g\sub{0}}\;\lbar\cs^i.
\ee
The symplectic form \eqref{symplectic-form} allows us to introduce the Poisson bracket \cite{Papadimitriou:2010as} 
\be\label{PB}
\hskip-0.2cm\{\ca,\cb\}_{\text{\tiny PB}}=\int d^dx\(\frac{\d \ca}{\d e_{(0)i}^a}\frac{\d \cb}{\d\P_a^i}+\frac{\d \ca}{\d A_{(0)i}}\frac{\d \cb}{\d\P^i}+\cb\frac{\overleftarrow\d }{\d\P_{\lbar\J}^i}\frac{\d \ca}{\d \lbar\J_{(0)+i}}+ \ca\frac{\overleftarrow\d}{\d \J_{(0)+i}}\frac{\d \cb}{\d\P_{\J}^i}-\ca\leftrightarrow \cb\),
\ee
for any functions $\ca$ and $\cb$ on the space of sources and local operators. It should be emphasized that, although this result is obtained here holographically from the bulk theory, it applies to any local quantum field theory where the local operators are defined as in the local renormalization group \cite{Osborn:1991gm}. Indeed, the subsequent analysis in this appendix applies to any local quantum field theory \cite{Papadimitriou:2016yit}, independently of whether it admits a holographic dual. 

The Ward identities \eqref{WIDs} imply that the local functions 
\bsub
\bal
\cw_{Mi}=&\; D_{(0)j}(e^{a}_{(0)i}\P_a^j+\P_{\J}^j\J_{(0)+i}+\lbar\J_{(0)+i}\P_{\lbar\J}^j)-\P_\J^j\cd_{(0)i}\J_{(0)+j}-\lbar\J_{(0)+j}\overleftarrow{\cd }_{\hskip-0.08cm(0)i}\P_{\lbar\J}^j-F_{(0)ij}\P^j\NO\\
&+\sqrt{-g\sub{0}}\;\ca_{M(0)i},\hskip-0.0cm\\
\cw_{R}=&\;D_{(0)i}\P^i +i\frak g(\P_\J^i\J_{(0)+i}-\lbar\J_{(0)+i}\P_{\lbar\J}^i)-\sqrt{-g\sub{0}}\;\ca_{R(0)},\\
\cw_{S}=&\; \cd_{(0)i}\P_{\lbar\J}^i-\frac12\P^i_a\G^a\J_{(0)+i}-\frac{ic_2}{2(d-2)}\P^i(\Hat\G_{(0)ij}-(d-2)g_{(0)ij})\Hat\G^{jpq}_{(0)}\cd_{(0)p}\J_{(0)+q}\NO\\
&-\sqrt{-g\sub{0}}\;\ca_{S(0)},\\
\cw_{W}=&\;\frac1\ell\(-e_{(0)i}^a\P_a^i-\frac12\lbar\J_{(0)+i}\P_{\lbar\J}^i-\frac12\P_{\J}^i\J_{(0)+i}-\sqrt{-g\sub{0}}\;\ca_{W(0)}\),\\
\cw_{sW}=&\;\frac1\ell\(-\Hat\G_{(0)i}\P_{\lbar\J}^i+\frac{i(d-1)c_2}{2}\P^i\J_{(0)+i}+\sqrt{-g\sub{0}}\;\ca_{sW(0)}\),\\
\cw_{L}^{ab}=&\;-e^{i[a}_{(0)}\P^{b]}_i+\frac14(\P_\J^i\G^{ab}\J_{(0)+i}-\lbar\J_{(0)+i}\G^{ab}\P_{\lbar\J}^i),
\eal
\esub
vanish identically on the symplectic space of sources and local operators. It is straightforward to show that the Poisson bracket of the generating function 
\bal\label{PBH-gen}
\cc[\s,\x_o,\th_o,\l_o,\e_{o+},\e_{o-}]=&\;\int d^dx\Big(\s\cw_W+\x_o^i\cw_{Mi}+\th_o\cw_R+\l_{o\;ab}\cw_L^{ab}\NO\\
&\hskip0.6in+\lbar\e_{o+}\cw_S+\lbar\e_{o-}\cw_{sW}+\lbar\cw_S\e_{o+}+\lbar\cw_{sW}\e_{o-}\Big),
\eal
with the field theory sources reproduces the PBH transformations \eqref{PBH-sources}. Specifically, one finds  
\bsub
\label{PBH-sources-PB}
\bal
 \{\cc[\s,\x_o,\th_o,\l_o,\e_{o+},\e_{o-}],e_i^a\sub{0}\}_{\text{\tiny PB}}=&\;-\frac{\d\cc}{\d\P^i_a}=\d_{\x_o,\l_o',\th_o',\e_o}e_i^a\sub{0},\\
\{\cc[\s,\x_o,\th_o,\l_o,\e_{o+},\e_{o-}],A_{(0)i}\}_{\text{\tiny PB}}=&\;-\frac{\d\cc}{\d\P^i}=\d_{\x_o,\l_o',\th_o',\e_o}A_{(0)i},\\
\{\cc[\s,\x_o,\th_o,\l_o,\e_{o+},\e_{o-}],\J_{(0)+i}\}_{\text{\tiny PB}}=&\;-\frac{\d\cc}{\d\P_{\J}^i}=\d_{\x_o,\l_o',\th_o',\e_o}\J_{(0)+i},
\eal
\esub
where
\be\label{PBH-shifted}
\l_{o}'^{ab}=\l_o^{ab}-\x_o^k\o_{(0)k}^{ab},\qquad \th_o'=\th_o-A_{(0)k}\x_o^k.
\ee
It follows that the Ward identities are first class constraints on the space of sources and local operators generating the corresponding local symmetries through the Poisson bracket \eqref{PB}.

This observation can be utilized in order to determine the transformation of the local operators,\footnote{The transformation of the local operators can also be obtained by using the explicit form of the Fefferman-Graham expansions as in \cite{Imbimbo:1999bj}, but the Poisson bracket derivation has the advantages that it applies to any local field theory, since it only requires knowledge of the Ward identities, and it is practically simpler.} which are given by the Poisson brackets 
\bsub
\label{PBH-momenta}
\bal
\d_{\s,\x_o,\l_o',\th_o',\e_{o+},\e_{o-}} \P^i_a &=\{\cc[\s,\x_o,\th_o,\l_o,\e_{o+},\e_{o-}],\P^i_a\}_{\text{\tiny PB}}=\frac{\d\cc}{\d e^a_{(0)i}},\\
\d_{\s,\x_o,\l_o',\th_o',\e_{o+},\e_{o-}} \P^i &=\{\cc[\s,\x_o,\th_o,\l_o,\e_{o+},\e_{o-}],\P^i\}_{\text{\tiny PB}}=\frac{\d\cc}{\d A_{(0)i}},\\
\d_{\s,\x_o,\l_o',\th_o',\e_{o+},\e_{o-}}\P_{\lbar\J}^i &=\{\cc[\s,\x_o,\th_o,\l_o,\e_{o+},\e_{o-}],\P_{\lbar\J}^i\}_{\text{\tiny PB}}=\frac{\d\cc}{\d \lbar\J_{(0)+i}}.
\eal
\esub
Since these transformations are obtained by evaluating the functional derivatives of the generating function \eqref{PBH-gen} with respect to the sources, they get contributions from the anomalies! In particular, the {\em anomalous} contributions to the operator transformations are given by
\be\label{current-anomalies}
\d_{\s,\x_o,\l_o',\th_o',\e_o}^{\ca} \ct^i_a =\frac{-1}{e\sub{0}}\frac{\d\cc^\ca}{\d e^a_{(0)i}},\quad
\d_{\s,\x_o,\l_o',\th_o',\e_o}^{\ca} \cj^i =\frac{1}{e\sub{0}}\frac{\d\cc^\ca}{\d A_{(0)i}},\quad
\d_{\s,\x_o,\l_o',\th_o',\e_o}^{\ca}\cs^i =\frac{1}{e\sub{0}}\frac{\d\cc^\ca}{\d \lbar\J_{(0)+i}},
\ee
where $e\sub{0}\equiv\sqrt{-g\sub{0}}$ and 
\bal\label{PBH-anomalies}
\cc^{\ca}[\s,\x_o,\th_o,\e_{o+},\e_{o-}]=&\;\int d^dx\sqrt{-g\sub{0}}\Big(-\frac{\s}{\ell}\ca_{W(0)}-\th_o\ca_{R(0)}\NO\\
&\hskip0.6in-\lbar\e_{o+}\ca_{S(0)}+\frac1\ell\lbar\e_{o-}\ca_{sW(0)}-\lbar\ca_{S(0)}\e_{o+}+\frac1\ell \lbar\ca_{sW(0)}\e_{o-}\Big).
\eal
These anomalous contributions to the transformation of the local operators are central to the analysis of the symmetries preserved by the quantum field theory vacua.   

Finally, the Poisson bracket of the generating functions \eqref{PBH-gen} can be used to determine the infinite dimensional superalgebra of generalized PBH transformations, as is done e.g. in eq.~(3.2.35) of \cite{Papadimitriou:2010as} for local Weyl transformations and diffeomorphisms. However, in order for the algebra to close in the present context, it is necessary to know the bulk supergravity action and the corresponding supersymmetry transformations to all orders in the gravitino.

\section{Supersymmetry transformation of the bulk action}
\label{bulk-susy}
\setcounter{equation}{0}

In this appendix we show that the supersymmetry variations \eqref{SUSY-trans} leave the action \eqref{action} invariant up to boundary terms that we determine explicitly. We emphasize again that throughout our analysis we keep terms up to quadratic order in the gravitino only. For the purpose of this appendix it is convenient to keep the constants $c_1$, $c_2$, $c_3$, $c_4$ and $\frak g$ in the action \eqref{action} and in the supersymmetry transformations \eqref{SUSY-trans} arbitrary. As we shall see, invariance of \eqref{action} under the supersymmetry transformations \eqref{SUSY-trans} determines all these parameters, up to an overall sign. It is also useful to note that, since there is no torsion, the Christoffel symbol in the gravitino covariant derivative in \eqref{action} can be omitted, thus simplifying the calculation.

\paragraph{Bosonic terms} The supersymmetry variation of the bosonic terms in the action \eqref{action} is: 
\bsub
\bal
\d\(\sqrt{-g}\;2\L\)&=\sqrt{-g}\;\L\lbar\e\G^\m\J_\m+\tx{h.c.},\\
\d\(\sqrt{-g}R\)&=-\sqrt{-g}(R_{\m\n}-\frac12Rg_{\m\n})\lbar\e\G^\m\J^\n\NO\\
&\hskip0.5cm+\pa_\m\(\sqrt{-g}\nabla_\n(\lbar\e\G^{(\m}\J^{\n)})-\sqrt{-g}\nabla^\m(\lbar\e\G^\n\J_\n)\)+\tx{h.c.},\\
\d\(\sqrt{-g}F^2\)&=\sqrt{-g}\(\frac12F^2g^{\m\n}-2F^{\m\r}F^\n{}_\r\)\lbar\e\G_\m\J_\n-4ic_3\sqrt{-g}\;F^{\m\n}\nabla_\m(\lbar\e\J_\n)+\tx{h.c.},\\
\hskip-0.1cm\e^{\m\n\r\s\l}\d\(F_{\m\n}F_{\r\s}A_\l\)& =-ic_3\e^{\m\n\r\s\l}\(3 F_{\m\n} F_{\r\s}\lbar\e\J_\l+4\nabla_\m(F_{\r\s}A_\l\lbar\e\J_\n)\)+\tx{h.c.}.
\eal
\esub

\paragraph{Fermionic terms} The variation of the fermionic terms requires extensive use of the gamma matrix identities \eqref{Gamma-IDs}. After some algebra these variations take the form 
\bsub
\bal
\d\(\sqrt{-g}\;\lbar\J_\m\G^{\m\n\r}\overleftrightarrow\nabla_\n\J_\r\)&=\sqrt{-g}\;\d\lbar\J_\m\G^{\m\n\r}\nabla_\n\J_\r-\sqrt{-g}\;\nabla_\n(\d\lbar\J_\m)\G^{\m\n\r}\J_\r+\tx{h.c.}\NO\\
&=2\sqrt{-g}\;\d\lbar\J_\m\G^{\m\n\r}\nabla_\n\J_\r-\pa_\n(\sqrt{-g}\;\d\lbar\J_\m\G^{\m\n\r}\J_\r)+\tx{h.c.}\NO\\
&=\pa_\m(2\sqrt{-g}\;\lbar\e\G^{\m\n\r}\nabla_\n\J_\r-\sqrt{-g}\;\d\lbar\J_\n\G^{\n\m\r}\J_\r)+\frac{(d-1)}{\ell}\sqrt{-g}\;\lbar\e\G^{\m\n}\nabla_\m\J_\n\NO\\
&\hskip0.5cm-\sqrt{-g}\;(R_{\m\n}-\frac12Rg_{\m\n})\lbar\e\G^\m\J^\n-2i\frak g \sqrt{-g}\;A_\m\lbar\e\G^{\m\n\r}\nabla_\n\J_\r\NO\\
&\hskip0.5cm+2(d-1)ic_4\sqrt{-g}\;F^{\s\t}\lbar\e(-\G_{\t\s}{}^{\n\r}+2\d_{[\t}{}^{[\r}\d_{\s]}{}^{\n]})\nabla_\n\J_\r+\tx{h.c.},\\
\d(\sqrt{-g}\;\lbar\J_\m\G^{\m\n}\J_\n)&=\pa_\m(\sqrt{-g}\;\lbar\e\G^{\m\n}\J_\n)-\sqrt{-g}\;\lbar\e\G^{\m\n}\nabla_\m\J_\n+\frac{d}{2\ell}\sqrt{-g}\;\lbar\e\G^\m\J_\m\\
&\hskip0.5cm-i\frak g\sqrt{-g}\; A_\m\lbar\e\G^{\m\n}\J_\n+ic_4\sqrt{-g}\;F^{\s\t}\lbar\e\((d-2)\G_{\s\t}{}^\n-2\G_\s\d_\t^\n\)\J_\n+\tx{h.c.},\NO\\
\d(\sqrt{-g}\;\lbar\J_\m\G^{\m\n\r}A_\n\J_\r) &=\pa_\m(\sqrt{-g}\;\lbar\e\G^{\m\n\r}A_\n\J_\r)-\frac12\sqrt{-g}\;F_{\m\n}\lbar\e\G^{\m\n\r}\J_\r+\sqrt{-g}\;\lbar\e\G^{\m\n\r}A_{\m}\nabla_\n\J_\r\NO\\
&\hskip0.5cm+\frac{d-1}{2\ell}\sqrt{-g}\;\lbar\e\G^{\m\n}A_\m\J_\n\NO\\
&\hskip0.5cm+i(d-1)c_4\sqrt{-g}\;F^{\s\t}\lbar\e(-\G_{\t\s}{}^{\n\r}+2\d_{[\t}{}^{[\r}\d_{\s]}{}^{\n]})A_\n\J_\r+\tx{h.c.},\\
\d(\sqrt{-g}\;F^{\m\n}\lbar\J_\m\J_\n)&=\sqrt{-g}\;F^{\m\n}(\lbar\e \overleftarrow\nabla_\m)\J_\n+\frac{1}{2\ell}\sqrt{-g}\;F^{\m\n}\lbar\e(\G_\m-2i\ell\frak g A_\m)\J_\n\NO\\
&\hskip0.5cm+ic_4\sqrt{-g}\;F^{\m\n}F_{\r\s}\lbar\e\(\G^{\s\r}{}_\m-2(d-2)\d^\r_\m\G^\s\)\J_\n+\tx{h.c.},\\
\d\(\sqrt{-g}\;\lbar\J_\m\G^{\m\n\r\s}\J_\n F_{\r\s}\)&=\pa_\m(\sqrt{-g}\;\lbar\e\G^{\m\n\r\s}\J_\n F_{\r\s})
-\sqrt{-g}\;\lbar\e\G^{\m\n\r\s}(\nabla_\m\J_\n) F_{\r\s}\NO\\
&\hskip0.5cm+\frac{d-2}{2\ell}\sqrt{-g}\;\lbar\e\G^{\n\r\s}\J_\n F_{\r\s}-i\frak g\sqrt{-g}\;\lbar\e\G^{\m\n\r\s}\J_\n A_\m F_{\r\s}\\
&\hskip0.5cm-ic_4\sqrt{-g}\;\lbar\e(d\G_{\l\k}{}^{\n\r\s}+2\d_{\k}^{\n}\G_{\l}{}^{\r\s}-6(d-2)\d_{\k}^{[\n}\d_{\l}^{\r}\G^{\s]})\J_\n F_{\r\s}F^{\k\l}+\tx{h.c.}.\NO
\eal
\esub

Combining these transformations we find that the total supersymmetry variation is a pure boundary term provided the constants $c_1$, $c_2$, $c_3$, $c_4$ take the values
\be
c_1=\frac{\mp2\ell}{3\sqrt{3}}\d_{d,4},\quad c_2=\frac{\pm\ell}{\sqrt{2(d-1)(d-2)}},\quad c_3=\pm\sqrt{\frac{d-1}{8(d-2)}},\quad c_4=\frac{\pm1}{\sqrt{8(d-1)(d-2)}},
\ee
while the gauge coupling is given by
\be
\frak g=\pm\frac1\ell\sqrt{\frac{(d-1)(d-2)}{2}}.
\ee
The signs in these expressions are correlated so that there are only two possible choices: either the upper sign or the lower sign must be chosen for all constants. Choosing the upper sign leads to the action \eqref{action} and the supersymmetry variations \eqref{SUSY-trans}. Note also that the constant $c_1$ is non-zero only for $d=4$, i.e. for $D=5$.

\addcontentsline{toc}{section}{References}


\bibliographystyle{jhepcap}
\bibliography{references}

\end{document}